\documentclass[fleqn,usenatbib]{mnras}
\usepackage{newtxtext,newtxmath} 
\usepackage[T1]{fontenc}
\DeclareRobustCommand{\VAN}[3]{#2}
\let\VANthebibliography\thebibliography
\def\thebibliography{\DeclareRobustCommand{\VAN}[3]{##3}\VANthebibliography}
\usepackage{graphicx}	
\usepackage{amsmath}	
\usepackage{graphicx,color}
\usepackage{comment}
\usepackage{amsmath}
\usepackage{pifont}
\usepackage{afterpage}
\usepackage{xcolor}
\usepackage{caption}
\DeclareCaptionLabelFormat{bf}{\textbf{Table #2}}
\usepackage{natbib,twoopt}
\usepackage[hyphenbreaks]{breakurl}
\newcommand {\magg}{\hphantom{>}} 
\newcommand{\xmark}{\ding{55}}%
\DeclareMathAlphabet\mathzapf       {T1}{pzc} {mb} {it}
\bibpunct{(}{)}{;}{a}{}{,}             
\definecolor{cobalt}{rgb}{0.06, 0.2, 0.65}
\hypersetup{
  colorlinks,
  citecolor=cobalt,
  linkcolor=cobalt,
  urlcolor=cobalt,
}
\makeatletter
  \newcommandtwoopt{\citeads}[3][][]{\href{http://adsabs.harvard.edu/abs/#3}%
    {\def\hyper@linkstart##1##2{}%
     \let\hyper@linkend\@empty\citealp[#1][#2]{#3}}}
  \newcommandtwoopt{\citepads}[3][][]{\href{http://adsabs.harvard.edu/abs/#3}%
    {\def\hyper@linkstart##1##2{}%
     \let\hyper@linkend\@empty\citep[#1][#2]{#3}}}
  \newcommandtwoopt{\citetads}[3][][]{\href{http://adsabs.harvard.edu/abs/#3}%
    {\def\hyper@linkstart##1##2{}%
     \let\hyper@linkend\@empty\citet[#1][#2]{#3}}}
  \newcommandtwoopt{\citeyearads}[3][][]%
    {\href{http://adsabs.harvard.edu/abs/#3}
    {\def\hyper@linkstart##1##2{}%
     \let\hyper@linkend\@empty\citeyear[#1][#2]{#3}}}
\makeatother

\title{Characterization of a sample of $\gamma$-ray active galactic nuclei}

\author[A. Ulgiati et al.]{Alberto Ulgiati$^{1,2,3}$\thanks{E-mail:
ulgiati.alberto@gmail.com}, Paolo Padovani$^{3,4}$\thanks{E-mail:
ppadovan@eso.org}, Paolo Giommi$^{5,6,7}$, Simona Paiano$^{1}$, Ciro Pinto$^{1}$\\
$^{1}$INAF - IASF Palermo, via Ugo La Malfa, 153, I-90146, Palermo, Italy \\ 
$^{2}$Universit\`a degli Studi di Palermo, Dipartimento di Fisica e Chimica, via Archirafi 36, I-90123 Palermo, Italy\\
$^{3}$European Southern Observatory, Karl-Schwarzschild-Stra{\ss}e 2, D-85748 Garching bei M\"unchen, Germany\\
$^{4}$Associated to INAF - Osservatorio di Astrofisica e Scienza dello Spazio, Via Piero 
Gobetti 93/3, I-40129 Bologna, Italy\\
$^{5}$Institute for Advanced Study, Technische Universit\"at M\"unchen, Lichtenbergstrasse 2a, D-85748 Garching bei München, Germany\\
$^{6}$Center for Astrophysics and Space Science (CASS), New York University Abu Dhabi, PO Box 129188 Abu Dhabi, United Arab Emirates\\
$^{7}$Associated to INAF, Osservatorio Astronomico di Brera, via Brera, 28, I-20121 Milano, Italy\\
}

\date{Accepted 2025 September 1. Received 2025 August 30; in original form 2025 February 5}

\pubyear{\the\year{}}

\begin{document}
\label{firstpage}
\pagerange{\pageref{firstpage}--\pageref{lastpage}}
\maketitle

\begin{abstract}
We analyse 77 \textit{Fermi} sources and their potential low-energy counterparts previously proposed in the literature. These sources were classified as active galactic nuclei, mainly blazars, based on optical spectroscopy. The main goals of this work are to examine these associations, classify the blazars based on their multi-wavelength spectral energy distributions (SEDs), and identify potential masquerading BL Lac objects. Through SED analysis, we assess whether the multi-wavelength emission follows the characteristic double-peaked curve of blazars. Additionally, we propose the region of origin of the emission at different wavelengths, investigate the correlation between $\gamma$-ray and lower-energy emission, and classify objects as low-, intermediate-, high- or extreme high synchrotron peaked (LSP, ISP, HSP, E-HSP) blazars. We search for masquerading BL Lacs, a class of flat-spectrum radio quasars where broad emission lines are swamped by non-thermal jet emission. The multi-wavelength analysis revealed that the 64 radio-loud sources in our sample exhibit an SED with a double-peak structure, typically ascribed to jet activity. Based on the synchrotron peak, 46 are HSP, 11 as ISP, and 7 as LSP. We also found 9--18 masquerading BL Lac candidates ($\approx$15--30\% of the radio-loud sample). For the 13 radio-quiet UGSs, the SEDs do not exhibit the double-peak structure typical of jetted AGN. Further analysis ruled out star formation as the origin of the observed $\gamma$-ray emission, making its reconciliation with lower-energy emission challenging. We explored alternative counterparts, identifying low-energy matches for 7 sources, with no plausible counterparts found for the others.

\end{abstract}




\section{Introduction}
\label{sec:introduction}
Active Galactic Nuclei (AGN) are astrophysical sources, in which a supermassive black hole (SMBH) accretes matter at the center of a galaxy, producing strong and variable emission over the entire electromagnetic spectrum, from the radio band to very high energy (VHE) $\gamma$-rays \citep[][]{Padovani_2017}. This gives us different windows on the physics of the various AGN sub-structures.
In fact, the infrared (IR) band is mostly sensitive to obscuring material and dust, the optical/ultraviolet (UV) band is related to emission from the accretion disk, while the X-ray band traces the emission of a (putative) corona. $\gamma$-ray and (high flux density) radio samples, on the other hand, preferentially select jetted AGN emitting strong non-thermal (jet [or associated lobe] related) radiation \citep[e.g.][and references therein]{Padovani_2017}, although emission from these relativistic jets extends over the whole electromagnetic spectrum. 

Blazars are a subclass of AGN where the jet is aligned toward the observer \citep[][]{Urry_1995,Giommi_2013}. Their emission is affected by Doppler boosting and is largely jet-dominated, while their spectral energy distribution (SED) is characterized by a two-bump structure: the first peak extends from the radio to the X-ray band and is due to synchrotron emission; the second peak dominates in the X-rays and $\gamma$-rays but its origin is still under debate. The main theoretical models predict a leptonic (inverse Compton) or hadronic origin (synchrotron emission of protons or decays of neutral pions generated by $p-p$ or $p-\gamma$ interactions: e.g. \citealt{Cerruti_2015, costomante2018, Rodrigues2019, Gao2019, Cerruti2020}). Because of the non-thermal emission, the SED can be mainly described by a power-law, f($\nu$) $\propto$ $\nu^{-\alpha}$ (in the f($\nu$) vs $\nu$ diagram), with a high-energy cut-off, where $\alpha$ varies across the spectrum. 

Traditionally, blazars are divided in two main classes, based on differences in their optical spectra: Flat Spectrum Radio Quasars (FSRQs), characterised by prominent emission lines (Equivalent Width, EW $ > \, 5~\mathring{A}$), and BL Lacertae objects (BLL) which have weak or absent spectral lines \citep[e.g.][]{Falomo_2014}. However, there are some FSRQs whose very bright, Doppler boosted jet continuum washes out their emission lines. These objects, which therefore appear to be BLLs, are instead intrinsically FSRQs and have been called masquerading BLL \citep[][]{Giommi_2013}.
These can be identified using a number of parameters, as detailed in \cite{Padovani_2019,Padovani_2022,Paiano_2023} (see also Section \ref{sec:masq}).

According to the position of the synchrotron peak ($\nu_p^s$) blazars can also be divided in four sub-classes: low-synchrotron peaked (LSP), when $\nu_p^s$ < 10$^{14}$ Hz, intermediate-synchrotron peaked (ISP), when 10$^{14}$ Hz < $\nu_p^s$ < 10$^{15}$ Hz, high-synchrotron peaked (HSP), when 10$^{15}$ Hz < $\nu_p^s$ < 10$^{17}$ Hz and extreme high-synchrotron peaked (E-HSP), when $\nu_p^s$ > 10$^{17}$ Hz \citep[][]{Padovani_1995,Abdo_2010_b,Costamante_2020}. Fig. \ref{fig:SED_LSP_HSP} shows a template of the typical SEDs of LSPs and HSPs.

\afterpage{
\begin{figure}
\hspace{-0.5cm}
\centering
\includegraphics[width=9truecm]{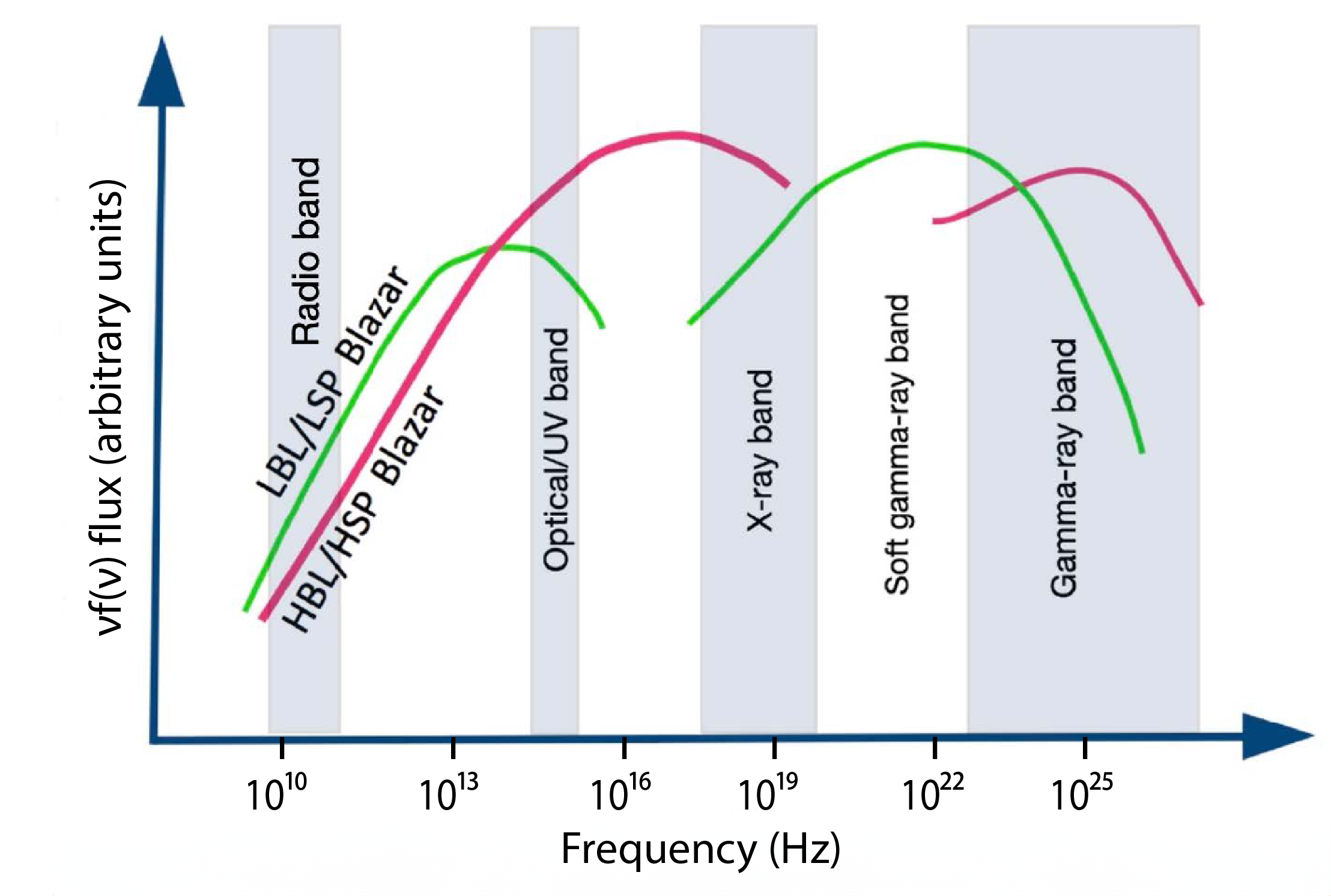}
\caption[Caption for LOF]{Template of SEDs for LSPs (green) and HSPs (red), highlighting the different energy bands. Adapted from a figure reported in the online software \textit{Firmamento}\footnotemark \citep[see][for details on the software]{Tripathi_2024}.}
\label{fig:SED_LSP_HSP}
\end{figure}
\footnotetext{https://firmamento.hosting.nyu.edu/home}
}


Given their characteristics, blazars represent the largest extra-galactic population of the $\gamma$-ray sky. This energy band is therefore the most efficient to search for new blazars, and the {\it Fermi} observatory is ideally suitable to detect blazars in large numbers thanks to its Large Area Telescope (LAT), a $\gamma$-ray detector for photons from 20 MeV up to 300 GeV. Over the years the LAT has produced several catalogues, reporting in them the sources revealed with increasingly longer observation time. On July 2023, the \textit{Fermi} collaboration published the latest version of its catalogue \citep[4FGL-DR4][]{4FGL_cat,4FGL_DR4}, based on 14 years of observations and containing 7195 sources. 4765 of them have already been associated or identified with targets at other wavelengths thanks to a positional overlap in the sky, measurements of correlated variability at other wavelengths, and/or multi-wavelength spectral properties.

Blazars, however, are not the only subclass of AGN capable of emitting $\gamma$-ray photons. The $\gamma$-ray sky is also populated by a small, but possibly astrophysically relevant, fraction of Seyfert galaxies and radio galaxies \citep[][]{Cheung_2010,Abdo_2010,Ackermann_2011,Grandi_2012,Paliya_2015,Angioni_2017,Rieger_2017,Jarvela_2021,Ye_2023,4FGL_DR4}. $\gamma$-ray emission can also come from non-active galaxies. For example, in star-forming and starburst galaxies, cosmic rays accelerated by supernova remnants can produce $\gamma$-ray radiation through inelastic collisions with ambient gas particles and subsequent $\pi^0$-decay \citep[][and references therein]{Peng_2019}. In this case $\gamma$-ray emission appears to be relatively steady over time, at variance with blazars and jetted AGN in general, where instead it appears to be quite variable \citep[][]{4FGL_DR4}.

Within the 4FGL-DR4 catalogue, there is still a significant fraction of sources, $\sim$ 30\%, that are not yet associated and classified, the so called Unassociated Gamma-ray Sources (UGSs). Their association is however important since they could hide new blazars and/or new AGN  capable of emitting $\gamma$-ray photons in the energy range probed by {\it Fermi} (such as radio galaxies or Seyfert-like objects, for which only a few sources are known to be $\gamma$-ray emitters). Furthermore, they appear to have lower $\gamma$-ray fluxes on average, and therefore could represent higher redshift objects whose study is particularly important for evolutionary analyses \citep[e.g.][]{Ajello_2014,Ghisellini_2017}. 

In the last decade a strong effort has been done in order to unveil new AGN  among the UGSs of the different {\it Fermi} catalogues published over the years. 
Since the $\gamma$-ray sky is dominated by blazars, much research has been directed at this class of objects. Statistical algorithms \citep[][]{Ackermann_2012,Mirabal_2012,Mao_2013} and machine learning \citep[][]{Doert_2014,Salvetti_2017b,Kaur_2019,Kaur_2019b,Kerby_2021,Kaur_2023} searches have been developed in order to associate the UGS, based on $\gamma$-ray spectral features and variability information, with the aim of distinguishing between blazar-like objects and pulsar-like ones (the two most numerous classes of sources in the catalogues). Other works were focused on the study of lower energy sources found in the $\gamma$-ray uncertainty regions, such as \citet[][]{Falcone_2011,Takahashi_2012,Landi_2015,Salvetti_2017a,Kaur_2019,Kaur_2019b,Kerby_2021,Kaur_2023} and \citet[][]{Marchesini_2020} in the X-ray band, \citet[][]{Dabrusco_2013} and \citet[][ and references therein]{Massaro_2015} in the infrared (IR) band and \citet[][]{Fujinaga_2014,Fujinaga_2016,Takahashi_2013} in the radio band, while a multi-wavelength approach has been adopted by \citet[][]{Paiano_SED,Fronte_2023}. 

In any case, the firm classification of UGSs comes from the study of the optical spectra of their counterparts. \citet[][]{Shaw_2012,Shaw_2013} have studied the spectra of about 500 sources of the second {\it Fermi} catalogue, confirming the BLL classification and estimating for them a redshift or a redshift lower limit. Based on the selection made using the colours of the IR counterparts in the Wide-field Infrared Survey Explorer (WISE) survey, \citet[][]{Dabrusco_2013} and \citet[][ and references therein]{Massaro_2015}, have studied the optical spectra of a sample of AGN candidates, finding that most of them were blazars. 

The association and classification of the UGSs is a fundamental goal of our research group. 
As the years have passed and deeper {\it Fermi} catalogues have been published, we have increasingly focused on identifying AGN emitting $\gamma$-ray radiation within the low energy counterparts of the UGSs. 
Since we are interested in searching for AGN, we focused on UGSs with high galactic latitude ($|b|>10^{\circ}$), which have a greater probability to be extra-galactic.
This systematic investigation began with \cite{ paiano2017_ufo1, Paiano_SED, paiano2019_ufo2}, using the the 2FGL and 3FGL catalogues, and has continued with the 4FGL-DR4 catalogue, as discussed in \citet[][]{Ulgiati_2024b}, Paiano et al., in prep., \cite{Ulgiati_2024}.
In 2015, we started an (still on-going) observational spectroscopic campaign, aimed at the study and classification of $\gamma$-ray blazars and other AGN classes conducted with large-aperture (8-10 m class) telescopes that, due to their high efficiency and spectral resolution, allow us to have the sensitivity needed to detect faint features, expected particularly in BLLs.

In this paper, we provide a detailed characterization of a sample of 77 $\gamma$-ray sources from the second, third and fourth \textit{Fermi} catalogues \citep[2FGL, 3FGL, 4FGL-DR4:][]{Nolan_2012,Acero_2015,4FGL_DR4}, that were initially classified as UGSs. Our analysis is based primarily on the association and classification results presented by our group in \citet[][]{paiano2017_ufo1,paiano2019_ufo2,Ulgiati_2024}. These works include sources newly associated and classified for the first time by us, as well as others whose associations and classifications largely agree with independent analyses available in the literature. Specifically, we investigate the emission of these objects across the entire electromagnetic spectrum through a comprehensive analysis of their SEDs. This allows us to study the association between the low-energy counterparts and the $\gamma$-ray detections, infer the locations of the emissions at different wavelengths, and classify blazars into the sub-classes LSP, ISP, HSP and E-HSP. 
Additionally, we search for potential masquerading BL Lacs, a peculiar class of blazars that exhibit a BL Lacs optical spectrum, while being high-excitation galaxies. These objects benefit from several external radiation fields, which provide additional targets for protons and could enhance neutrino production compared to the low-excitation galaxies.
This is performed by comparing the absolute and relative emission intensities to threshold values designed to identify this subclass of sources, following the criteria in \citet[][]{Padovani_2019}.

The paper is structured as follows: in Sec. \ref{sec:associations_P}, the association and classification criteria adopted in \citet[][]{paiano2017_ufo1,paiano2019_ufo2,Ulgiati_2024} are described; in Sec. \ref{sec:SED_building}, the SEDs are constructed; in Sec. \ref{sec:gamma_origin}, the dependence of $\gamma$-ray luminosity and 1.4 GHz radio power is tested to investigate the possibility that a jet is the origin of the $\gamma$-ray emission; the results are discussed in Sec. \ref{sec:discussion_P}, where the origin of the $\gamma$-ray emission is debated; finally, the conclusions are reported in Sec. \ref{sec:conclusion_P}.

\section{Source multi-wavelength associations}
\label{sec:associations_P}

The process of searching for lower energy (compared to the $\gamma$-ray band) counterparts started from the identification of an X-ray source within the 3$\sigma$ \textit{Fermi} error boxes of the UGSs \citep[see][for more details]{paiano2017_ufo1,paiano2019_ufo2,Ulgiati_2024,Ulgiati_2024b}. In this framework, the semi-major and semi-minor axes of the 4FGL-DR4 95\% confidence ellipses were increased by 50\%, corresponding to a $\sim$99\% confidence level region. For this research we analysed \textit{Swift}/XRT data covering the $\gamma$-ray positions. Starting from the X-ray coordinates and uncertainties, we looked for radio and optical counterparts that are positionally coincident \citep[see details in][]{Paiano_SED,Ulgiati_2024b}. 

For the search for radio counterparts, three main catalogues  were used: the NRAO VLA Sky Survey \citep[NVSS][]{Condon_1998}, the Very Large Array Sky Survey (VLASS:  \citealt{Lacy_2020}) and the Rapid Australian SKA Pathfinder (ASKAP) Continuum Survey (RACS: \citealt{Hale_2021}). Radio images provided by the LOw-Frequency ARray (LOFAR) Two-metre Sky Survey \citep[LoTSS, ][]{LoTSS_2022} in the northern hemisphere and RACS in the southern hemisphere were also analyzed, with the aim of identifying any uncatalogued sources \citep[][]{Ulgiati_2024}. Dedicated observations were carried out using the Australian Telescope Compact Array (ATCA) on a sample of 20 potential X-ray counterparts of extra-galactic UGSs ($|b|$ > 10$^{\circ}$), not coincident with a radio source \citep[][]{Ulgiati_2024}. Full details on the ATCA program, are presented in \citet[][]{Ulgiati_2024b}. 
The search for optical sources involved the United States Naval Observatory \citep[USNO,][]{Monet_2003}, the Sloan Digital Sky Survey \citep[SDSS, ][]{Ahumada_2020}, the Panoramic Survey Telescope and Rapid Response System \citep[PanSTARRS, ][]{Chambers_2016} database, the Dark Energy Survey \citep[DES, ][]{Abbott_2021} and the SuperCOSMOS Sky Survey  \citep[SSS, ][]{Hambly_2004} catalogues. 

\subsection{Radio-loud and radio-quiet source in our UGS sample}

A characteristic that distinguishes blazars and radio galaxies is their brightness in the radio band. The radio-loudness parameter (R) is defined as the ratio between the radio flux (in the range 2 - 4 GHz) and the optical flux of an object. Sources with R~$>$~10 are considered radio-loud \citep[][]{Kellermann_1989}. Blazars and radio galaxies are radio-loud sources \citep[e.g.][]{Padovani_2017, 4FGL_DR4}, while Seyfert galaxies and quasi-stellar objects (QSOs) are typically radio-quiet \citep[][and reference therein]{Padovani_2017}. 

We estimate R values for our objects (Table \ref{tab:MW_par}). For sources without a cataloged radio counterpart and lacking a flux estimate in the literature, we established an upper limit on their flux. For sources with a declination greater than -40$^{\circ}$, we used the detection threshold given by the VLASS  catalogue (0.345 mJy at 5$\sigma$), while for sources below this declination, we used the detection threshold from the RACS  catalogue (1.5 mJy at 5$\sigma$). A total of 13 sources are found to be radio-quiet (see Tab. \ref{tab:MW_par}). This is an unexpected and potentially interesting result since they could represent a type of $\gamma$-ray objects that is still relatively unexplored. Having a multi-wavelength view of these objects is bound to shed light on the matter. Hence, we constructed the SEDs for all objects in the sample. 

\begin{table*}
\begin{center}
\caption{Multi-wavelength parameters for the objects under analysis.} 
\resizebox{16cm }{!}{
\begin{tabular}{llcrrcrrrcclc}
\hline 
4FGL Name & Counterpart & Association & $f^{\rm radio}_{\nu}$ & $f^{\rm opt}_{\nu}$ & Fractional variability & $R$ & L$_{\gamma}~~~$ & P$_{\rm 1.4GHz}~$ & Classification & Classification & Redshift & Redshift\\ 
 & & Reference & & & $\gamma$-ray & & erg s$^{-1}$ & W Hz$^{-1}$ & & Reference & & Reference \\
\hline
 & & &  & [$\times$ 10$^{-28}$] & \\
\hline
4FGL J0004.0+0840 & SDSS J000359.23+084138.2 & PS19 & \magg16.5 & \magg3.0 & 0.5 $\pm$ 0.5 & 545 & >46.2 & >26.0 & BLL & PS19 & >1.5035 & PS19\\
4FGL J0006.4+0135 & XRT J000626.92+013610.3 & CY17 & \magg\magg5.8 & \magg2.8 & - & 212 & 45.6 & 25.0 & BLL & PS19 & \magg0.787 & PS19\\
4FGL J0023.6$-$4209 & XRT J002303.5-420509.6 & UA24 & \magg\magg1.4 & 208.9 & 0.7 $\pm$ 0.3 & 1 & 42.8 & 22.2 & SY2 & UA24 & \magg0.053 & JD09, UA24\\
3FGL J0031.6+0938 & XRT J003159.86+093618.4 & PS19 & \magg<0.4 & \magg7.6 & - & <5 & 44.9 & <22.9 & NLSY1 & PS19 & \magg0.2207 & PS19\\
4FGL J0049.1+4223 & XRT J004859.10+422351.0 & PS17 & \magg\magg4.5 & \magg4.8 & - & 95 & 44.7 & 24.0 & BLL & PS17 & \magg0.302 & PS17\\
4FGL J0102.4+0942 & XRT J010217.10+094409.5 & TY13 & \magg10.0 & \magg6.9 & 0.5 $\pm$ 0.3 & 145 & 45.2 & 24.7 & BLL & PS17b & \magg0.42 & PS17\\
4FGL J0112.0+3442 & XRT J011124.86+344154.1 & UA24 & \magg41.3 & \magg6.3 & - & 655 & 44.9 & 25.2 & BLL & UA24 & \magg0.3997 & SDSS16, UA24\\
4FGL J0117.9+1430 & XRT J011804.79+143159.6 & UA24 & \magg<1.8 & 12.0 & 0.5 $\pm$ 0.2 & <15 & 44.1 & <23.1 & NLSY1 & AS03 & \magg0.129 & SDSS16, UA24\\
4FGL J0158.8+0101 & XRT J015852.77+010132.8 & MF13, PL13 & \magg45.8 & \magg1.0 & - & 4583 & 45.5 & 25.4 & BLL & PS19 & \magg0.4537 & PS19\\
4FGL J0202.7+3133 & XRT J020242.13+313211.4 & UA24 & \magg15.1 & 12 & - & 126 & >44.6 & >24.7 & BLL & UA24 & \textbf{>0.35$^*$} & this work\\
4FGL J0234.3$-$0628 & XRT J023410.27-062825.6 & CY19 & \magg\magg4.8 & \magg4.8 & - & 101 & >45.6 & >24.7 & BLL & PS19 & >0.63 & PS19\\
4FGL J0238.7+2555 & XRT J023853.80+255407.1 & PS17 & \magg10.6 & \magg3.3 & - & 321 & 45.3 & 25.0 & BLL &  PS17 & \magg0.584 & PS17\\
4FGL J0251.1$-$1830 & XRT J025111.70-183111.1 & CY19 & \magg\magg8.8 & \magg3.3 & 0.2 $\pm$ 0.4 & 291 & >45.4 & >25.0 & BLL & PS19 & >0.615 & PS19\\
4FGL J0259.0+0552 & XRT J025857.56+055244.4 & PS19 & \magg\magg5.8 & 13.2 & 0.5 $\pm$ 0.1 & 44 & >46.3 & >24.9 & BLL & PS19 & >0.7$^{*}$ & PS19\\
4FGL J0305.1$-$1608 & XRT J030515.00-160816.6 & TY13 & \magg35.6 & 10.0 & 0.3 $\pm$ 0.2 & 356 & 45.0 & 25.0 & BLL & PS17 & \magg0.312 & PS17\\
4FGL J0338.5+1302 & XRT J033829.20+130215.7 & DR14 & \magg13.4 & 16.0 & 0.4 $\pm$ 0.1 & 85 & >45.9 & >24.7 & BLL & ME16 & >0.382 & PS17\\
4FGL J0409.8$-$0359 & XRT J040946.50-040003.5 & MF13 & \magg55.6 & \magg4.8 & 0.5 $\pm$ 0.1 & 1162 & >46.0 & >25.9 & BLL & TY13 & >0.7$^{*}$ & PS17\\
4FGL J0414.6$-$0842 & XRT J041433.08-084206.7 & PS19 & \magg28.4 & \magg7.6 & 0.6 $\pm$ 0.4 & 374 & >44.8 & >25.0 & BLL & PS19 & >0.35$^{*}$ & PS19\\
4FGL J0506.9+0323 & XRT J050650.14+032358.6 & CY19 & \magg18.4 & 11.0 & - & 168 & >43.9 & >23.7 & BLL & PS19 & >0.1$^{*}$ & PS19\\
4FGL J0641.4+3349 & XRT J064111.24+334502.0 & UA24 & \magg\magg1.2 & 52.5 & - & 2 & 44.1 & 23.2 & SY1 & UA24 & \magg0.1657 & MT16\\
4FGL J0644.6+6039 & XRT J064435.70+603851.3 & MF13 & \magg20.4 & \magg8.3 & 0.5 $\pm$ 0.1 & 245 & >45.8 & >25.3 & BLL & PS17 & >0.581 & PS17\\
4FGL J0838.5+4013 & XRT J083902.98+401546.9 & UA24 & \magg24.0 & 17.4 & - & 138 & 43.9 & 24.4 & BLG & ME15 & \magg0.1945 & SDSS9, UA24\\
4FGL J0848.7+7017 & XRT J084839.52+701728.0 & PS19 & \magg26.1 & \magg4.0 & - & 656 & >46.3 & >26.0 & BLL & PS19 & >1.2435 & PS19\\
4FGL J0930.5+5132 & XRT J093033.36+513214.6 & PS19 & \magg\magg7.3 & \magg3.3 & - & 220 & 43.9 & 23.8 & BLL & PS19 & \magg0.1893 & PS19\\
4FGL J0937.9$-$1434 & XRT J093754.70-143350.4 & MF13, PL13 & \magg15.3 & 13.0 & 0.6 $\pm$ 0.2 & 116 & 44.8 & 24.5 & BLL & PS19 & \magg0.287 & PS19\\
4FGL J0938.8+5155 & XRT J093834.50+515454.8 & KS21 & \magg\magg0.1 & \magg2.8 & 0.3 $\pm$ 0.4 & 3 & 44.9 & 22.9 & QSO & AF17 & \magg0.4168 & AF17, UA24\\
4FGL J0952.8+0712 & XRT J095249.50+071329.9 & PS17 & \magg25.9 & \magg6.9 & 0.5 $\pm$ 0.3 & 374 & 45.2 & 25.4 & BLL & PS17 & \magg0.574 & PS17\\
4FGL J1016.1$-$4247 & XRT J101620.78-424723.2 & PL13 & \magg\magg7.8 & \magg5.2 & 0.0 $\pm$ 1.5 & 149 & >45.8 & >25.0 & BLL & RM23 & \textbf{>0.65$^*$} & this work\\
4FGL J1039.2+3258 & XRT J103852.17+325651.9 & UA24 & \magg\magg6.1 & \magg4.8 & - & 128 & 45.0 & 24.2 & BLL & PR10 & \textbf{\magg0.32$^*$} & this work\\
4FGL J1049.5+1548 & XRT J104939.30+154837.6 & PA14 & \magg30.0 & 63.0 & 0.2 $\pm$ 0.1 & 48 & 45.4 & 24.9 & BLL & PA14 & \magg0.326 & PA14\\
4FGL J1049.8+2741 & XRT J104938.70+274212.1 & CY19 & \magg\magg7.2 & 19.1 & 0.1 $\pm$ 1.1 & 38 & 43.9 & 23.6 &  BLG & dR19 & \magg0.144 & dR19\\
4FGL J1125.1+4811 & XRT J112526.01+480922.8 & UA24 & \magg<0.3 & \magg2.8 & 0.5 $\pm$ 0.9 & <13 & 45.9 & <24.2 & QSO & UA24 & \magg1.649 & PI18, UA24\\
4FGL J1128.8+3757 & SDSS J112903.20+375656.7 & AF13 & \magg28.3 & \magg1.7 & 0.8 $\pm$ 0.2 & 1631 & >46.6 & >26.1 & BLL & PS17b & >1.211 & PS17b\\
4FGL J1131.6+4657 & XRT J113142.36+470009.2 & UA24 & \magg73.9 & 36.3 & 0.6 $\pm$ 0.3 & 204 & 43.8 & 24.5 & BLG & UA24 & \magg0.1255 & SDSS16, UA24\\
4FGL J1146.0$-$0638 & XRT J114600.87-063853.9 & CY19, PS19 & \magg\magg5.8 & \magg5.8 & - & 100 & 45.7 & 24.8 & BLL & PS19 & \magg0.6407 & PS19\\
4FGL J1223.5+0818 & XRT J122327.49+082030.4 & CY19, PS19 & \magg\magg9.5 & 10.0 & 0.6 $\pm$ 0.2 & 95 & >45.8 & >25.1 & BLL & PS19 & >0.7187 & PS19\\
4FGL J1223.9+7954 & XRT J122358.10+795328.6 & MF13 & \magg37.2 & \magg6.9 & - & 538 & 44.8 & 25.1 & BLL & MF15 & \magg0.375 & PS17\\
4FGL J1234.7$-$0434 & XRT J123448.00-043246.2 & CY19, PS19 & \magg<0.4 & 13.1 & 0.4 $\pm$ 0.5 & <3 & 44.7 & <23.2 & SY2 & CM01 & \magg0.2765 & CM01, PS19\\
4FGL J1256.8+5329 & XRT J125630.54+533202.3 & KS21 & \magg<0.3 & \magg2.1 & 0.4 $\pm$ 0.3 & <17 & 46.2 & <24.2 & QSO & UA24 & \magg0.996 & SDSS16, UA24\\
3FGL J1258.4+2123 & XRT J125821.45+212351.0 & MF15 & \magg15.1 & \magg2.3 & - & 661 & 46.0 & 25.2 & BLL & PS19 & \magg0.6265 & PS19\\
4FGL J1308.7+0347 & XRT J130832.27+034405.4 & UA24 & \magg<0.3 & 47.9 & 0.2 $\pm$ 0.4 & <1 & 45.6 & <23.3 & QSO & HC08 & \magg0.6193 & SDSS9, UA24\\
4FGL J1340.8$-$0409 & XRT J134042.00-041007.0 & MF13, TY13 & \magg14.6 & 48.0 & 0.3 $\pm$ 0.2 & 31 & 44.8 & 24.3 & BLL & RF15 & \magg0.223 & PS17\\
4FGL J1346.5+5330 & XRT J134545.15+533252.5 & UA24 & 248.1 & 57.5 & 0.3 $\pm$ 0.1 & 431 & 44.3 & 25.1 & FRI & PL11 & \magg0.1359 & SDSS9, UA24\\
4FGL J1410.7+7405 & XRT J141045.66+740509.9 & LR15 & \magg2.2 & \magg7.6 & 0.3 $\pm$ 0.1 & 29 & >45.6 & >24.3 & BLL & ME16 & \textbf{>0.55$^*$} & this work\\
4FGL J1411.5$-$0723 & XRT J141133.30-072253.3 & PS17 & \magg56.0 & 25.0 & - & 223 & >45.7 & >25.9 & BLL & PS17 & >0.72$^{*}$ & PS17\\
4FGL J1430.6+1543 & XRT J143057.97+154556.1 & UA24 & \magg<0.3 & 39.8 & 0.7 $\pm$ 0.3 & <1 & 43.9 & <22.9 & SY1 & SY11 & \magg0.1633 & SDSS9, UA24\\
4FGL J1511.8$-$0513 & XRT J151148.50-051346.7 & TY13, PL13 & \magg15.1 & 36.0 & 0.2 $\pm$ 0.1 & 42 & >45.7 & >24.9 & BLL & CN16 & >0.45 & PS17\\
4FGL J1526.1$-$0831 & XRT J152603.17-083146.4 & CY19 & \magg24.0 & 17.0 & 0.5 $\pm$ 0.2 & 138 & >45.2 & >25.0 & BLL & PS19 & >0.40$^{*}$ & PS19\\
4FGL J1535.9+3743 & XRT J153550.56+374056.8 & UA24 &  \magg26.3 & \magg4.4 & 1.0 $\pm$ 0.2 & 603 & 46.0 & 25.5 & FSRQ & UA24 & \magg0.6255 & SDSS10, UA24\\ 
4FGL J1539.1+1008 & XRT J153848.51+101841.7 & UA24 & \magg<0.3 & 17.4 & - & <2 & 44.7 & <23.2 & SY1 & TY14 & \magg0.2345 & SDSS10, UA24\\
4FGL J1541.7+1413 & XRT J154150.16+141437.6 & CY19 & \magg27.8 & 16.0 & 0.1 $\pm$ 0.5 & 176 & 44.5 & 24.6 & BLL & PS19 & \magg0.223 & SDSS10, PS19\\
4FGL J1544.9+3218 & XRT J154433.15+322148.6 & CY19 & \magg12.0 & 12.0 & - & 100 & 44.6 & 24.5 & BLL & UA24 & \textbf{\magg0.3$^*$} & this work\\
4FGL J1554.2+2008 & XRT J155424.17+201125.5 & MA15 & \magg42.0 & 20.9 & - & 201 & 44.6 & 24.7 & BLG & SJ91, MS91 & \magg0.2225 & MS91\\
4FGL J1555.3+2903 & XRT J155513.01+290328.0 & UA24 & \magg21.7 & 19.1 & 0.5 $\pm$ 0.4 & 114 & 44.0 & 24.2 & BLG & UA24 & \magg0.1767 & SDSS9, UA24\\
4FGL J1631.8+4144 & XRT J163146.82+414631.8 & CY19 & \magg\magg0.8 & \magg2.3 & 0.4 $\pm$ 0.3 & 34 & >45.3 & >24.0 & BLL & UA24 & \textbf{>0.65$^*$} & this work\\
4FGL J1648.7+4834 & XRT J164900.56+483409.2 & KS21 & \magg\magg2.5 & \magg6.3 & - & 40.0 & >45.3 & >24.4 & BLL & UA24 & \textbf{>0.6$^*$} & this work\\
4FGL J1704.2+1234 & XRT J170409.60+123421.3 & MF13 & \magg18.9 & 10.0 & 0.7 $\pm$ 0.2 & 189 & 45.5 & 25.0 & BLL & CN16b & \magg0.452 & CN16b\\
4FGL J1704.5$-$0527 & XRT J170433.80-052841.1 & KS21 & \magg\magg9.7 & 13.0 & 0.2 $\pm$ 0.1 & 74 & 46.4 & 25.1 & BLL & PS17 & >0.7$^{*}$ & PS17\\
4FGL J2030.0$-$0310 & XRT J203014.34-030721.9 & UA24 & \magg\magg0.3 & 69.2 & - & 1 & 42.5 & 21.2 & SY2 & UA24 & \magg0.036 & JD09, UA24\\
4FGL J2115.2+1218 & XRT J211522.00+121802.8 & MF13 & \magg10.2 & 48.0 & 0.6 $\pm$ 0.2 & 21 & >45.5 & >24.8 & BLL & PS17b & >0.497 & PS17\\
4FGL J2150.7$-$1750 & XRT J215046.43-174954.5 & PS19 & \magg13.1 & 33.0 & 0.1 $\pm$ 0.2 & 40 & 44.6 & 24.1 & BLL & PS19 & \magg0.1855 & PS19\\
4FGL J2207.1+2222 & XRT J220704.18+222231.9 & CY19 & \magg\magg6.5 & \magg2.5 & 0.7 $\pm$ 0.3 & 258 & 45.2 & 24.6 & BLL & UA24 & \textbf{\magg0.45$^*$} & this work\\
4FGL J2209.7$-$0451 & XRT J220941.70-045110.3 & CY17 & \magg15.1 & 11.0 & - & 138 & 45.4 & 24.8 & BLL & PS19 & \magg0.3967 & PS19\\
4FGL J2212.4+0708 & XRT J221230.98+070652.5 & PS19 & \magg<0.3 & \magg2.1 & - & <13 & 46.5 & <24.4 & QSO & PS19 & \magg1.0 & PS19\\
4FGL J2228.6$-$1636 & XRT J222830.18-163643.0 & MF13 & \magg15.8 & 11.0 & 0.9 $\pm$ 0.2 & 144 & 45.5 & 25.1 & BLL & PS19 & \magg0.525 & PS19\\
4FGL J2229.1+2254 & XRT J222911.18+225500.0 & PS19 & \magg\magg3.3 & 10.0 & - & 33 & 45.0 & 24.2 & BLL & PS19 & \magg0.440 & PS19\\
4FGL J2240.3-5241 & XRT J224017.55-524112.3 & CY19 & \magg31.0 & 22.9 & 0.4 $\pm$ 0.2 & 135 & >46.1 & >25.6 & BLL & UA24 & \textbf{>0.65$^*$} & this work\\
4FGL J2244.6+2502 & XRT J224436.70+250342.6 & CY19 & \magg\magg9.2 & 10.0 & 0.1 $\pm$ 1.0 & 92 & 45.7 & 25.0 & BLL & PS19 & \magg0.650 & PS19\\
4FGL J2245.9+1544 & XRTJ224604.90+154435.5 & MF13 & \magg\magg6.2 & \magg8.3 & 0.4 $\pm$ 0.1 & 75 & 45.9 & 24.8 & BLL & PS17b & \magg0.5965 & PS19\\
4FGL J2250.4+1748 & XRT J225032.88+174914.8 & PS19 & \magg36.7 & \magg3.0 & 0.5 $\pm$ 0.2 & 1217 & 45.3 & 25.1 & BLL & PS19 & \magg0.3437 & PS19\\
4FGL J2317.7+2839 & XRT J231740.15+283955.4 & KS21 & \magg\magg4.5 & \magg5.2 & 0.6 $\pm$ 0.2 & 86 & >45.4 & >24.5 & BLL & UA24 & \textbf{>0.5$^*$} & this work\\
4FGL J2321.5$-$1619 & XRT J232137.01-161928.5 & CY19 & \magg\magg7.3 & 63.0 & 0.8 $\pm$ 0.2 & 12 & 45.8 & 25.0 & BLL & PS19 & \magg0.6938 & PS19\\
4FGL J2323.1+2040 & XRT J232320.30+203523.6 & UA24 & 160.4 & 631 & - & 25 & 42.9 & 23.7 & BLG & UA24 & \magg0.038 & MM96\\
4FGL J2346.7+0705 & XRT J234639.80+070506.8 & PS17 & 184.4 & 44.0 & 0.4 $\pm$ 0.1 & 423 & 44.9 & 25.1 & BLL & PS17 & \magg0.171 & PS17\\
4FGL J2353.2+3135 & XRT J235319.12+313613.4 & UA24 & \magg60.6 & \magg2.3 & 0.5 $\pm$ 0.2 & 2646 & >46.2 & >26.1 & BLL & UA24 & >0.8809 & UA24\\
4FGL J2358.3+3830 & XRT J235825.17+382856.4 & MS18 & \magg24.9 & 16.0 & 0.5 $\pm$ 0.1 & 157 & 44.8 & 24.4 & \textbf{FSRQ} & \textbf{this work} & \magg0.2001 & MS18\\
4FGL J2358.5$-$1808 & XRT J235836.72-180717.4 & MF13 & \magg15.8 & 30.0 & 0.2 $\pm$ 0.1 & 52 & >45.1 & >24.4 & BLL & MF13 & >0.25$^{*}$ & PS19\\
\hline
\end{tabular}   
}
\label{tab:MW_par}
\end{center}
\footnotesize{
\raggedright
\textbf{Note.} Column 1: 4FGL name of the target; Column 2: Proposed counterpart; Column 3: Reference of the proposed association; Column 4: 2 - 4 GHz radio density flux (mJy); Column 5: g-band optical density flux (erg cm$^{-2}$ s$^{-1}$ Hz$^{-1}$); Column 6: Fractional variability in $\gamma$-ray band; Column 7: \textit{radio-loudness} R; Column 8: Log of the 0.1-100 GeV $\gamma$-ray luminosity (erg s$^{-1}$); Column 9: Logarithm of the 1.4 GHz radio power (W Hz$^{-1}$); Column 10: Counterpart classification from optical spectroscopy; Column 11: Reference of the classification from optical spectroscopy;  Column 12: Redshift (* = Photometric redshift; Boldface indicates new photometric redshift); Column 13: Reference of the redshift \\
References are listed using the following abbreviation: AF13 for \citet[][]{Acero_2013}, AM15 for \citet[][]{Ackermann_2015}, AJ07 for \citet[][]{Adelman_2007}, SDSS9 for \citet[][]{Ahn_2012}, SDSS10 for \citet[][]{Ahn_2014}, SDSS16 for \citet[][]{Ahumada_2020}, AF17 for \citet[][]{Albareti_2017}, CN16 for \citet[][]{Crespo_2016}, CN16b for \citet[][]{Crespo_2016b}, AS03 for \citet[][]{Anderson_2003}, CY17 for \citet[][]{Chang_2017}, CY19 for \citet[][]{Chang_2019}, CM01 for \citet[][]{Colless_2001}, DR14 for \citet[][]{Dabrusco_2014}, dR19 for \citet[][]{demenezes_2019}, HC08 for \citet[][]{Hu_2008}, JD09 for \citet[][]{Jones_2009}, KS21 for \citet[][]{Kerby_2021}, LR15 for \citet[][]{Landi_2015}, MM96 for \citet[][]{Marcha_1996}, MS18 for \citet[][]{Marchesi_2018}, ME16 for \citet[][]{Marchesini_2016}, ME15 for \citet[][]{EMassaro_2015} MF13 for \citet[][]{Massaro_2013}, FM14 for \citet[][]{Massaro_2014} MF15 for \citet[][]{Massaro_2015}, MT16  for \citet[][]{monroe_2016}, MS91 for \citet[][]{Morris_1991}, PA14 for \citet[][]{Paggi_2014}, PS17 for \citet[][]{paiano2017_ufo1}, PS17b for \citet[][]{Paiano_SED}, PS19 for \citet[][]{paiano2019_ufo2}, PI18 for \citet[][]{Paris_2018}, PL11 for \citet[][]{Petrov_2011}, PL13 for \citet[][]{Petrov_2013}, PR10 for \citet[][]{Plotkin_2010} RM23 for \citet[][]{Rajagopal_2023}, RF15 for \citet[][]{Ricci_2015}, SY11 for \citet[][]{Shen_2011}, SS96 for \citet[][]{Shectman96}, SJ91 for \citet[][]{Stocke_1991}, TY13 for \citet[][]{Takahashi_2013}, TY14 for \citet[][]{Toba_2014}, UA24 for \citet[][]{Ulgiati_2024}\\
} 
\end{table*}

\section{Construction and properties of the Broad band SED of UGSs}
\label{sec:SED_building}

We employed the online tool VOU-Blazars V2.00 \citep[][]{Chang_2020}, to gather multifrequency data for each UGS and build the multi-wavelength SED, exploiting its capability to scan a variety of  catalogues  (detailed in Appendix \ref{sec:appendice}) for flux measurements that span a broad spectrum of the electromagnetic range. To complete the SED we also included the radio flux upper limits, for those sources without a catalogued radio counterpart, and X-ray spectral points deriving from our XRT spectra analysis \citep[see][for more details]{Ulgiati_2024b}. Given the temporal variability of AGN, it is possible to have multiple spectral points at the same frequency. However, we want to analyze the average behavior of the objects. For this reason, we decided to average the spectral points for each frequency. Following the approach used in \citet{Paiano_SED}, we overlaid the SED with a curve that emulates the typical double-peaked shape of blazars. 
These two peaks are modelled through an analytic form \citep[Eq. 1 of ][]{Paiano_SED} that combines two power laws with exponential cutoffs to match the distinct rise and decline of each component. Specifically, the seven model parameters allow for adjustment of the amplitude, width, and peak frequency of each component, following the SED of a generic blazar \citep[][for more details]{Paiano_SED}. 
Additionally, we incorporate the template of a giant elliptical host galaxy at the redshift of the object \citep[][]{Coleman_1980}. The SEDs are reported in the Appendix (Fig. \ref{fig:SED} for radio-loud sources, and  Fig. \ref{fig:SED_RQ_noalt} and \ref{fig:SED_alt_all} for radio-quiet sources). 

\section{The jet as the source of $\gamma$-ray radiation}
\label{sec:gamma_origin}

For blazars, $\gamma$-ray emission is generated by non-thermal processes and the jet is the dominant structure for radiation production in this band. There is a correlation between $\gamma$-ray luminosity and 1.4 GHz radio power in jetted AGN  \citep[e.g. see][]{Healey_2007,Kharb_2010,Ghirlanda_2011}. We then compared the $\gamma$-ray luminosities and radio powers of our UGSs.
For the objects for which the redshift measurement is lacking, that is the case of BLL with featureless optical spectra, we can estimate a photometric redshift by overlaying a host elliptical galaxy template at different redshifts onto the SEDs, using the \textit{Firmamento} software. 
Two scenarios can appear. When the multi-wavelength SED exhibits a signature of host galaxy emission, we can set the redshift by aligning the elliptical galaxy template with the IR/optical photometric data. When instead no evidence of host galaxy is observed, additional considerations are required. Since blazar emission is highly variable, while that of the host galaxy remains constant, we analyse the IR/optical light curves to constrain the host galaxy flux. Specifically, we assume that the host galaxy flux cannot exceed the minimum observed flux, providing a lower limit on the photometric redshift. For objects lacking IR/optical light curves, we establish an upper limit on the redshift using the method proposed by \citet{Landt_2002}. These authors studied how the shape of the optical spectrum changes depending on the ratio between the non-thermal flux and the host galaxy flux. They observed that the host galaxy’s contribution becomes negligible in the optical spectrum when the non-thermal flux exceeds that of the galaxy by a factor $\gtrsim$ 10. Therefore, when a source’s optical spectrum is described by a featureless power-law, we can place an upper limit on the host galaxy’s flux, and thus derive a lower limit for the redshift
\citep[see][for details]{Chang_2019}. 

\begin{figure}
\hspace{-0.5cm}
\centering
   \includegraphics[width=9.5truecm]{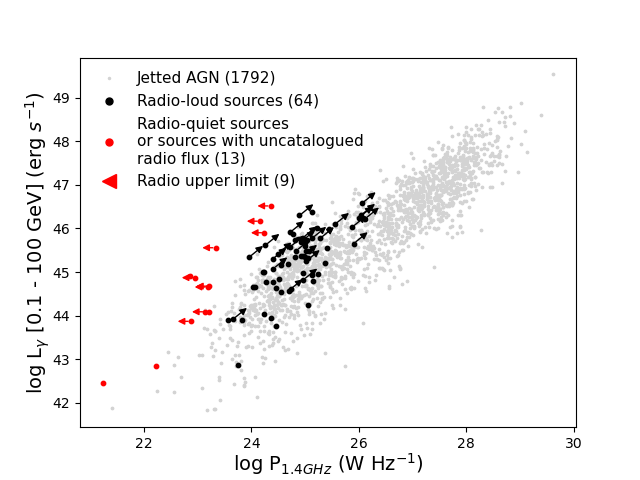}
\caption{L$_{\gamma-ray}$ vs. P$_{1.4GHz}$ for our sample (black and red points) and the comparison sample, represented by 4FGL-DR4 jetted AGN (light-grey points). The red points are either radio-quiet sources or sources with uncatalogued radio flux. Diagonal arrows denote lower limits on redshift and therefore powers, horizontal arrows represent radio luminosity upper limits.}
\label{fig:diagn_Lgamma_Lradio}
\end{figure}

We report in Table \ref{tab:MW_par} the $\gamma$-ray luminosity (L$_{\gamma}$), the radio power (P$_{\rm 1.4GHz}$), and the redshift of the sources in our sample, and in Fig. \ref{fig:diagn_Lgamma_Lradio} the L$_{\gamma}$ vs. P$_{\rm 1.4GHz}$ diagram, where we compare our sources (black and red points) with the jetted AGN from the 4FGL-DR4 catalogue (grey points). Since not all sources have flux measurements at 1.4 GHz, we estimate the flux at this frequency by interpolating from the available data points at $\nu_{VLASS}$ = 3 GHz, $\nu_{RACS}$ = 887.5 MHz, and $\nu_{LOFAR}$ = 144 MHz using a power-law model. We assume a spectral index of $\alpha$ = 0 for blazars, due to their flat radio spectrum, and $\alpha$ = 0.7 for other AGN classes.
All radio-loud sources (black points) are placed in the locus defined by the \textit{Fermi} jetted AGN, while the UGSs with radio flux upper limits or classified as radio-quiet are clearly offset, as their radio power is too low relative to their $\gamma$-ray power. Therefore, this sub-sample is very unlikely to contain jetted AGN.

\section{Discussion}
\label{sec:discussion_P}

\subsection{The case of radio-loud sources}

The analysis of the multi-wavelength SEDs for the radio-loud objects in our sample (64 out of 77 sources; see Fig. \ref{fig:SED}), indicates that the spectral properties of the proposed counterparts are consistent with $\gamma$-ray emission. The radio emission appears to be flat ($\alpha$ < 0.5, according to the SEDs template of jetted AGN reported in Fig. \ref{fig:SED_LSP_HSP}) and there is evidence of jet activity as indicated by the characteristic double-peak structure in the SEDs. This is likely the origin of the high-energy (HE, E>100 MeV) spectrum. The emission in the X-ray band is consistent with the $\gamma$-ray regime, suggesting a common emission mechanism between these two components. Specifically, the LSPs have a flat ($\alpha$ $\lesssim$ 1) X-ray spectrum, and a steep ($\alpha$ $\gtrsim$ 1) $\gamma$-ray spectrum; for the HSPs, the $\gamma$-ray spectrum is instead flat ($\alpha$ $\lesssim$ 1) and the X-ray spectrum is steep ($\alpha$ $\gtrsim$ 1). Following the approach presented in \citet[][]{Paiano_SED}, we tested the compatibility of the SEDs with typical blazar emission by using a set of 47 well-known blazar templates. This comparison was quantified through a $\chi^2$ minimization, allowing us to verify the presence of the characteristic double-peaked shape and to classify the sources as LSP, ISP, HSP or E-HSP. The results of this analysis are reported in Table \ref{tab:class_LIHSP}. The overall SED properties of the radio-loud AGN  in our sample support their association with blazars as counterparts of the \textit{Fermi} $\gamma$-ray sources. Moreover 6 objects reveal a prominent host galaxy that stands out from the non-thermal component (e.g. see the case of XRT J083902.98+401546.9/4FGL J0838.5+4013 in Fig. \ref{fig:SED}).

It is worth noting the case of XRT J235825.17+382856.4/PAN J235825.17+382856.4, that shows a Seyfert 2\textit{-like} optical spectrum \citep[][]{Marchesi_2018}, and the typical HSP blazar behaviour in its multi-wavelength SED. Its strong radio emission indicates that the jet is likely aligned with the observer, consistent with its blazar\textit{-like} SED. Probably this object is a low-luminosity FSRQ\footnote{In Table \ref{tab:MW_par}, the new classification has already been reported.}, given its absolute magnitude $\rm M_{\rm g}$ $\sim -21.6$, derived by the analysis of the PanSTARRS imaging.


To provide an independent classification of these sources as LSP, ISP, HSP or E-HSP based on the synchrotron peak frequency ($\nu^{s}_{p}$), we employed the \textit{BlaST} tool \citep[][]{Glauch_2022}. \textit{BlaST} is a machine-learning algorithm that provides an automatic measure of this parameter from the binned SED without any need for manual data preparation, taking also into account the possible presence of emission from dust in the host galaxy (IR bump) and the accretion disk (blue bump) \citep{Glauch_2022}.
The results of the classification are also reported in Table \ref{tab:class_LIHSP}. We compare the classification results from the blazar recognizing tool \citet[][]{Paiano_SED}, from the BlaST-based approach, and from the optical spectroscopy. The three methods agree in confirming the blazar nature of the radio-loud sources, the consistency of the $\gamma$-ray emission with the proposed counterparts, and the realistic shape of the SEDs compared to known blazars. As for the sub-classification, the results from the two SED-based methods are consistent within the uncertainties on the synchrotron peak frequencies. The proposed classifications are consistent, when present, with those found in the literature \citep[e.g.][]{Paiano_SED, Chang_2019, Chiaro_2019, Zhao_2024,Liang_2025}. As expected \citep[e. g. see][]{4lac}, HSPs are the dominant class, which is typical for \textit{Fermi} catalogues, especially among BL Lac objects.

\begin{table*}
\begin{center}
\caption{Classification of sources according to the shape of their SEDs and the position of the synchrotron peak.} 
\resizebox{13cm }{!}{
\begin{tabular}{llccccc}
\hline 
Name & Counterpart & Classification & $\chi^2_{\nu,min}$ & $\log~\nu^s_p$ & Classification & Classification \\
     & & from SED templates & & (Hz)     & from Blast &  from spectroscopy              \\
\hline

4FGL J0004.0+0840 & SDSS J000359.23+084138.2 & ISP/HSP & 0.3 & 14.4 $\pm$ 0.6 & ISP & BLL\\
4FGL J0006.4+0135 & XRT J000626.92+013610.3 & HSP & 0.2 & 15.1 $\pm$ 0.4 & HSP & BLL\\
4FGL J0049.1+4223 & XRT J004859.10+422351.0 & HSP & 0.5 & 15.7 $\pm$ 0.3 & HSP & BLL\\
4FGL J0102.4+0942 & XRT J010217.10+094409.5 & ISP/HSP & 0.3 & 14.7 $\pm$ 0.3 & ISP & BLL\\
4FGL J0112.0+3442 & XRT J011124.86+344154.1 & ISP & 0.2 & 14.2 $\pm$ 0.9 & ISP & BLL\\
4FGL J0158.8+0101 & XRT J015852.77+010132.8 & ISP & 0.2 & 13.0 $\pm$ 0.5 & LSP & BLL\\
4FGL J0202.7+3133 & XRT J020242.13+313211.4 & HSP & 0.2 & 15.3 $\pm$ 0.4 & HSP & BLL\\
4FGL J0234.3$-$0628 & XRT J023410.27-062825.6 & HSP & 0.4 & 15.1 $\pm$ 0.4 & HSP & BLL\\
4FGL J0238.7+2555 & XRT J023853.80+255407.1 & HSP & 0.7 & 16.5 $\pm$ 0.4 & HSP & BLL\\
4FGL J0251.1$-$1830 & XRT J025111.70-183111.1 & HSP & 0.3 & 16.1 $\pm$ 0.5 & HSP & BLL\\
4FGL J0259.0+0552 & XRT J025857.56+055244.4 & HSP & 0.6 & 15.6 $\pm$ 0.5 & HSP & BLL\\
4FGL J0305.1$-$1608 & XRT J030515.00-160816.6 & LSP & 1.0 & 15.7 $\pm$ 1.0 & HSP & BLL\\
4FGL J0338.5+1302 & XRT J033829.20+130215.7 & HSP & 0.6 & 15.7 $\pm$ 0.3 & HSP & BLL\\
4FGL J0409.8$-$0359 & XRT J040946.50-040003.5 & ISP/HSP & 0.4 & 14.1 $\pm$ 0.4 & ISP & BLL\\
4FGL J0414.6$-$0842 & XRT J041433.08-084206.7 & ISP/HSP & 0.6 & 14.1 $\pm$ 0.6 & ISP & BLL\\
4FGL J0506.9+0323 & XRT J050650.14+032358.6 & HSP & 0.3 & 15.4 $\pm$ 0.3 & HSP & BLL\\
4FGL J0644.6+6039 & XRT J064435.70+603851.3 & HSP & 0.5 & 16.2 $\pm$ 0.5 & HSP & BLL\\
4FGL J0838.5+4013 & XRT J083902.98+401546.9 & HSP & 1.0 & 15.8 $\pm$ 0.4 & HSP & BLG\\
4FGL J0848.7+7017 & XRT J084839.52+701728.0 & ISP/HSP & 0.7 & 13.7 $\pm$ 0.6 & LSP & BLL\\
4FGL J0930.5+5132 & XRT J093033.36+513214.6 & HSP & 0.2 & 15.1 $\pm$ 0.4 & HSP & BLL\\
4FGL J0937.9$-$1434 & XRT J093754.70-143350.4 & HSP & 0.4 & 15.3 $\pm$ 0.3 & HSP & BLL\\
4FGL J0952.8+0712 & XRT J095249.50+071329.9 & HSP & 0.3 & 15.4 $\pm$ 1.1 & HSP & BLL\\
4FGL J1016.1$-$4247 & XRT J101620.78-424723.2 & HSP & 0.5 & 15.8 $\pm$ 0.4 & HSP & BLL\\
4FGL J1039.2+3258 & XRT J103852.17+325651.9 & HSP & 0.5 & 15.1 $\pm$ 0.7 & HSP & BLL\\
4FGL J1049.5+1548 & XRT J104939.30+154837.6 & ISP/HSP & 0.3 & 14.8 $\pm$ 0.4 & ISP & BLL\\
4FGL J1049.8+2741 & XRT J104938.70+274212.1 & HSP & 0.5 & 15.8 $\pm$ 0.4 & HSP & BLG\\
4FGL J1128.8+3757 & SDSS J112903.20+375656.7 & ? & ? & 13.9 $\pm$ 0.5 & LSP & BLL\\
4FGL J1131.6+4657 & XRT J113142.36+470009.2 & ISP/HSP & 0.6 & 15.6 $\pm$ 0.6 & HSP & BLG\\
4FGL J1146.0$-$0638 & XRT J114600.87-063853.9 & HSP & 0.4 & 16.1 $\pm$ 0.4 & HSP & BLL\\
4FGL J1223.5+0818 & XRT J122327.49+082030.4 & HSP & 0.4 & 16.1 $\pm$ 0.6 & HSP & BLL\\
4FGL J1223.9+7954 & XRT J122358.10+795328.6 & HSP & 0.5 & 14.9 $\pm$ 0.7 & ISP & BLL\\
3FGL J1258.4+2123 & XRT J125821.45+212351.0 & HSP & 0.4 & 16.6 $\pm$ 0.4 & HSP & BLL\\
4FGL J1340.8$-$0409 & XRT J134042.00-041007.0 & HSP & 0.3 & 15.2 $\pm$ 0.4 & HSP & BLL\\
4FGL J1346.5+5330 & XRT J134545.15+533252.5 & ISP & 1.0 & 13.6 $\pm$ 0.9 & LSP & FRI\\
4FGL J1410.7+7405 & XRT J141045.66+740509.9 & HSP & 0.8 & 15.2 $\pm$ 0.4 & HSP & BLL\\
4FGL J1411.5$-$0723 & XRT J141133.30-072253.3 & HSP & 0.3 & 15.6 $\pm$ 0.5 & HSP & BLL\\
4FGL J1511.8$-$0513 & XRT J151148.50-051346.7 & HSP & 0.6 & 17.1 $\pm$ 0.5 & HSP & BLL\\
4FGL J1526.1$-$0831 & XRT J152603.17-083146.4 & HSP & 0.4 & 15.5 $\pm$ 0.4 & HSP & BLL\\
4FGL J1535.9+3743 & XRT J153550.56+374056.8 & ISP & 0.5 & 13.5 $\pm$ 0.6 & LSP & FSRQ\\
4FGL J1541.7+1413 & XRT J154150.16+141437.6 & HSP & 0.4 & 15.5 $\pm$ 0.4 & HSP & BLL\\
4FGL J1544.9+3218 & XRT J154433.15+322148.6 & HSP & 0.5 & 15.5 $\pm$ 0.4 & HSP & BLL\\
4FGL J1554.2+2008 & XRT J155424.17+201125.5 & HSP & 0.4 & 17.4 $\pm$ 0.7 & HSP & BLG\\
4FGL J1555.3+2903 & XRT J155513.01+290328.0 & HSP & 0.7 & 15.8 $\pm$ 0.6 & HSP & BLG\\
4FGL J1631.8+4144 & XRT J163146.82+414631.8 & HSP & 0.4 & 16.1 $\pm$ 0.4 & HSP & BLL\\
4FGL J1648.7+4834 & XRT J164900.56+483409.2 & HSP & 0.7 & 16.0 $\pm$ 0.8 & HSP & BLL\\
4FGL J1704.2+1234 & XRT J170409.60+123421.3 & HSP & 0.7 & 16.1 $\pm$ 0.6 & HSP & BLL\\
4FGL J1704.5$-$0527 & XRT J170433.80-052841.1 & HSP & 0.6 & 16.0 $\pm$ 0.5 & HSP & BLL\\
4FGL J2115.2+1218 & XRT J211522.00+121802.8 & HSP & 0.3 & 15.4 $\pm$ 0.3 & HSP & BLL\\
4FGL J2150.7$-$1750 & XRT J215046.43-174954.5 & ISP/HSP & 1.0 & 14.6 $\pm$ 0.5 & ISP & BLL\\
4FGL J2207.1+2222 & XRT J220704.18+222231.9 & HSP & 0.2 & 16.0 $\pm$ 0.4 & HSP & BLL\\
4FGL J2209.7$-$0451 & XRT J220941.70-045110.3 & HSP & 1.4 & 15.4 $\pm$ 0.4 & HSP & BLL\\
4FGL J2228.6$-$1636 & XRT J222830.18-163643.0 & ISP/HSP & 0.1 & 14.6 $\pm$ 0.3 & ISP & BLL\\
4FGL J2229.1+2254 & XRT J222911.18+225500.0 & HSP & 0.3 & 17.1 $\pm$ 0.5 & HSP & BLL\\
4FGL J2240.3$-$5241 & XRT J224017.55-524112.3 & HSP & 0.5 & 14.5 $\pm$ 0.5 & ISP & BLL\\
4FGL J2244.6+2502 & XRT J224436.70+250342.6 & HSP & 0.4 & 15.5 $\pm$ 0.3 & HSP & BLL\\
4FGL J2245.9+1544 & XRTJ224604.90+154435.5 & HSP & 0.8 & 15.8 $\pm$ 0.7 & HSP & BLL\\
4FGL J2250.4+1748 & XRT J225032.88+174914.8 & ISP & 0.5 & 13.6 $\pm$ 0.7 & LSP & BLL\\
4FGL J2317.7+2839 & XRT J231740.15+283955.4 & HSP & 0.7 & 15.0 $\pm$ 0.3 & HSP & BLL\\ 
4FGL J2321.5$-$1619 & XRT J232137.01-161928.5 & HSP & 0.3 & 16.0 $\pm$ 0.4 & HSP & BLL\\
4FGL J2323.1+2040 & XRT J232320.30+203523.6 & ISP & 0.9 & 14.6 $\pm$ 0.5 & ISP & BLG\\
4FGL J2346.7+0705 & XRT J234639.80+070506.8 & HSP & 0.6 & 15.3 $\pm$ 0.3 & HSP & BLL\\
4FGL J2353.2+3135 & XRT J235319.12+313613.4 & LSP & 0.7 & 12.9 $\pm$ 0.5 & LSP & BLL\\
4FGL J2358.3+3830 & XRT J235825.17+382856.4 & HSP & 0.5 & 16.3 $\pm$ 0.5 & HSP & FSRQ\\
4FGL J2358.5$-$1808 & XRT J235836.72-180717.4 & HSP & 0.2 & 15.6 $\pm$ 0.3 & HSP & BLL\\
\hline
\end{tabular}
}
\label{tab:class_LIHSP}
\end{center}
\raggedright
\footnotesize{
\textbf{Note.} Column 1: 4FGL name; Column 2: Name of the proposed counterpart; Column 3 and 4: Classification obtained using the blazar recognition tool described in \citet{Paiano_SED}; Column 4: $\chi^2_{\nu,min}$ value derived by the blazar SED template fitting of the blazar recognition tool; Column 5: Logarithm of the synchrotron peak frequency, $\nu^s_p$ (Hz) $\pm1\sigma$, derived by BlaST tool \citep[][]{Glauch_2022}; Column 6: Classification from BlaST tool; Column 7: Classification of the UGS counterpart from optical spectroscopy.}
\end{table*}

\subsection{Unveiling masquerading BL Lacs among radio-loud UGS}\label{sec:masq}

\begin{figure}
\hspace{-0.3cm}
\includegraphics[width=0.55\textwidth]{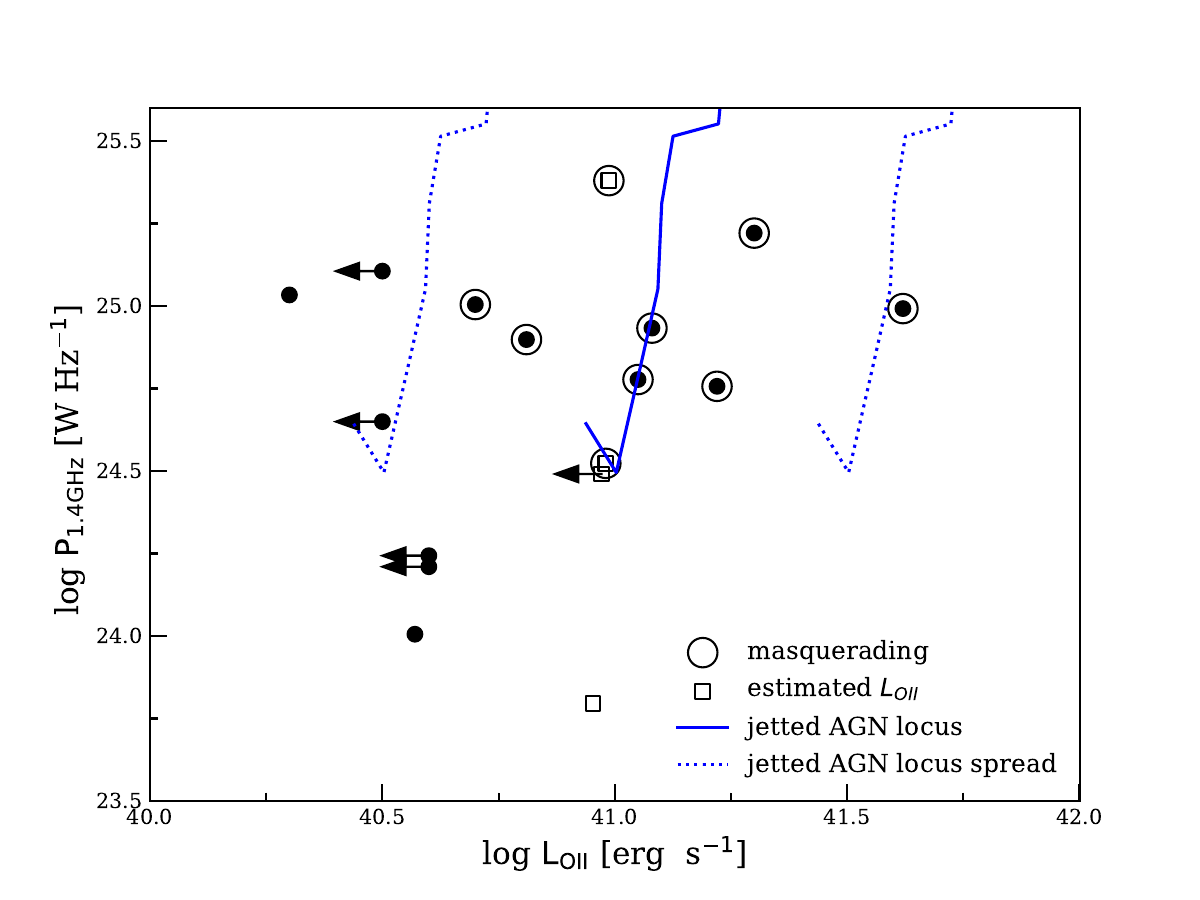}
\caption{$P_{\rm 1.4GHz}$ vs. $L_{\rm [\ion{O}{II}]}$ for 
 the objects in our sample 
having [O~II] information (black
  filled circles), with masquerading sources highlighted (larger empty
  circles). Sources for which $L_{\rm [\ion{O}{II}]}$ has been estimated
  from $L_{\rm [\ion{O}{III}]}$ are denoted by black empty squares. 
  The solid blue line is the locus of jetted quasars, with
  the two dotted lines indicating a spread of 0.5 dex, which includes most
  of the points in Figure 4 of \protect\cite{Kalfountzou_2012} (converted from 
  radio powers in W Hz$^{-1}$ sr$^{-1}$ and line powers in W). 
  The arrows denote upper limits on $L_{\rm [\ion{O}{II}]}$.}
\label{fig:Lr_LOII}
\end{figure}

\cite{Padovani_2019,Padovani_2022} and \cite{Paiano_2023} used four parameters for the classification of masquerading BLLs. These are:
(1) location on the radio power -- [\ion{O}{II}] emission line power, $P_{\rm 1.4GHz}$ -- $L_{\rm [\ion{O}{II}]}$, diagram, which defines the locus of jetted (radio-loud) quasars; 
(2) a radio power $P_{\rm 1.4GHz} > 10^{26}$ W Hz$^{-1}$, since high-excitation galaxies 
(HEGs), the class to which FSRQs belong, become the dominant population in the radio sky 
above this value; (3) an Eddington ratio, i.e. the ratio between the (accretion-related) 
observed luminosity, $L_{\rm acc}$, and the Eddington luminosity\footnote{The Eddington luminosity is $L_{\rm Edd}$ = $1.26 \times 10^{46} (M$ / $10^8 \, \rm M_{\odot})$ erg s$^{-1}$, where $\rm M_{\odot}$ is one solar mass.}, $L_{\rm acc}/L_{\rm Edd} \gtrsim 0.01$, which is 
again typical of HEGs \citep{Padovani_2017}; (4) a $\gamma$-ray Eddington ratio 
$L_{\gamma}$/$L_{\rm Edd} \gtrsim 0.1$, where $L_{\gamma}$ is the rest-frame, k-corrected, 
$\gamma$-ray power between 0.1 and 100 GeV. 

Since $L_{\rm acc}$ cannot be directly determined in these sources, an indirect 
approach has to be used instead. This is done by using relationships between 
$L_{\rm acc}$ and $L_{\rm [\ion{O}{II}]}$ and $L_{\rm [\ion{O}{III}]}$ 
(\citealt{Punsly_2011}; see also \citealt{Padovani_2019} for more details.). 
As stressed by \cite{Paiano_2023} parameters (3) and (4) are the least certain 
given their dependence on $L_{\rm acc}$ and $M_{\rm BH}$. This is even truer in our case
since, without individual measurements,  we had to assume the typical value for blazars 
of $M_{\rm BH}$ = $6.3 \times 10^8 M_{\odot}$ (e.g., \citealt{Padovani_2022} and 
references therein). 

Fig. \ref{fig:Lr_LOII} shows the location of the sources with 
[\ion{O}{II}]\footnote{As done in \cite{Paiano_2023} for four objects 
for which only [\ion{O}{III}] was available we converted $L_{\rm [\ion{O}{III}]}$ to
$L_{\rm [\ion{O}{II}]}$ using Figure 7 of \cite{Kalfountzou_2012}.}
information on the $P_{\rm 1.4GHz}$ -- $L_{\rm [\ion{O}{II}]}$ diagram. 
Nine objects are close to the locus of jetted quasars and are therefore
``bona fide'' masquerading BLLs, six of which 
also have $L_{\rm acc}$/$L_{\rm Edd} \ge 0.01$\footnote{One more source, SDSS J011124.86+344154.6 (4FGL J0112.0+3442), has $L_{\rm acc}$/$L_{\rm Edd} $=$ 0.009$ 
but falls very close to the locus of jetted AGN.}. 
One source (USNOB0754$-$0223141 counterpart of 4FGLJ0937.9$-$1434)
has an estimated $L_{\rm [\ion{O}{II}]}$ upper limit right on the locus. However,
given that this object has no other HEG-like property we are not including it with the masquerading sources.  

Table \ref{tab:masq} lists the 18 sources which qualify as masquerading 
according to at least one parameter. Three BLLs without emission line
information have $P_{\rm 1.4GHz} > 10^{26}$ W Hz$^{-1}$ (and 
$L_{\gamma}$/$L_{\rm Edd} > 0.1$), while for the remaining 6 
the classification is based only on $L_{\gamma}$/$L_{\rm Edd}$, as $L_{\rm acc}$/$L_{\rm Edd}$ requires emission lines. Based on the discussion above, we regard these nine objects as candidate masquerading BLLs. Therefore, of the 61 sources with redshift
information, $\sim 15\%$ (possibly $\sim 30\%$), are of the masquerading type, which is somewhat smaller than the value of $\gtrsim 34\%$ derived by \cite{Paiano_2023} for a sample
of candidate neutrino sources. However, due to the lower spectral quality of $\approx 1/3$ of the sources studied in this work compared to those in the latter paper, and the widely different selection criteria, a direct comparison between the results for the two samples is not straightforward. 


\begin{table*}
\caption{Masquerading BLL properties.}
 \begin{center}
 \begin{tabular}{lllcccc}
   \hline
    Name & 4FGL Name & ~~~~~$z$ & $P_{\rm 1.4GHz} - L_{\rm {[\ion{O}{II}]}}$ & $P_{\rm 1.4GHz}$ & 
    $L_{\rm acc}/L_{\rm Edd}$ & $L_{\gamma}/L_{\rm Edd}$ \\
      \hline
 SDSS J000359.23+08413 & 4FGL J0004.0+0840  & >1.5035 & --  & I & --  & \checkmark \\
 SDSS J011124.86+34415 & 4FGL J0112.0+3442  &   \magg0.3997 & \checkmark & I & \xmark & \xmark \\
 SDSS J015852.77+01013 & 4FGL J0158.8+0101  &   \magg0.4537 & \checkmark & I & \xmark & \xmark \\
 SDSS J025857.55+05524 & 4FGL J0259.0+0552  & >0.6000 & --  & I & --  & \checkmark \\
 SDSS J030515.00$-$16081 & 4FGL J0305.1$-$1608  &   \magg0.3120 & \checkmark & I & \checkmark & \xmark \\
 USNOB1$-$1602-0082223  & 4FGL J0848.7+7017  & >1.2435 & --  & \checkmark & --  & \checkmark \\
 SDSS J104939.30+15483 & 4FGL J1049.5+1548  &   \magg0.3260 & \checkmark & I & \checkmark &\xmark\\
 SDSS J112903.20+37565 & 4FGL J1128.8+3757  & >1.2110 & --  & \checkmark & --  & \checkmark \\
 SDSS J125821+212351   & 3FGL J1258.4+2123  &   \magg0.6265 & --  & I & --  & \checkmark \\
 SDSS J154150.16+14143 & 4FGL J1541.7+1413  &   \magg0.2230 &\checkmark& I &\xmark&\xmark\\
 SDSS J170409.60+12342 & 4FGL J1704.2+1234  &   \magg0.4520 & \checkmark & I & \checkmark &\xmark\\
 USNOB0845$-$0308445    & 4FGL J1704.5$-$0527  & >0.3000 & --  & I & --  & \checkmark \\
 SDSS J220941.70-04511 & 4FGL J2209.7-0451  &   \magg0.3967 & \checkmark & I & \checkmark &\xmark\\
 DES J224017.71$-$524113 & 4FGL J2240.3$-$5241  & >0.6500 & --  & I & --  & \checkmark \\
 SDSS J224436.70+25034 & 4FGL J2244.6+2502  &   \magg0.6500 & \checkmark & I & \checkmark &\xmark\\
 NVSS J224604.90+15443 & 4FGL J2245.9+1544  &   \magg0.5965 & \checkmark & I & \checkmark &\xmark\\
 SDSS J235319.54+31361 & 4FGL J2353.2+3135  & >0.8809 & --  & \checkmark & --  & \checkmark \\
 USNOB1$-$0718$-$1032041  & 4FGL J2358.5$-$1808  & >0.7000 & --  & I & --  & \checkmark \\
 \hline
  \end{tabular}
  \end{center}
\footnotesize {
\raggedright
\textit{Notes.} `\checkmark' implies that the condition is met, `I' that the condition is not met but this does not mean this is not a masquerading BLL,  `\xmark' that the condition is not met}, and `--' that no information is available.
 \label{tab:masq}
\end{table*}

\subsection{The case of the radio-quiet sources}
\label{sec:radio_quiet}

Regarding the radio-quiet sources in the sample, most of them do not exhibit radio emission (thus, we had to set upper limits), making it difficult to confirm the presence of a jet structure. 
This is strengthened by the analysis conducted in Sec. \ref{sec:gamma_origin}, where we observed that these objects deviate from the behavior of jetted AGN-like sources. The multi-wavelength SED analysis confirms for these objects an emission that differs from a double-peak trend. We compared the multi-wavelength emission of these objects with the templates of 47 well-known blazars, following the method described in \citet[][]{Paiano_SED}, and found that they cannot be classified into any of the standard blazar sub-classes. The $\chi^2$ test yields unphysical or poorly constrained values, indicating that their SED shapes do not follow a double-peaked structure.

\begin{figure}
\hspace{-0.5cm}
\centering
\includegraphics[width=0.45\textwidth, angle=0]{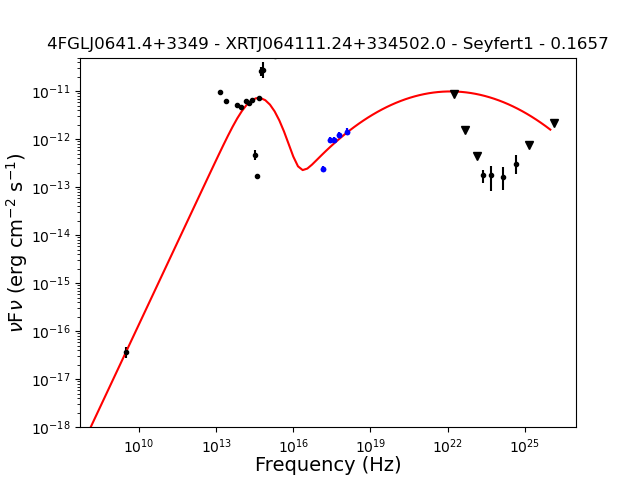}\\
\caption{SED of XRT J064111.24+334502.0, proposed counterpart in \citet[][]{Ulgiati_2024} for the UGS 4FGL J0641.4+3349. Black spectral point are from VOU-Blazar, blue points from our analysis. The points with a triangle shape are upper limits.}  
\label{fig:SED_RQ_EX}
\end{figure}

When a double-peak curve is overlaid on the SEDs, it is observed that the spectral slope in the X-ray band is also not compatible with that in the $\gamma$-ray band (see an example in Fig. \ref{fig:SED_RQ_EX}, to be compared with Fig. \ref{fig:SED_LSP_HSP}; all cases are discussed in subsection \ref{sec:alter}). Due to the absence of a jet, the $\gamma$-ray emission cannot obviously be associated to it. As for the IR-X-ray emission, however, it appears to have a thermal origin.  
Having ruled out the presence of a jet in these sources, we explore possible alternatives for the origin of their $\gamma$-ray emission.

\subsubsection{Gamma-ray emission from star-formation-related processes?}

As mentioned in Section \ref{sec:introduction}, $\gamma$-ray photons can be produced through star-formation-related (SFR) processes. Since $\gamma$-ray emission from starburst galaxies is constant over time, we decided to investigate the fractional variability (FV) of the radio-quiet objects. The values are extracted from the 4FGL-DR4  catalogue \citep{4FGL_DR4} and are available for $\sim$ 70\% of the sample (Table \ref{tab:MW_par}). 
We notice that the mean value of the FV parameter for our entire sample is consistent with that of the 4FGL AGN ($\langle$FV$_{4FGL-DR4}^{AGN}$$\rangle$ = 0.6 $\pm$ 0.4, compared to $\langle$FV$_{sample}$$\rangle$ = 0.5 $\pm$ 0.2).
Regarding radio-quiet objects, while three of them (4FGL J0023.6$-$4209, 4FGL J0117.9+1430 and 4FGL J1430.6+1543) show variability, in other three cases (4FGL J0938.8+5155, 4FGL J1125.1+4811 and 4FGL J1234.7$-$0434) the FV parameter is compatible with 0 (for the remaining objects the parameter is not available). Therefore, in these three objects $\gamma$-ray emission could stem from SFR processes. However, even $\sim$15\% of the 4FGL-DR4 AGN have an FV parameter compatible with 0, which demonstrates that this condition is necessary but not sufficient to identify starburst galaxies. We therefore decided to perform a different test.

An alternative approach is to examine the $L_{\gamma}$ vs. $P_{\rm 1.4GHz}$ relationship reported in \citet[][]{Peng_2019} for star-forming galaxies, in order to infer the level of $\gamma$-ray emission due to SFR process (within the hypothesis that all radio emission is due to the latter). The formula reported in \citet[][]{Peng_2019} involves the use of monochromatic radio continuum power at 1.4 GHz. We report in Table \ref{tab:confr_gamma} the results of this analysis, comparing the predicted $\gamma$-ray emission with the the catalogued one. This results to be significantly lower, typically by $\sim 3$ orders of magnitude.
This suggests that SFR processes are unlikely to be the source of the \textit{Fermi} emission. 

\begin{table*}
\begin{center}
\caption{Comparison between the predicted $\gamma$-ray luminosity due to SFR processes and the estimated luminosity from the fluxes reported in the 4FGL-DR4 for the radio-quiet sample.} 
\begin{tabular}{lllrrc}
\hline 
ID & 4FGL Name & Optical counterpart   & P$_{\rm 1.4GHz}$  & L$_{1-500GeV}^{SFP}$ & L$_{1-500GeV}^{Fermi}$ \\
          & &                       &    (W Hz$^{-1}$)       &      (erg s$^{-1}$)    &   (erg s$^{-1}$)\\ 
\hline

1 & 4FGL J0023.6$-$4209 & DES J002303.74-420508.4 & 22.2 & 40.3$^{+0.6}_{-0.6}$ & 42.7 \\
2 & 3FGL J0031.6+0938 & SDSS J003159.86+093618.4 & <22.9 & 41.2$^{+0.6}_{-0.6}$ & 44.2 \\
3 & 4FGL J0117.9+1430 & SDSS J011804.83+143158.6 & <23.1 & <41.4$^{+0.6}_{-0.6}$ & 43.6 \\
4 & 4FGL J0641.4+3349 & PAN J064111.22+334459.7 & 23.2 & 41.6$^{+0.6}_{-0.6}$ & 44.0 \\
5 & 4FGL J0938.8+5155 & SDSS J093834.72+515452.3 & 22.9 & 41.2$^{+0.6}_{-0.6}$ & 44.8 \\
6 & 4FGL J1125.1+4811 & SDSS J112526.27+480922.0 & <24.2 & <42.9$^{+0.6}_{-0.6}$ & 46.0 \\
7 & 4FGL J1234.7$-$0434 & PAN J123448.05-043245.2 & 23.2 & 41.6$^{+0.6}_{-0.6}$ & 44.6\\
8 & 4FGL J1256.8+5329 & SDSS J125630.43+533204.3 & <24.2 & <42.9$^{+0.6}_{-0.6}$ & 45.6 \\
9 & 4FGL J1308.7+0347 & SDSS J130832.10+034403.9 & <23.3 & <41.7$^{+0.6}_{-0.6}$ & 45.5 \\
10 & 4FGL J1430.6+1543 & SDSS J143058.03+154555.6$^*$ & <22.9 & <41.1$^{+0.6}_{-0.6}$ & 43.6 \\
11 & 4FGL J1539.1+1008 & SDSS J153848.47+101843.2$^*$ & <23.2 & <41.6$^{+0.6}_{-0.6}$ & 44.2 \\
12 & 4FGL J2030.0$-$0310 & PAN J203014.27-030722.5 & 21.2 & 38.9$^{+0.6}_{-0.6}$ & 42.4 \\
13 & 4FGL J2212.4+0708 & SDSS J221230.98+070652.5 & <24.4 & <43.0$^{+0.6}_{-0.6}$ & 46.4\\
\hline
\end{tabular}
\label{tab:confr_gamma}
\end{center}
\raggedright
\footnotesize{\textbf{Note.} Column 1: Identification number; Column 2: 4FGL Name; Column 3: Optical counterpart;  Column 4: Logarithm of the 1.4 GHz radio power (W Hz$^{-1}$); Column 5: Logarithm of the $\gamma$-ray luminosity predicted by the $L_{\gamma}$ vs. $P_{\rm 1.4GHz}$ relationship for SFR processes (erg s$^{-1}$). The reported uncertainty is at a 1 $\sigma$ confidence level and was estimated from Figure 3 of \citet{Peng_2019}; Column 6: Logarithm of the $\gamma$-ray luminosity estimated from the 4FGL-DR4 catalogued flux (erg s$^{-1}$).\\
(*) L$_{radio}$ is measured at 1.4 GHz.}
\end{table*}


\subsubsection{Notes on individual sources and possible alternative counterparts}
\label{sec:alter}

Given the challenges in attributing the $\gamma$-ray emission in our sample of radio-quiet sources to either jet activity or star formation processes, we have delved deeper into the analysis to explore alternative counterparts. 
The original proposed counterparts are selected to be the unique X-ray sources with SNR $\geq$ 3$\sigma$ contained within the 3$\sigma$ \textit{Fermi} error regions of these UGSs. We decide to decrease the detection threshold (SNR $<$ 3$\sigma$) for the X-ray counterparts, and to investigate potential radio counterparts regardless of the presence of an X-ray source. The radio catalogues primarily used for this research are VLASS and RACS.
Below we present detailed notes on each individual sources:

\begin{itemize}
\item \textit{4FGL J0023.6$-$4209}: The proposed counterpart is the Seyfert 2 XRT J002303.59-420509.6/DES J002303.74-420508.4 \citep[][]{Ulgiati_2024}.  Based on the SED presented in Fig. \ref{fig:SED_alt_all}, we can conclude that the multi-wavelength emission of this object is not consistent with that of a blazar (see Fig. \ref{fig:SED_LSP_HSP} for comparison). It does not exhibit a flat radio spectrum ($\alpha \, < \, 0.5$), and furthermore, both the X-ray and $\gamma$-ray spectra are flat, which contradicts the characteristics of jet-like emission. For this source, according to the spectroscopic classification, the likely origin of the X-ray component of XRT J002303.59-420509.6 is the disk, while that of the infrared one is the dusty torus. 
Within the error region of this UGS there are two other X-ray sources: XRT J002402.05-421851.6 and XRT J002336.81-420837.1, both with a SNR$\sim$1.6. XRT J002402.05-421851.6 has the same inconsistency as XRT J002303.59-420509.6, without the evident of a jet in the SED and a flat X-ray spectrum, while XRT J002336.81-420837.1 is a star.

In the radio band, we find that four sources are coincident with the $\gamma$-ray position (see Fig. \ref{fig:Xskymap}), but only two of them (RACS J002258.9$-$420624 and RACS J002402.90-421309.0) are also coincident with an optical source. \citet[][]{Zhang_2022} propose RACS J002258.9$-$420624 and RACS J002402.90-421309.0 as potential counterparts of the UGS using statistical methods.

From the multi-wavelength SED of the radio source RACS J002258.9$-$420624 (see Fig. \ref{fig:SED_alt_all}), coincident with the optical source DES J002259.06-420624.5 (g = 22.8), the radio-optical points appears consistent with the $\gamma$-ray data (see Fig. \ref{fig:SED_LSP_HSP} as comparison), even though this source is not detected in the X-ray band. The SED exhibits a hint of a double-peak shape and also shows a flat radio spectrum. This could be a clue for the presence of a jet. The analysis performed using the tabulated SED templates suggests that the object resembles an E-HBL ($\chi^2_{\nu,\mathrm{min}} = 0.7$). However, given the limited number of data points in the optical band and the lack of X-ray observations, the result should be considered highly speculative.
This can be a possible counterpart of 4FGL J0023.6$-$4209; however, the absence of X-ray detection introduces some uncertainty regarding the association. Optical spectroscopy may provide further confirmation.
Regarding the other radio source, RACS J002402.90-421309.0, it has poor spectral coverage, which does not allow for SED analysis. 

\item \textit{3FGL J0031.6+0938}: The proposed counterpart of this $\gamma$-ray emitter is XRT J003159.86+093618.4, which coincides with the optical source SDSS J003159.86+093618.4, exhibiting a narrow-line Seyfert 1 (NLSY1) optical spectrum \citep[][]{paiano2019_ufo2}. Although the SED shows a blazar\textit{-like} double-peak trend (see Fig. \ref{fig:SED_RQ_noalt}), the source is not detected in the radio band. Moreover, the SED recognizing method fails to identify a blazar sub-class with a spectral shape similar to that of the object. 
There are no other possible X-ray counterparts in the 3$\sigma$ \textit{Fermi} error region (see Fig. \ref{fig:Xskymap}).
Investigating possible alternative counterparts among the radio sources coincident with the $\gamma$-ray emission, we find that the radio source VLASS1QLCIR J003119.56+094033.3, coincident with the optical source SDSS J003119.60+094033.0 (g=20.6), which has a spectrum published in the SDSS DR16 archive, is identified as a FSRQ (M$_g$ $\sim$ -25.8) 
at z=2.40604 \citep{Ahumada_2020}. 
From the multi-wavelength SED, given the low-energy emission trend, it is unlikely that this HSP blazar ($\log \nu^s_p$ = 15.1 $\pm$ 1.2) is the counterpart of the $\gamma$-ray emission, as such a system would require a very strong Compton dominance\footnote{The Compton dominance is defined as the ratio between the powers at the two peaks in the SED.}. The SED recognizing test also yields a chaotic solution, reinforcing the idea that this is not the true counterpart of 3FGL J0031.6+0938.
Five additional radio sources are located within the error box of this UGS (see Fig. \ref{fig:Xskymap}), but only two show optical emission, VLASS1QLCIR J003213.25+093703.8 and RACS J003147.4+094145. Both present a radio to optical spectrum consistent with a non-active galaxy, so incompatible with a $\gamma$-ray emission.  

Therefore, there are no plausible alternative counterparts for 3FGL J0031.6+0938. 


\item  \textit{4FGL J0117.9+1430}: Two X-ray sources are within the \textit{Fermi} error region of this UGS: the NLSY1 SDSS J011804.83+143158.6 with a SNR = 4.7 and XRT J011756.53+143059.8 with a SNR = 1.8. We proposed in \citet[][]{Ulgiati_2024} the NLSY1 as the possible low energy counterpart of this UGS. In this case as well, the SED shows a double-peak trend (see Fig. \ref{fig:SED_alt_all}), although the source is not detected in the radio band. We are also searching for potential alternative counterparts for this source.
Nine radio source are present within the $\gamma$-ray error ellipse (see Fig. \ref{fig:Xskymap}), but only three of them are also coincident with an optical source. One of them (VLASS1QLCIR J011734.73+141910.2) is a star. 
VLASS1QLCIR J011808.08+143504.7 is coincident with the optical source SDSS J011808.07+143504.7, which appears to be a non-active elliptical galaxy at redshift 0.427 \citep[][]{Ahumada_2020} with a steep radio spectrum ($\alpha$ > 0.5) and the low-energy emission appears too low compared with the $\gamma$-ray emission. 
%
The other radio source is VLASS1QLCIR J011810.42+142217.9, which is coincident with the optical source SDSS J011810.39+142218.2. No X-ray detection is revealed from the XRT image (see Fig. \ref{fig:Xskymap}). The IR and optical emissions are consistent with a giant elliptical galaxy at z $\sim$ 0.5. The SED is consistent with an LSP source ($\log \nu^s_p$ = 12.8 $\pm$ 0.6) with a high Compton dominance (see Fig. \ref{fig:SED_alt_all}). The SED recognizing method supports the interpretation that this object has a blazar-like shape and classifies it as an LSP/ISP ($\chi^2_{\nu,\text{min}}$ = 0.8). This could potentially represent a counterpart of 4FGL J0117.9+1430.

\item \textit{4FGL J0641.4+3349}: This UGS contain two X-ray sources within its error region: XRT J064111.2+334502.0 with SNR = 13.8 and XRT J064142.06+334930.5 with SNR = 1.7.
XRT J064111.2+334502.0 was proposed as counterpart in \citet[][]{Ulgiati_2024}, and from the optical spectroscopy, it is a Seyfert 1 galaxy at z=0.1657 \citep[][]{monroe_2016}. From its multi-wavelength SED,  the $\gamma$-ray emission could be too low with the rest of the spectrum considering a jetted source (see Fig. \ref{fig:SED_alt_all}, and Fig. \ref{fig:SED_LSP_HSP} for comparison). 
The trend of the SED up to the X-ray band is consistent with that of a classical Seyfert for which the X-ray emission is of thermal origin.
The other X-ray source, XRT J064142.06+334930.5 is coincident with the radio source VLASS1QLCIR J064142.03+334927.5 and the optical source PAN J064142.10+334927.7. The low energy emission is consistent with the $\gamma$-ray one (Fig. \ref{fig:SED_alt_all}), in agreement with the findings reported by \citet[][]{Fronte_2023}. Superimposing the template of an elliptical galaxy on IR/optical data (see \citealt{Chang_2019} for details), a photometric redshift z $\sim$ 0.4 can be set. According to the position of the synchrotron frequency peak ($\log \nu^s_p$ = 14.4 $\pm$ 0.5), this source can be classified as an ISP blazar \citep[see also][]{Fronte_2023}, as stated also from the SED recognizing method ($\chi^2_{\nu,min}$ = 0.3). Since no optical spectra are available in the literature, spectroscopic observations will be performed in order to determine the classification of this object and verify a possible association with the $\gamma$-ray detection.
Within the \textit{Fermi} error box of this UGS, eight additional radio sources are present (see Fig. \ref{fig:Xskymap}). However, only one of these (VLASS1QLCIR J064144.15+334913.1.) is spatially coincident with an optical source, which is classified as a star, making its association with the $\gamma$-ray emission unlikely.

\item \textit{4FGL J0938.8+5155}: The counterpart proposed in \citet[][]{Ulgiati_2024} for this $\gamma$-ray emitter, according with \citet[][]{Kerby_2021}, is the QSO SDSS J093834.72+515452.3. Observing its multi-wavelength SED (Fig. \ref{fig:SED_alt_all}), we note that both X-ray and $\gamma$-ray spectra appear flat, a trend not compatible with a jetted AGN.
There are no other possible X-ray counterparts within the $\gamma$-ray error region.
Five radio sources are found in the UGS error box (Fig. \ref{fig:Xskymap}), but none, except VLASS1QLCIR J093807.91+515103.7, exhibit X-ray emission or a low-energy SED consistent with the $\gamma$-ray flux based on our analysis.
VLASS1QLCIR J093807.91+515103.7 is the only source exhibiting both radio (with flat spectrum) and optical emissions compatible with the high-energy data (Fig. \ref{fig:SED_alt_all}). Although no X-ray emission is detected, the multi-wavelength SED seems to display a potential double-peak structure.
It coincides with the optical source SDSS J093807.91+515103.8, which exhibits a featureless optical spectrum as reported in the literature \citep[][]{Ahumada_2020}, a characteristic of BLL. The multi-wavelength SED is compatible with an ISP blazar, with $\log \nu^s_p $=$ 14.6 \pm 0.6$, and its photometric redshift can be at approximately z $\sim$ 0.3. The SED recognizing method suggests a classification of this object as an ISP/HSP ($\chi^2_{\nu,\text{min}}$ = 0.8), thus in agreement with the classification provided by the BlaST tool. \citet{Bruzewski_2021} had also observed that the radio emitter VLASS1QLCIR J093807.91+515103.7 was contained within the $\gamma$-ray error box of 4FGL J0938.8+5155, identifying it as a potential counterpart.


\item \textit{4FGL J1125.1+4811}: The proposed counterpart for this $\gamma$-ray emitter is the X-ray source XRT J112526.01+480922.8, coincident with the optical source SDSS J112526.27+480922.0 \citep{Ulgiati_2024}. From the multi-wavelength SED (see Fig. \ref{fig:SED_alt_all}), the X-ray and $\gamma$-ray spectra appear flat, so inconsistent with each other assuming a blazar\textit{-like} double-peak SED.
New \textit{Swift} observations were made in the region of this UGS after the study by \citet[][]{Ulgiati_2024}, significantly increasing the statistics and revealing two new sources: XRT J112508.81+481032.4, with a SNR = 4.2 $\sigma$, and XRT J112531.47+480832.1, with a SNR = 2.6 $\sigma$ (see Fig. \ref{fig:Xskymap}).
While XRT J112531.47+480832.1 is not coincident with any optical source, XRT J112508.81+481032.4 coincides with the VLASS1QLCIR J112508.90+481033.7 radio source and the optical source SDSS J112508.91+481033.4, for which a SDSS spectrum is available \citep[][]{Ahumada_2020}. The spectrum is featureless with a power-law shape, indicating a BLL nature. From the multi-wavelength analysis, the multi-wavelength SED shows a double-peaked ISP blazar shape ($\log \nu^s_p$ = 14.4 $\pm$ 0.5), and since the redshift is unknown and no hint of host elliptical galaxy is evident, a photometric redshift of $z > 0.9$ was estimated using the IR/optical light curve (see details in Sec. \ref{sec:gamma_origin}). The SED recognizing method states that this object is a HSP ($\chi^2_{\nu,min}$ = 0.3).
On the light of these results, XRT J112508.81+481032.4 can be proposed as a plausible counterpart of this UGS. 

\item \textit{4FGL J1234.7$-$0434}: The X-ray source XRT J123448.00-043246.2, spatially coincident with the optical source PAN J123448.05-043245.2, was suggested as a potential counterpart in \citet{paiano2019_ufo2} and \citet[][]{Chang_2019}, exhibiting a Seyfert2\textit{-like} optical spectrum. The multi-wavelength SED of this object deviates from the typical double-peak trend of jetted AGN, and it is also not detected in the radio band, making difficult to reconcile the $\gamma$-ray emission with that at lower energies (Fig. \ref{fig:SED_alt_all}, and Fig. \ref{fig:SED_LSP_HSP}).
There are no other possible X-ray counterparts in the 3$\sigma$ \textit{Fermi} error region.
We find that there are four radio sources inside the $\gamma$-ray error box of this UGS (see Fig. \ref{fig:Xskymap}). 
VLASS1QLCIR J123449.72-043929.1 does not coincide with any optical source. RACS J123459.9-043522 and VLASS1QLCIR J123438.16-043631.7 are classified as a star and a passive galaxy, respectively, making them unlikely candidates for the counterpart of this UGS.
The radio emitter VLASS1QLCIR J123444.23-043622.2, coincident with the optical source PAN J123444.21-043622.4, has a radio-optical SED compatible with the $\gamma$-ray emission of this UGS (Fig. \ref{fig:SED_alt_all}), and despite the lack of X-ray detection, the multi-wavelength SED show a potential double-peak shape.
Furthermore, it is worth to note that 4FGL J1234.7$-$0434 is also contained in the 3HSP catalogue \citep[][]{Chang_2019} (it has $\log \nu^s_p$ = 14.5 $\pm$ 0.7), associated with the source 3HSP J123444.2-043622. The SED recognizing method confirms the HSP classification ($\chi^2_{\nu,\text{min}}$ = 0.4).


\item \textit{4FGL J1256.8+5329}: This UGS contains two X-ray sources within its error region: XRT J125630.5+533202.2 with a SNR=4.2 and XRT J125645.37+532150.4 with SNR=2.4. 
XRT J125630.5+533202.2 is coincident with the optical source SDSS J125630.43+533204.3 \citep[][]{Ulgiati_2024}. Its SED exhibit a double-peak shape, although the source is not detected in the radio band (Fig. \ref{fig:SED_alt_all}). 
XRT J125645.37+532150.4 is coincident with the optical source SDSS J125645.08+532148.3 and multi-wavelength SED is consistent with a non-active galaxy (it does not exhibit a flat radio spectrum, and it has a thermal IR/optical spectrum, typical of a non-active galaxy). 
Therefore it can not be the counterpart of the UGS.

Within the \textit{Fermi} error region of this UGS, four radio sources are detected.
Two of them (VLASS1QLCIR J125630.02+533514.2 and VLASS1QLCIR J125728.16+532553.2) lack optical counterparts, while VLASS1QLCIR J125739.13+533422.7 appears to be a non-active galaxy, making it an unlikely counterpart to the $\gamma$-ray emitter.
The remaining source CRATES J125637+533417 (VLASS1QLCIR J125638.62+533423.5) exhibits a SED (see Fig. \ref{fig:SED_alt_all}) with a double-peak shape with $\log \nu^s_p$ = 12.9 $\pm$ 0.4, characteristic of an LSP, (see Fig. \ref{fig:SED_LSP_HSP} for a typical LSP SED). According to the SED recognizing method, it is possible to classify it as an LSP/ISP ($\chi^2_{\nu,\mathrm{min}}$ = 0.9). In the IR/optical range, there is a suggestion of elliptical host galaxy emission, allowing an estimated photometric redshift of approximately z $\sim$ 0.45 (see details in Sec. \ref{sec:gamma_origin}). Spectroscopic observations would help to clarify the nature of this source, which can be considerate as a plausible counterpart of the 4FGL J1256.8+5329. In the 3LAC catalogue \citep[][]{Ackermann_2015}, CRATES J125637+533417 is reported in the \textit{ASSOC2} field as a possible counterpart to 4FGL J1256.8+5329, and is classified as a likely AGN.



\item \textit{4FGL J1308.7+0347}: The proposed counterpart for this $\gamma$-ray emitter is the X-ray source XRT J130832.27+034405.4, coincident with the optical source SDSS J130832.10+034403.9 \citep[][]{Ulgiati_2024}. Its multi-wavelength SED (see \ref{fig:SED_RQ_noalt}) does not show a double-peak shape, nor a flat radio spectrum, and considering the classification as QSO proposed in \citet[][]{Veron_2006}, the likely origin of the X-ray component of XRT J130832.27+034405.4 is the disk, while that of the infrared one is the dusty torus. 
There are no other possible X-ray counterparts in the 3$\sigma$ \textit{Fermi} error region (see Fig. \ref{fig:Xskymap}).
From the search for potential radio counterparts within the error region, two radio sources have been identified (see Fig. \ref{fig:Xskymap}). The first, VLASS1QLCIR J130906.79+034608.0, coincides with the optical source SDSS J130906.78+034607.5. Its optical spectrum \citep[][]{Ahumada_2020} appears featureless, indicating a likely BLL\textit{-like} nature. Based on the analysis of the multi-wavelength SED, the synchrotron frequency peak is set at $\log \nu^s_p$ = 12.9 $\pm$ 0.5 (see Fig. \ref{fig:SED_LSP_HSP}), classifying it as a LSP blazar. However, the low-energy emission trend suggests that this source is unlikely to be the counterpart of the $\gamma$-ray emission, as such a system would require a significant Compton dominance.
The second radio source, RACS J130856.2+035304, coincides with an optical source, which exhibit a \textit{Gaia} proper motion and has a stellar\textit{-like} SED. 

In conclusion, we find that no plausible alternative counterparts for this UGS can be proposed.

\item \textit{4FGL J1430.6+1543}: The X-ray source XRT J143057.9+154556.0, coinciding with the Seyfert 1 galaxy SDSS J143058.03+154555.6, was proposed as a potential counterpart (see \citealt{Ulgiati_2024}). 
However, assuming a double-peak blazar\textit{-like} shape for its multi-wavelength SED (Fig. \ref{fig:SED_RQ_noalt}), the observed radio and X-ray emissions are not consistent with the expected $\gamma$-ray flux, which appears too low.
Within the $\gamma$-ray error region, we find 10 radio sources (see  Fig. \ref{fig:Xskymap}), but none coincides with an optical counterpart. 
Consequently, no alternative counterparts can be proposed for this UGS.

\item \textit{4FGL J1539.1+1008}: The low energy counterpart proposed for this UGS is the X-ray source XRT J153848.5+101841.7, classified as a Seyfert 1 galaxy \citep[][]{Toba_2014}. 
The SED (see Fig. \ref{fig:SED_RQ_noalt}) exhibits a double-peak shape, although the source has no radio detection in the literature. In the absence of a flat radio spectrum, it is difficult to connect the $\gamma$-ray emission with this source. 
In the \textit{Fermi} error region of this UGS there is another X-ray source (see Fig \ref{fig:Xskymap}), that is not coincident with an optical source, and 10 radio objects.
Two of them have an optical spectrum available in the SDSS archive \citep[][]{Ahumada_2020}. One is VLASS1QLCIR J153917.29+101713.6, an AGN located at z=0.09692 \citep[][]{Ahumada_2020}. However, due to the lack of X-ray detection and a flat radio spectrum, this association is unlikely, and we cannot consider it a potential counterpart for the $\gamma$-ray detection.
The other one, VLASS1QLCIR J153938.94+101311.6, is coincident with the spiral galaxy SDSS J153938.96+101311.7 located at z=0.034. Further analysis is required to investigate these alternative associations. 
The remaining radio sources are not spatially coincident with optical counterparts, and there are insufficient spectral data points available to construct and analyze their SED. 
Therefore, no plausible alternative counterparts can be proposed for this UGS.

\item \textit{4FGL J2030.0$-$0310}: The proposed counterpart XRT J203014.34-030721.9, coincident with the Seyfert 2 galaxy PAN J203014.27-030722.56, displays a multi-wavelength SED (Fig. \ref{fig:SED_RQ_noalt}) where the $\gamma$-ray emission is comparatively weak relative to the radio and X-ray components, assuming a typical blazar\textit{-like} double-peak shape.
Within the \textit{Fermi} error box, three additional radio sources are present (see Fig. \ref{fig:Xskymap}), although none of them can be associated to any optical or X-ray sources. 
Consequently, no alternative counterpart candidates emerge from the analysis of this UGS.

\item \textit{4FGL J2212.4+0708}: The X-ray source XRT J221230.98+070652.5, coinciding with the optical source SDSS J221230.98+070652.5, has been suggested as the counterpart of this UGS and classified as a QSO \citep{paiano2019_ufo2}. From the multi-wavelength analysis (Fig. \ref{fig:SED_RQ_noalt}), the SED displays a double-peak shape with a pronounced $\gamma$-ray cut-off, though no radio data are currently available. 
No other X-ray sources can be detected in the UGS error box, and only one radio source, RACS J221229.1+070744, is found (see Fig. \ref{fig:Xskymap}), coinciding with the optical source SDSS J221229.11+070743.9. This source does not show a flat radio spectrum, typical of jet-dominated objects, and the $\gamma$-ray emission is too intense to be plausibly associated with the observed radio emission. This weakens the likelihood of an association between the radio source and the $\gamma$-ray emission and for this reason no other plausible counterparts can be proposed.


\end{itemize}

Table \ref{tab:new_ass} summarizes the results obtained from the search for new "alternative" counterparts for the 13 radio-quiet sources in the sample.
For six objects, no alternative counterparts were found, leaving the candidates proposed by \citet[][]{Ulgiati_2024, paiano2017_ufo1,paiano2019_ufo2} as the most plausible for these UGSs. Further analysis is needed to clarify their emission mechanisms and to interpret the observed SEDs.

\begin{figure}
\hspace{-0.5cm}
\centering
   \includegraphics[width=9.5truecm]{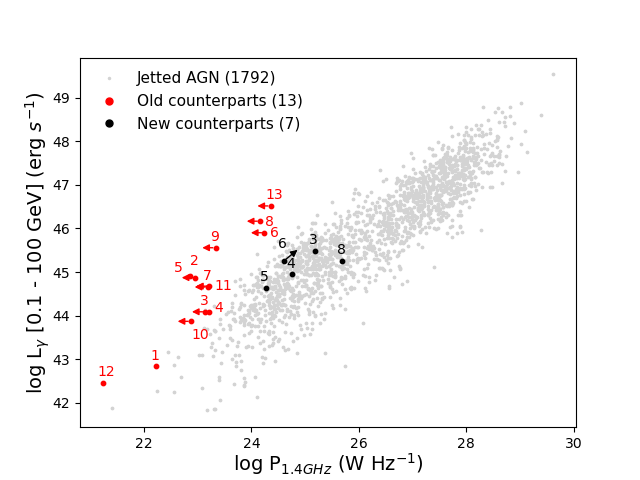}
\caption{L$_{\gamma-ray}$ vs. P$_{1.4GHz}$ for the alternative counterpart (black), the previously proposed counterparts (red), shown to track the UGS evolution in the plot, and for the comparison sample, the 4FGL-DR4 jetted AGN (light-grey points). Diagonal arrows denote lower limits on redshift and therefore powers, horizontal arrows represent radio luminosity upper limits. To each UGS an identifying number has been assigned, as shown in Table \ref{tab:confr_gamma}. It is worth noting that among the sources marked in red, some overlap. This is the case for the sources marked as 3 and 4, 7 and 11, 2 and 5.
}
\label{fig:diagn_Lgamma_Lradio_new}
\end{figure}

We have updated the L$_{\gamma}$ vs. P$_{1.4GHz}$ diagram (Fig. \ref{fig:diagn_Lgamma_Lradio_new}), including the 7 new "alternative" counterparts identified starting from the radio sources within the UGS region, regardless of X-ray emission presence or significance. 
These sources have all $R > 70$ and 
fall within the region typically occupied by jetted AGN, supporting the hypothesis that they are likely blazar-type objects and could be associated with the detected $\gamma$-ray emission. 
Regarding the UGSs 4FGL J0023.6-4209 and 4FGL J1234.7-0434, they do not appear in the plot because their new counterparts lack redshift information. This is due to the absence of an IR/optical light curve and insufficient optical coverage in the SEDs. 
Optical spectroscopy will certainly help in providing more conclusive classification for these sources, which in any case still have $R > 70$.

\begin{table*}
\begin{center}
\caption{Lower energy counterpart for the subset of radio-quiet $\gamma$-ray sources analysed in this paper} 
\resizebox{16cm }{!}{
\begin{tabular}{llllrrrcll}
\hline 
ID & 4FGL Name & Counterpart proposed & Alternative  & $f^{\rm radio}_{\nu}$ & $f^{\rm opt}_{\nu}$ & $R$ & Class & z & Reference\\   
          & & in previous works & counterpart  & & & &  &  & \\  
\hline          
          & &                   &              & & [$\times$ 10$^{-28}$] & & & & \\
\hline

1 & 4FGL J0023.6$-$4209 & XRT J002303.59$-$420509.6 & RACS J002258.9$-$420624 & 42.7 & 0.4 & 11020 & ISP & \magg? & this work\\
2 & 3FGL J0031.6+0938 & XRT J003159.86+093618.4 & - & <0.4 & 7.6 & <5 & NLSY1 & \magg0.2207 & \citet{paiano2019_ufo2}\\
3 & 4FGL J0117.9+1430 & XRT J011804.79+143159.6 & VLASS1QLCIR J011810.42+142217.9 & 22.1 & 0.9 & 2511 & LSP & \magg0.5$^*$ & this work\\
4 & 4FGL J0641.4+3349 & XRT J064111.24+334502.0 & VLASS1QLCIR J064142.03+334927.5 & 10.5 & 1.4 & 749 & ISP & \magg0.4$^*$ & \citet{Chang_2019}\\
5 & 4FGL J0938.8+5155 & XRT J093834.50+515454.8 & VLASS1QLCIR J093807.91+515103.7 & 6.8 & 9.5 & 71 & BLL & \magg0.3$^*$ & this work\\
6 & 4FGL J1125.1+4811 & XRT J112526.01+480922.8 & VLASS1QLCIR J112508.90+481033.7 & 1.8 & 2.1 & 83 & BLL & >0.9$^*$ & this work\\
7 & 4FGL J1234.7$-$0434 & XRT J123448.00$-$043246.2 & VLASS1QLCIR J123444.23$-$043622.2 & 18.4 & 5.3 & 347 & ISP & \magg? & this work\\
8 & 4FGL J1256.8+5329 & XRT J125630.54+533202.3 & VLASS1QLCIR J125638.62+533423.5 & 88.3 & 0.4 & 22645 & LSP & \magg0.45$^*$ & this work\\
9 & 4FGL J1308.7+0347 & XRT J130832.27+034405.4 & - & < 0.8 & 47.9 & <1 & QSO & \magg0.6193 & \citet{Ahumada_2020} \\
10 & 4FGL J1430.6+1543 & XRT J143057.97+154556.1 & - & <0.3 & 39.8 & <1 & Seyfert 1 & \magg0.1633 & \citet{Ahumada_2020} \\
11 & 4FGL J1539.1+1008 & XRT J153848.51+101841.7 & - & <0.3 & 17.4 & <2 & Seyfert 1 & \magg0.2345 & \citet{Ahumada_2020}\\
12 & 4FGL J2030.0$-$0310 & XRT J203014.34$-$030721.9 & - & 0.3 & 69.2 & 1 & Seyfert 2 & \magg0.036 & \citet{Jones_2009}\\
13 & 4FGL J2212.4+0708 & XRT J221230.98+070652.5 & - & <0.3 & 2.1 & <13 & QSO & \magg1 & \citet{paiano2019_ufo2}\\ 
\hline
\end{tabular}
}
\label{tab:new_ass}
\end{center}
\raggedright
\footnotesize{\textbf{Note.} Column 1: Identification number; Column 2: 4FGL Name; Column 3: Name of the counterpart proposed in \citet[][]{paiano2017_ufo1,paiano2019_ufo2,Ulgiati_2024};  Column 4: Name of the alternative counterpart proposed in this work; Column 5: 2 - 4 GHz radio density flux (mJy); Column 6: g-band optical density flux (erg cm$^{-2}$ s$^{-1}$ Hz$^{-1}$); Column 7: Fractional variability in $\gamma$-ray band; Column 8: \textit{radio-loudness} R defined as the ratio between radio flux density and optical flux density of the nuclear component; Column 9: Classification; Column 10: redshift (z); Column 11: Reference of the redshift.\\
(*) photometric redshift.
}
\end{table*}

\section{Conclusions}
\label{sec:conclusion_P}

We analyzed the multi-wavelength emission of a sample of 77 \textit{Fermi} sources, initially classified as unassociated in the 3FGL or 4FGL-DR4 catalogs. Their lower-energy counterparts were proposed and spectroscopically studied in \citet{paiano2017_ufo1, paiano2019_ufo2} and \citet{Ulgiati_2024}, or independently identified by other groups. Our objective was to characterize their multi-wavelength SEDs, identify potential new masquerading BL Lacs, and investigate the proposed association between the $\gamma$-ray flux and lower-energy emissions.

The multi-wavelength analysis reveals that the radio-loud sources in our sample exhibit SEDs consistent with the characteristic double-peak structure typical of blazars. This classification allows us to categorize them further as LSP, ISP, HSP or E-HSP sources, based on the location of the synchrotron emission peak in their SEDs. The majority are classified as HSP (46 objects), 11 as ISP and 7 as LSP.

In the search for masquerading BLL, 9 candidates are identified based on their absolute and relative emission intensities, representing approximately 15\% of the radio-loud sample, with a robust upper limit of 30\%. This proportion is lower than in previous studies, where \citet{Paiano_2023} found a percentage of masquerading BLL of 34\% among a sample of neutrino emitter blazars.

For the 13 radio-quiet objects (mainly classified Seyfert/QSO galaxies), the multi-wavelength SED cannot be attributed to jet emissions or star formation processes. This makes it challenging to reconcile the observed $\gamma$-ray emission with that of their lower-energy counterparts. Consequently, we explored potential alternative counterparts by examining the radio sources within the UGS error ellipses, independently of their X-ray detection or significance. Among these, we identified possible alternative counterparts for 7 out of the 13 sources, all of the jetted type, while no plausible alternatives were found for the remaining 6 UGS.
The investigation of these new counterparts will form the basis for an upcoming spectroscopic campaign, aimed at refining their classifications and further validating the proposed associations.

\section*{Data availability}
All data (except for 4FGLJ2030.0-0310 radio data, that are proprietary data) are publicly available from Swift Archive (https://www.swift.ac.uk/swift\_portal/), VizieR Database (https://vizier.cds.unistra.fr/viz-bin/VizieR), PanSTARRS-1 Image Archive (https://ps1images.stsci.edu/cgi-bin/ps1cutouts), ESO Digital Sky Survey (https://archive.eso.org/dss/dss), SDSS-DR16 Archive (https://skyserver.sdss.org/dr16/en/tools/chart/navi.aspx), CSIRO Data Access Portal (https://data.csiro.au/domain/casdaObservation), 6df Galaxy Redshift Database (http://www-wfau.roe.ac.uk/6dFGS), Asto Data Lab (https://datalab.noirlab.edu/sia.php), LoTSS Data Release 2 (DR2) Database (https://lofar-surveys.org/dr2\_release.html), VOU-Blazars V2.00 (https://github.com/ecylchang/VOU\_Blazars), \citet{paiano2017_ufo1}, \citet{paiano2019_ufo2}, \citet[][]{Ulgiati_2024b}, 4FGL-DR4 \textit{Fermi} catalogue (https://fermi.gsfc.nasa.gov/ssc/data/access/lat/14yr\_catalog/), Space Science Data Center (SSDC, https://www.ssdc.asi.it/).

\section*{Acknowledgements}

AU thanks the European Southern Observatory (ESO) for hosting him as a visiting PhD student at their facilities in Garching (Munich), during which this work was conceived and structured. CP acknowledges support from PRIN MUR 2022 SEAWIND 2022Y2T94C funded by NextGenerationEU and INAF Large Grant BLOSSOM.




\bibliographystyle{mnras}
\bibliography{biblio} 




\appendix

\section{VOU-BLAZARS catalogues used to built the multi-wavelength SED}
\label{sec:appendice}
We report the complete list of the catalogues queried in VOU-BLAZARS and used to create the multi-wavelength SED of the UGSs: NVSS, FIRST, SUMSS, VLASSQL, 2SXPS, SDS82, 1OUSX, RASS, XMMSL2, 4XMM-DR11, BMW, WGACAT, IPC2E, IPCSL, ChandraCSC2, MAXI, eROSITA-EDR, ZWCLUSTERS, PSZ2, ABELL, MCXC, 5BZCat, SDSSWHL, SWXCS, 3HSP, \textit{Fermi}GRB, MilliQuas, BROS, MST9Y, PULSAR, F2PSR, F357cat, XRTDEEP, WISH352, GLEAM, TGSS150, VLSSR, LoTSS, PMN, GB6, GB87, ATPMN, AT20G, NORTH20, CRATES, F357det, KUEHR, PCNT, PCCS44, PCCS70, PCCS100, PCCS143, PCCS217, PCCS353, PCCS2, ALMA, SPIRE, H-ATLAS-DR1, H-ATLAS-DR2, HATLAS-DR2NGP, H-ATLAS-DR2SGP, AKARIBSC, IRAS-PSC, WISE, WISEME, NEOWISE, 2MASS, USNO, SDSS, HSTGSC, PanSTARRS, GAIA, SMARTS, UVOT, GALEX, XMMOM, CMA, EXOSAT, XRTSPEC, OUSXB, OUSXG, OULC, BAT105m, BEPPOSAX, NuBlazar, 3FHL, 2FHL, 2BIGB, 4FGL-DR3, 2AGILE, FermiMeV, FMonLC.

\section{X-RAY SKYMAPS}

Here we provide the X-ray skymaps for the 13 radio-quiet UGSs analysed in this paper. 
The yellow and cyan ellipses are respectively the $2\sigma$ and $3\sigma$ \textit{Fermi} $\gamma$-ray error regions. X-ray detections, found through \textit{Swift}/XRT analysis, are reported as white circles. The X-ray counterpart proposed in \citet{Ulgiati_2024, paiano2017_ufo1, paiano2019_ufo2} are shown as green circles.

\setcounter{figure}{0}
\begin{figure*}
\center
   \includegraphics[width=5.5truecm]{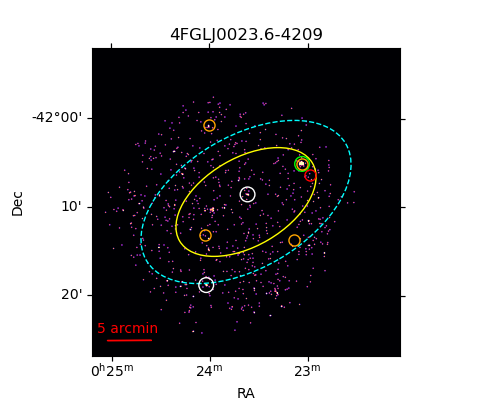}
   \includegraphics[width=5.5truecm]{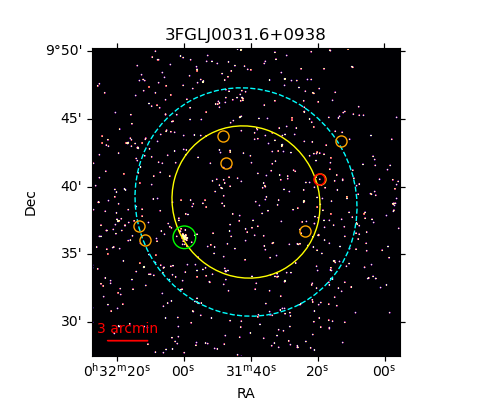}
   \includegraphics[width=5.5truecm]{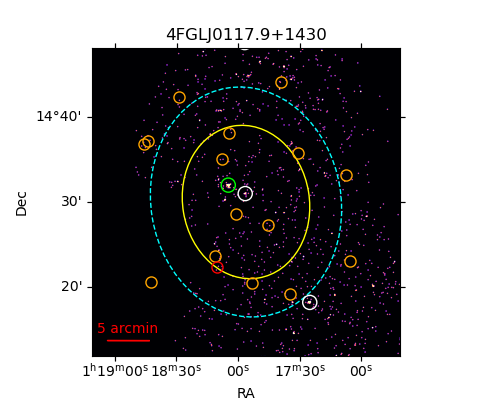}
   \includegraphics[width=5.5truecm]{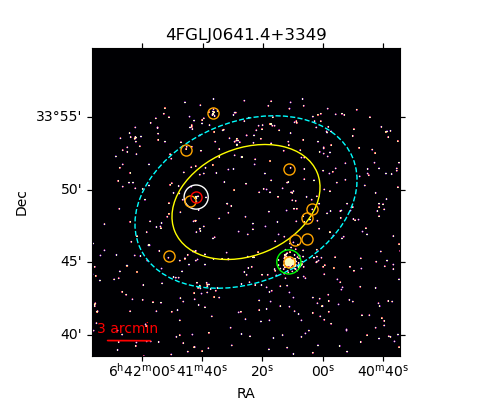}
   \includegraphics[width=5.5truecm]{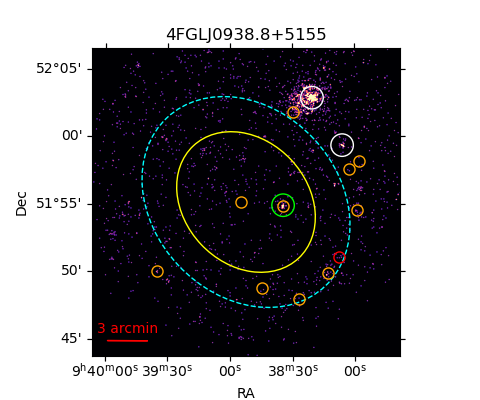}
   \includegraphics[width=5.5truecm]{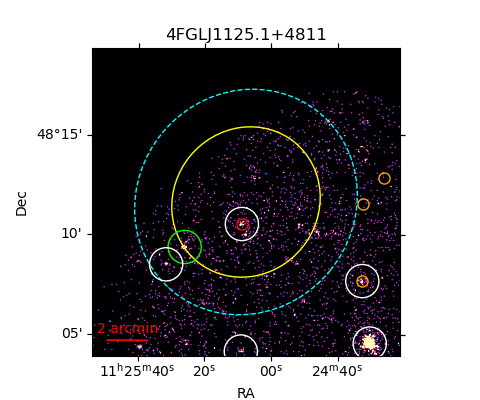}
\includegraphics[width=5.5truecm]{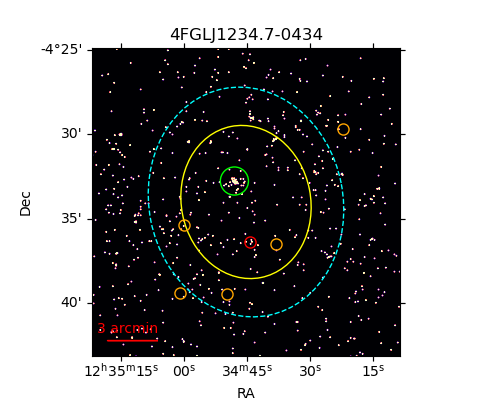}
   \includegraphics[width=5.5truecm]{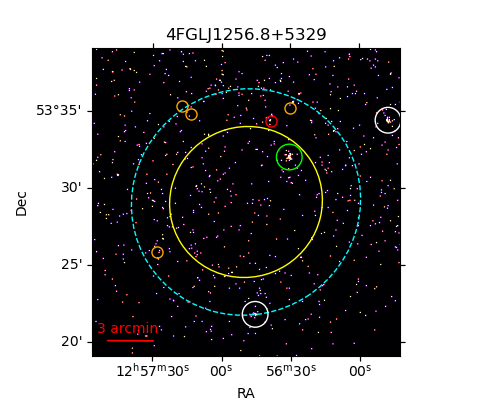}
   \includegraphics[width=5.5truecm]{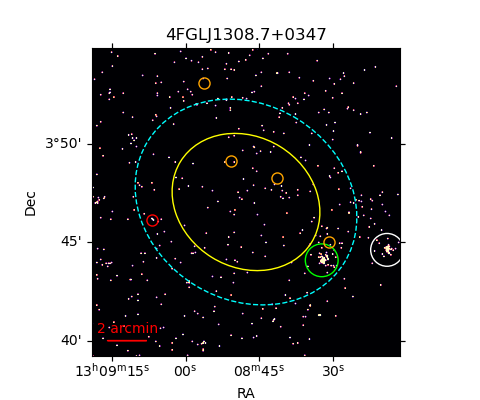}
   \includegraphics[width=5.5truecm]{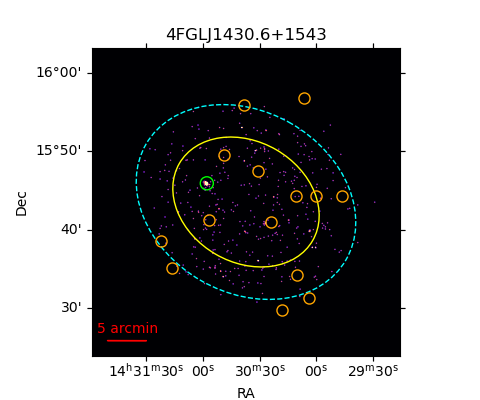}
   \includegraphics[width=5.5truecm]{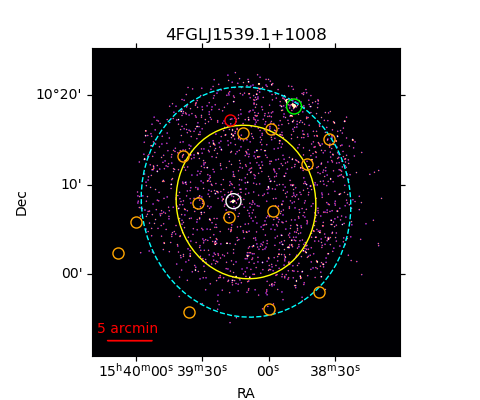}
   \includegraphics[width=5.5truecm]{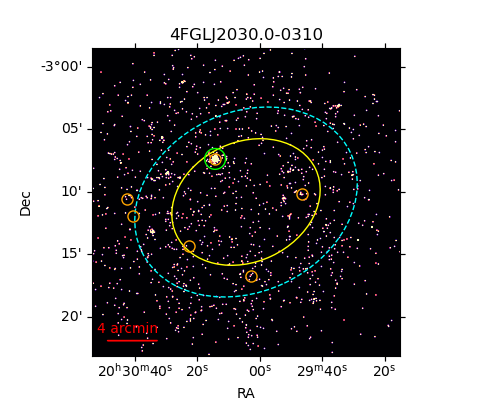}
   \includegraphics[width=5.5truecm]{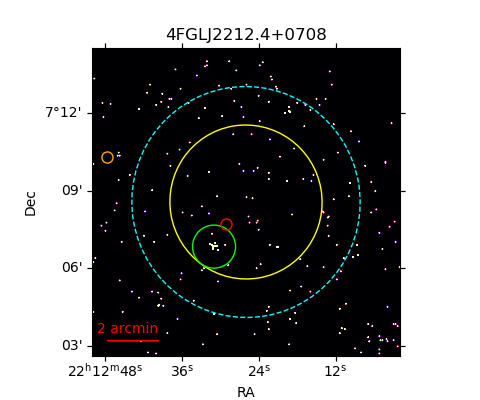}
\caption{Swift/XRT images of the 13 UGSs associated to radio-quiet counterparts. The yellow and cyan ellipses are respectively the 2$\sigma$ and 3$\sigma$ \textit{Fermi} $\gamma$-ray error regions. X-ray detections found through \textit{Swift}/XRT analysis, are reported as green circles (X-ray counterpart proposed in \citet[][]{paiano2017_ufo1}, \citet[][]{paiano2019_ufo2}, \citet{Ulgiati_2024} with detection significance $>$ 3) and white circles (other X-ray sources resulting from a new updated analysis and with detection significance $>$1). Radio detections are reported as orange circles (the red circles are the radio sources proposed as new counterparts in this work).} 
\label{fig:Xskymap}
\end{figure*}

\section{Spectral energy distributions of UGSs and their counterparts}

The Spectral Energy Distributions (SEDs) of the radio-loud (Fig. \ref{fig:SED}) and radio-quiet (Fig. \ref{fig:SED_RQ_noalt} and \ref{fig:SED_alt_all}) objects analyzed in this paper. 
At the top of each sub-figure, we provide the
name of the UGS, the name of the proposed X-ray or optical counterpart, the classification given by optical spectroscopy, and the redshifts (red-
shifts marked with a "*" are photometric, while the others are
spectroscopic). BLL stands for BL Lac, FSRQ for Flat Spectrum Radio Quasar, BLG for BLL galaxy dominated, FR 1 for Fanaroff-Riley 1, radio galaxies exhibiting an extended jet structure, with the emission from the core dominating over that of the lobes, QSO for quasi-stellar object, Seyfert 1 (2) for type 1 (2) Seyfert galaxy, NLSY1 for narrow line Seyfert 1 galaxy.

\setcounter{figure}{0}
\begin{figure*}
\center
\includegraphics[width=0.33\textwidth, angle=0]{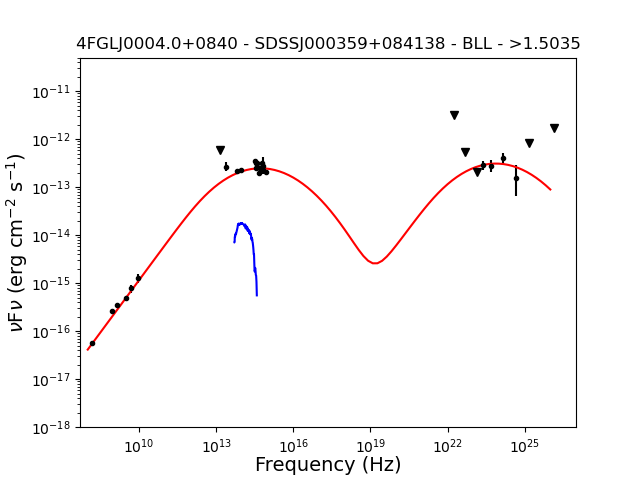}
\includegraphics[width=0.33\textwidth, angle=0]{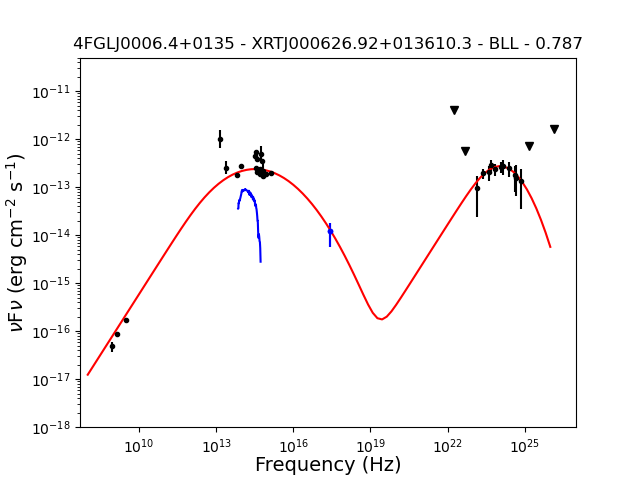}
\includegraphics[width=0.33\textwidth, angle=0]{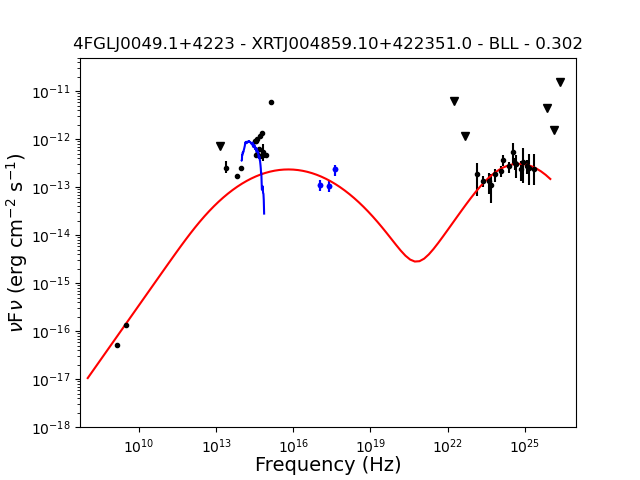}
\includegraphics[width=0.33\textwidth, angle=0]{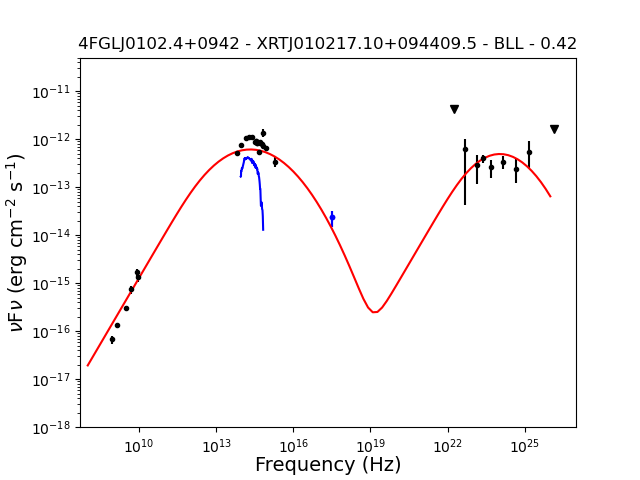}
\includegraphics[width=0.33\textwidth, angle=0]{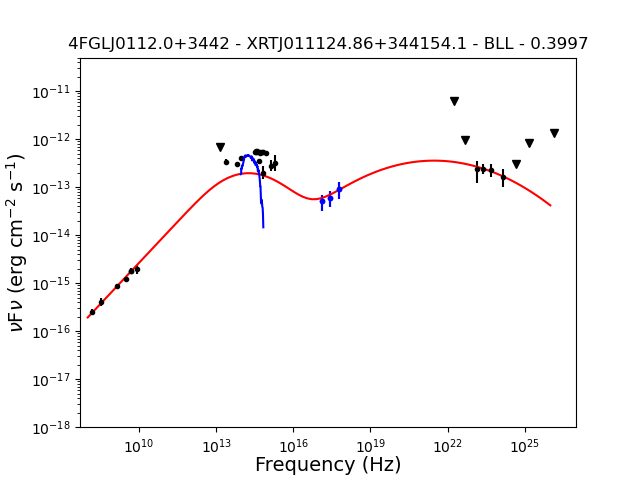}
\includegraphics[width=0.33\textwidth, angle=0]{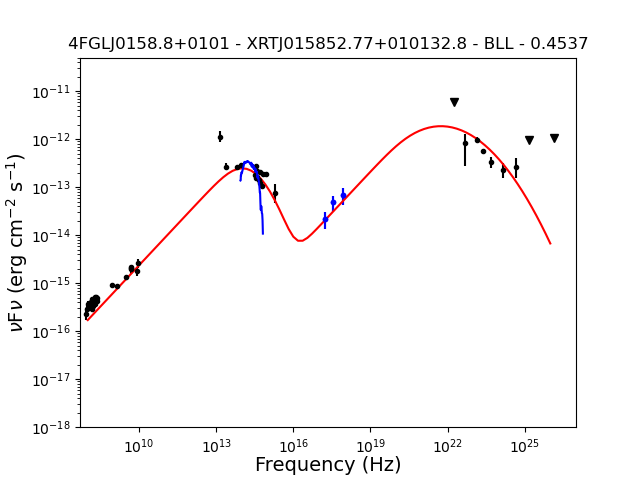}
\includegraphics[width=0.33\textwidth, angle=0]{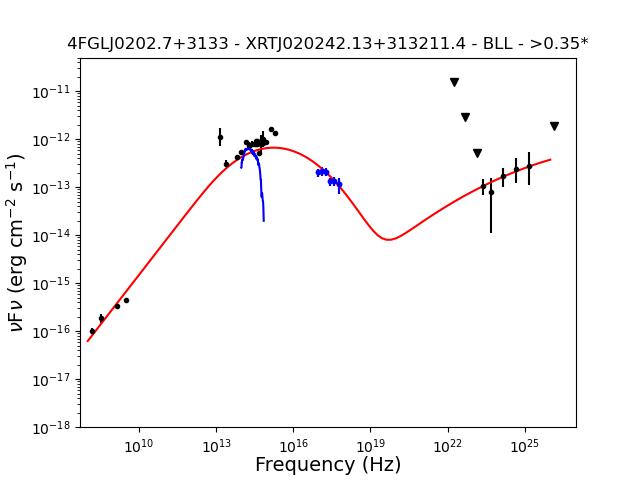}
\includegraphics[width=0.33\textwidth, angle=0]{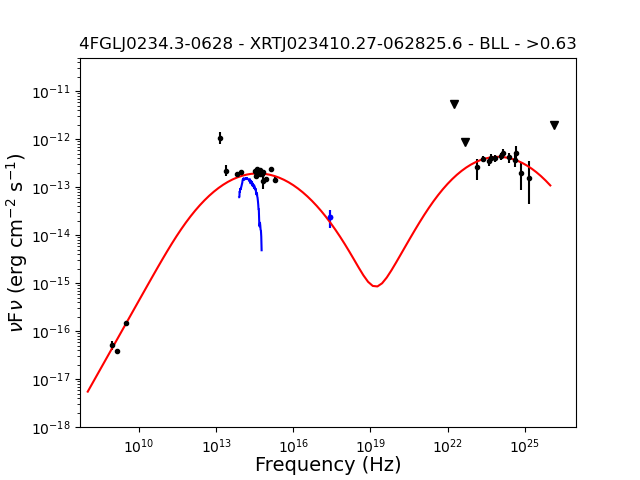}
\includegraphics[width=0.33\textwidth, angle=0]{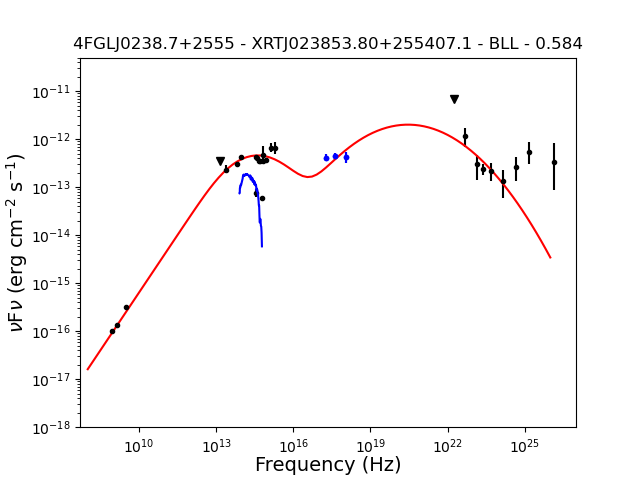}
\includegraphics[width=0.33\textwidth, angle=0]{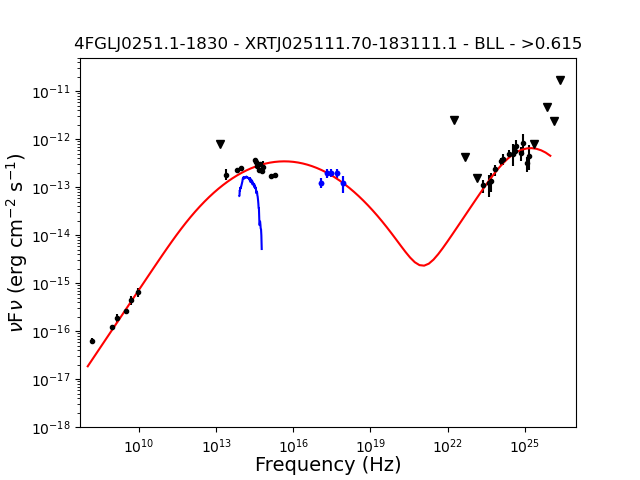}
\includegraphics[width=0.33\textwidth, angle=0]{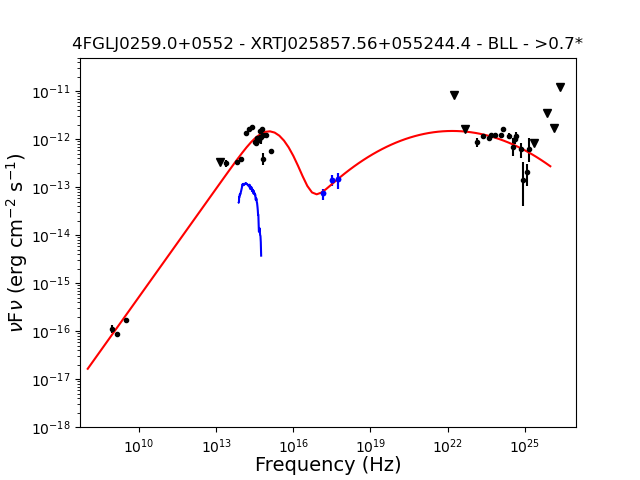}
\includegraphics[width=0.33\textwidth, angle=0]{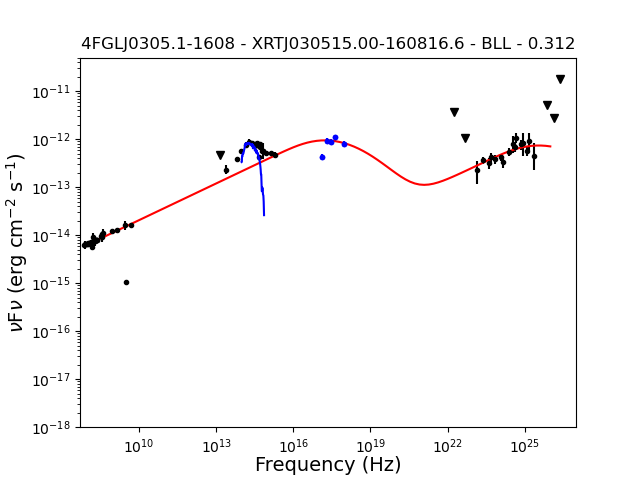}
\includegraphics[width=0.33\textwidth, angle=0]{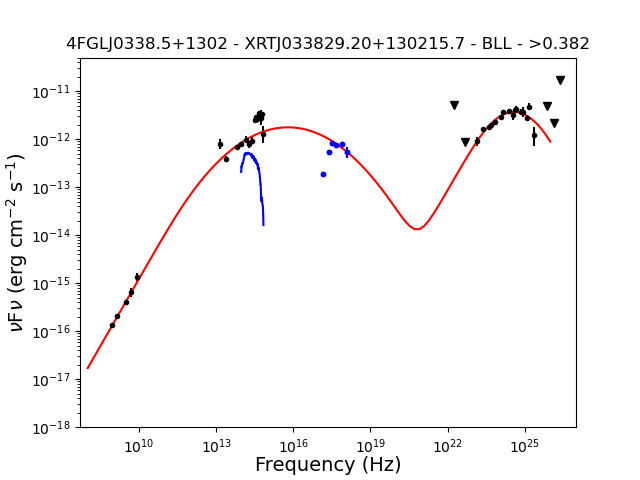}
\includegraphics[width=0.33\textwidth, angle=0]{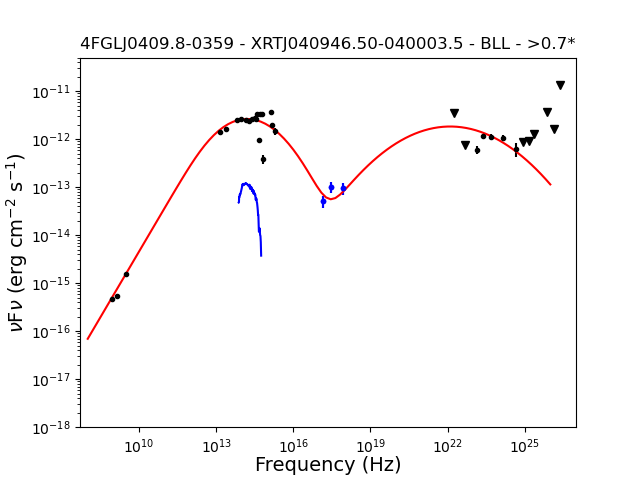}
\includegraphics[width=0.33\textwidth, angle=0]{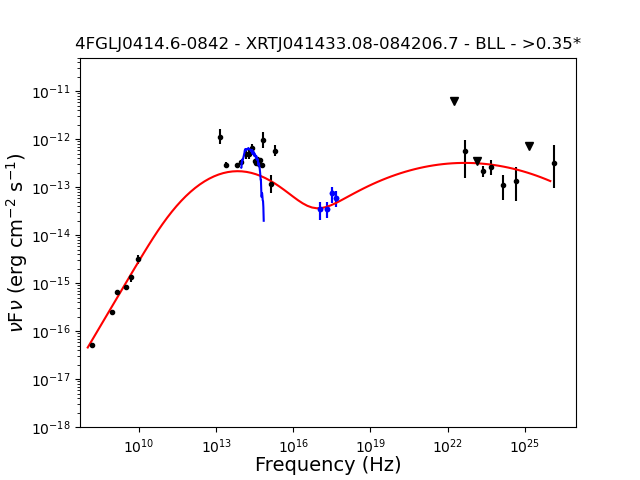}
\caption{Multiwavelenght SED of the UGS radio-loud counterparts proposed in \citet[][]{Ulgiati_2024, paiano2017_ufo1, paiano2019_ufo2} and analysed in this paper. At the top of each figure, there is the UGS and counterparts name, the classification, and the redshift (if marked with a '*' the redshift is photometric, otherwise it is spectroscopic). Black points are data from VOU-Blazar, the blue points are X-ray data from our analysis \citep[][]{Ulgiati_2024b}. The triangle points indicates upper limits. The red curve emulates the typical double-peaked shape of blazars. The blue one is the template of a giant elliptical galaxy at the object redshift.}  
\label{fig:SED}
\end{figure*}

\setcounter{figure}{0}
\begin{figure*}
\center
\includegraphics[width=0.33\textwidth, angle=0]{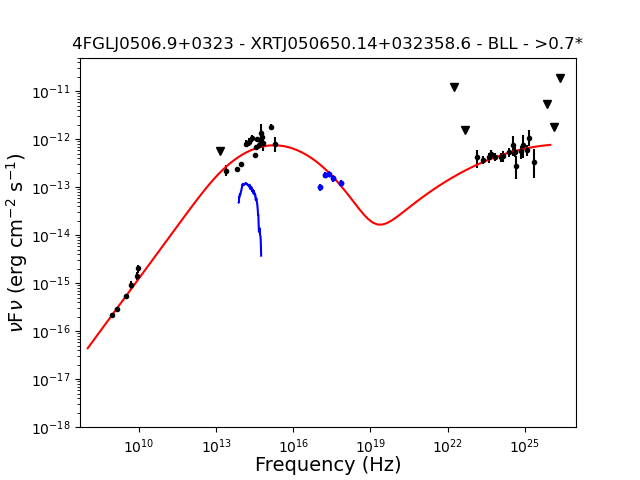}
\includegraphics[width=0.33\textwidth, angle=0]{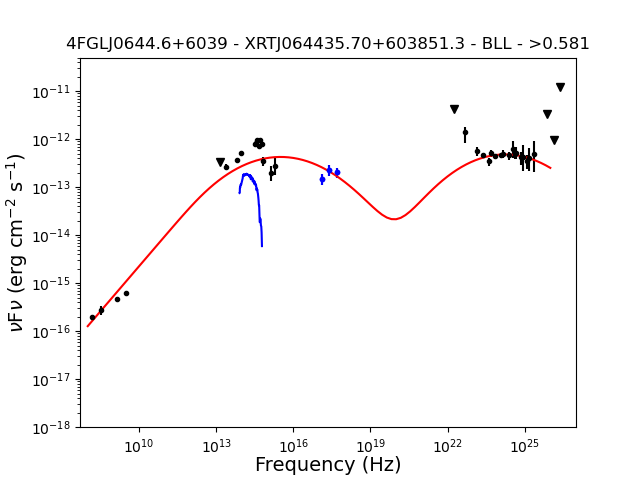}
\includegraphics[width=0.33\textwidth, angle=0]{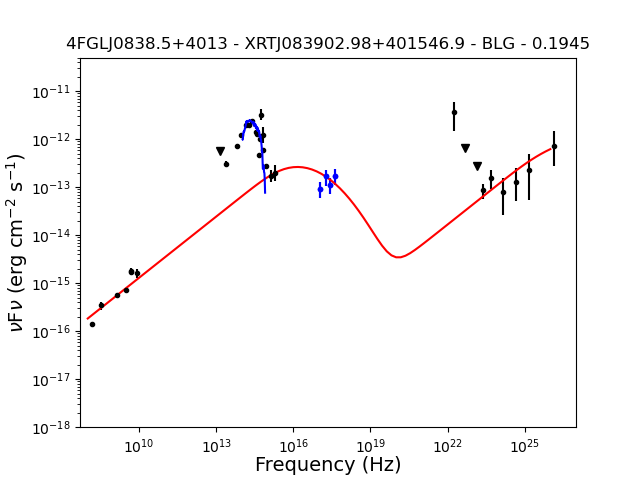}
\includegraphics[width=0.33\textwidth, angle=0]{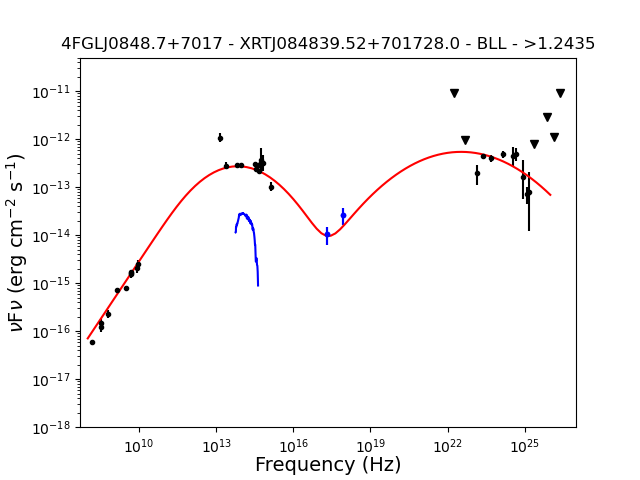}
\includegraphics[width=0.33\textwidth, angle=0]{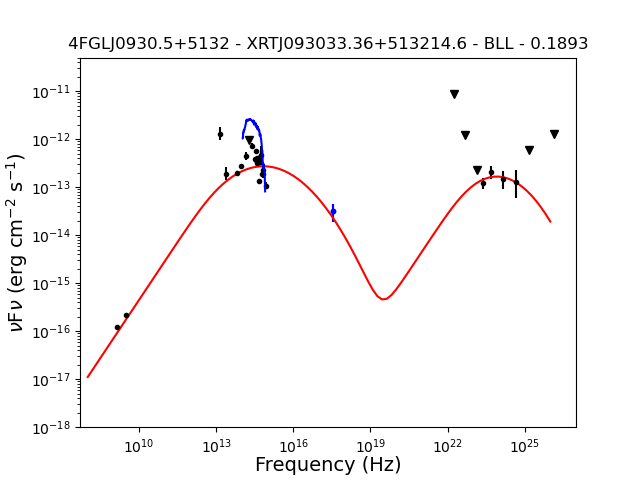}
\includegraphics[width=0.33\textwidth, angle=0]{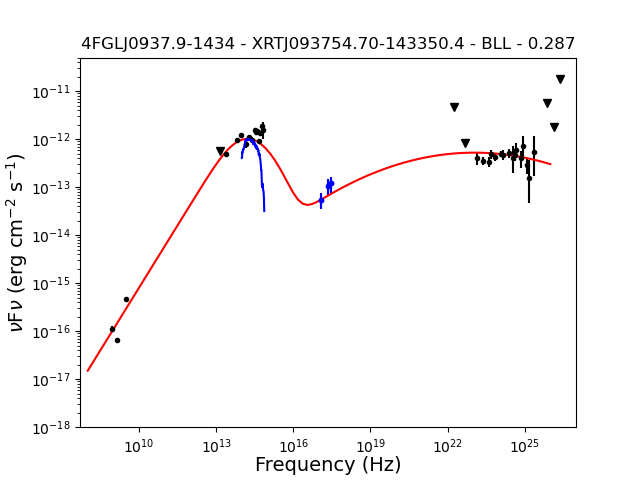}
\includegraphics[width=0.33\textwidth, angle=0]{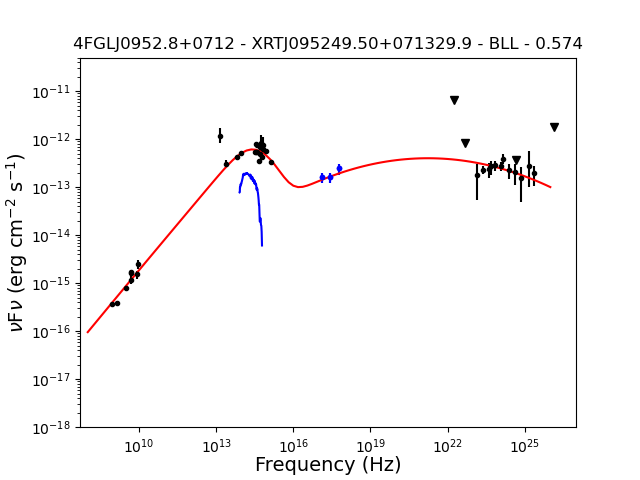}
\includegraphics[width=0.33\textwidth, angle=0]{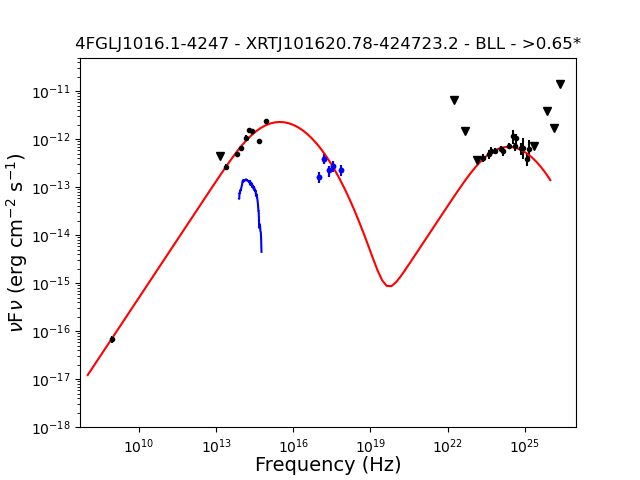}
\includegraphics[width=0.33\textwidth, angle=0]{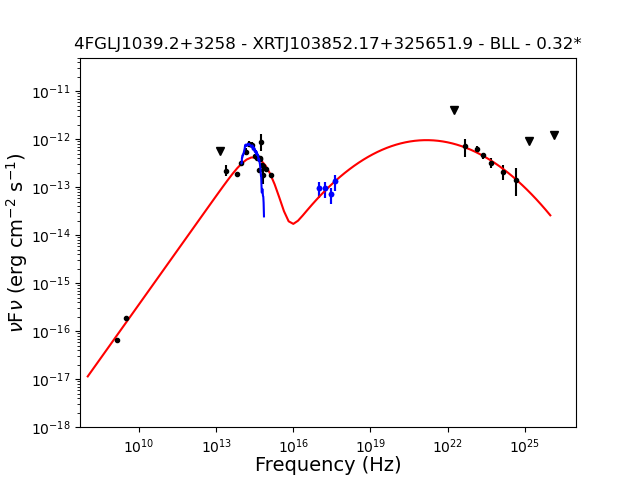}
\includegraphics[width=0.33\textwidth, angle=0]{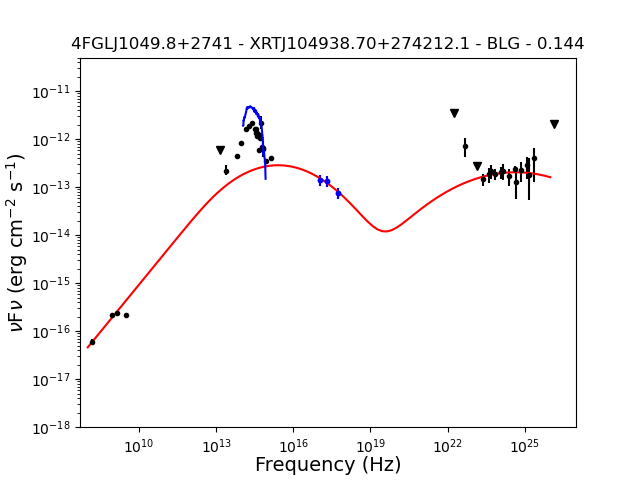}
\includegraphics[width=0.33\textwidth, angle=0]{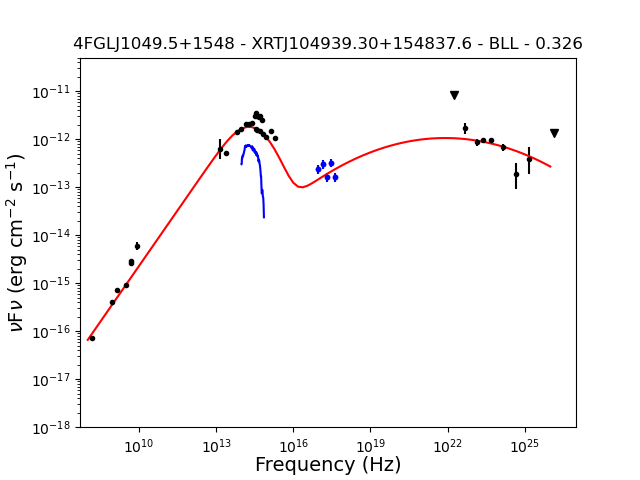}
\includegraphics[width=0.33\textwidth, angle=0]{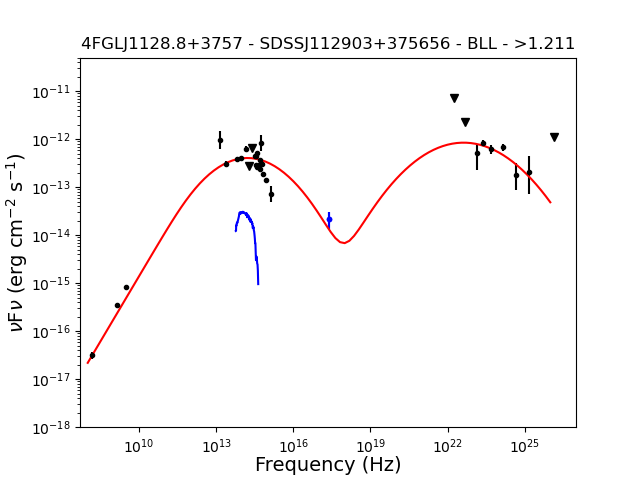}
\includegraphics[width=0.33\textwidth, angle=0]{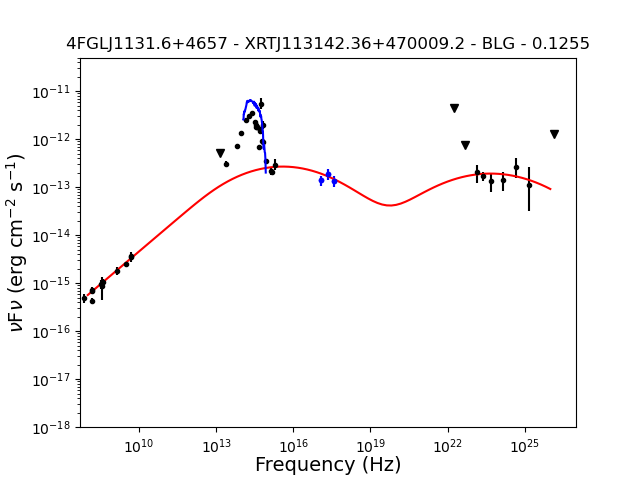}
\includegraphics[width=0.33\textwidth, angle=0]{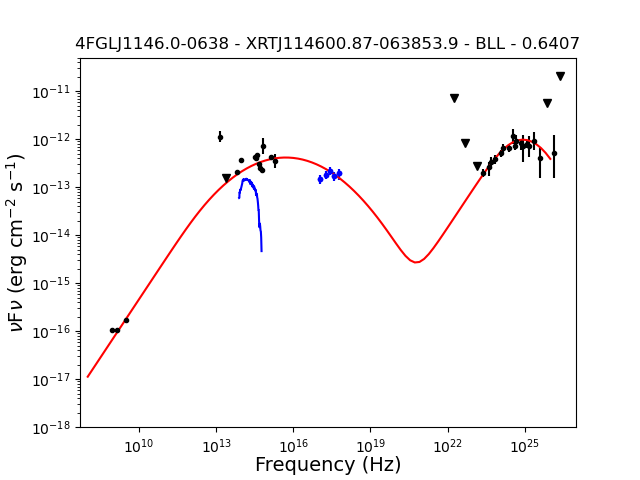}
\includegraphics[width=0.33\textwidth, angle=0]{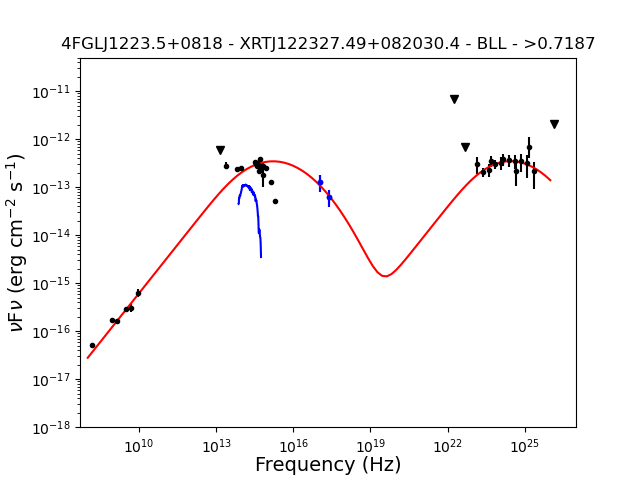}
\caption{Continued.}  
\label{fig:SED}
\end{figure*}

\setcounter{figure}{0}
\begin{figure*}
\center
\includegraphics[width=0.33\textwidth, angle=0]{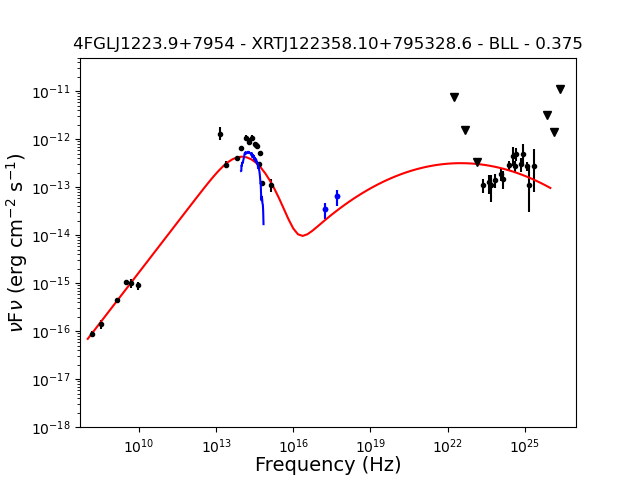}
\includegraphics[width=0.33\textwidth, angle=0]{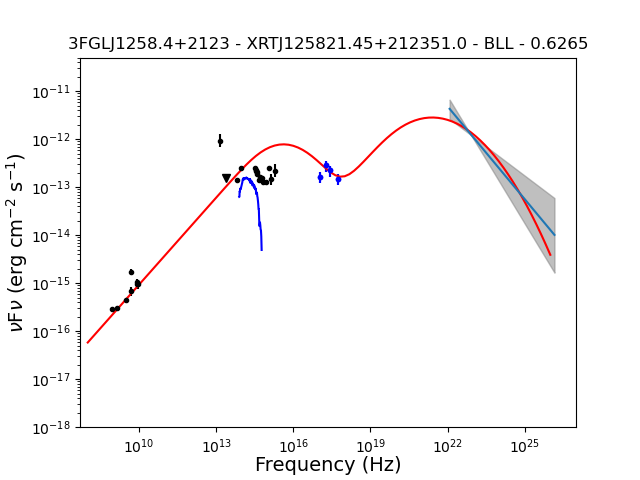}
\includegraphics[width=0.33\textwidth, angle=0]{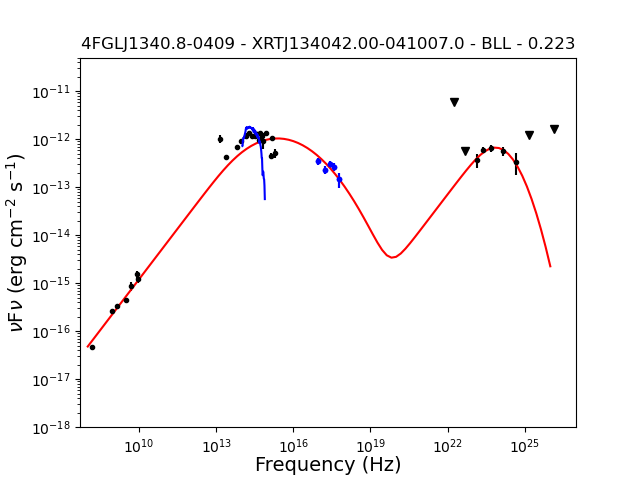}
\includegraphics[width=0.33\textwidth, angle=0]{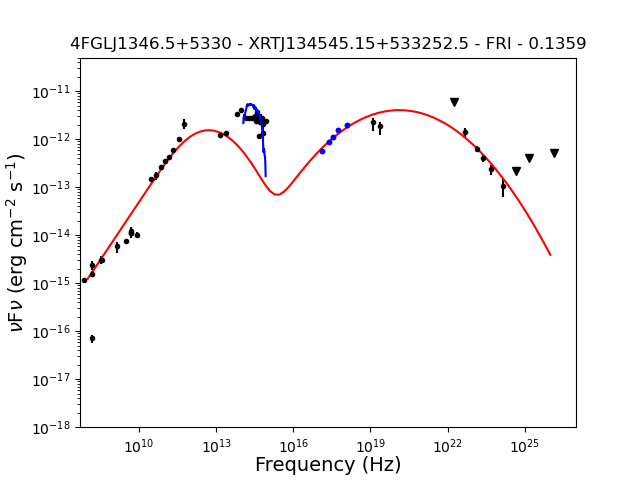}
\includegraphics[width=0.33\textwidth, angle=0]{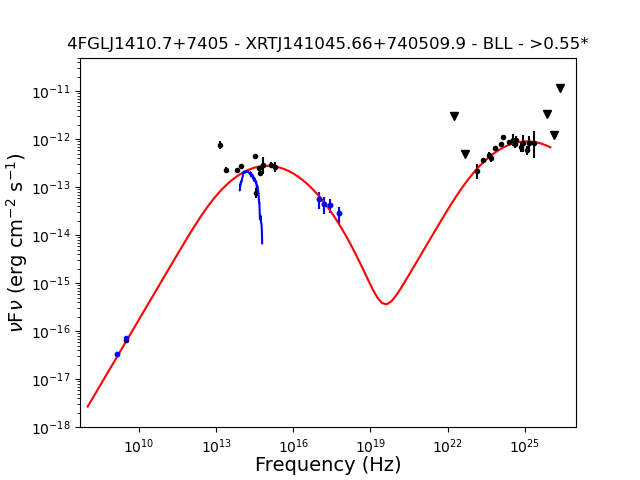}
\includegraphics[width=0.33\textwidth, angle=0]{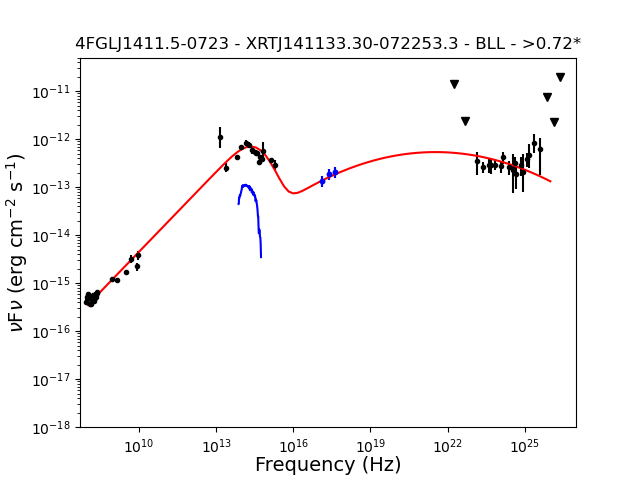}
\includegraphics[width=0.33\textwidth, angle=0]{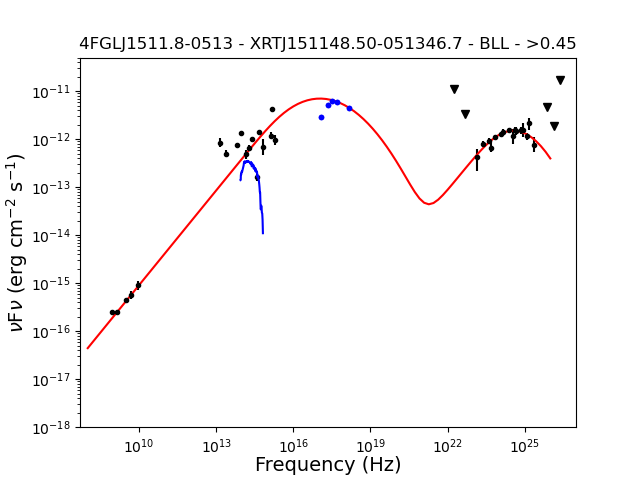}
\includegraphics[width=0.33\textwidth, angle=0]{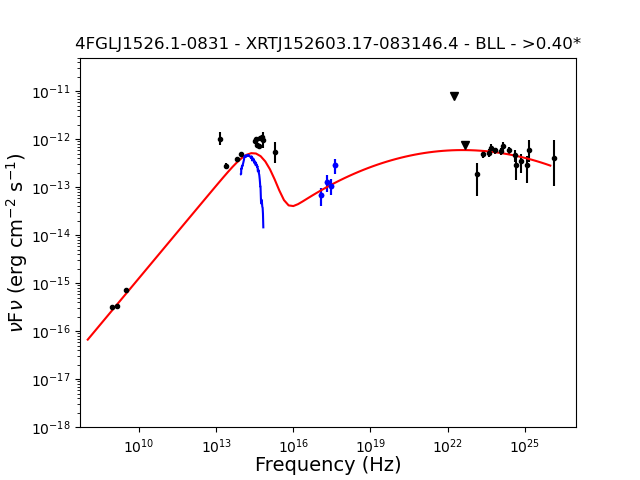}
\includegraphics[width=0.33\textwidth, angle=0]{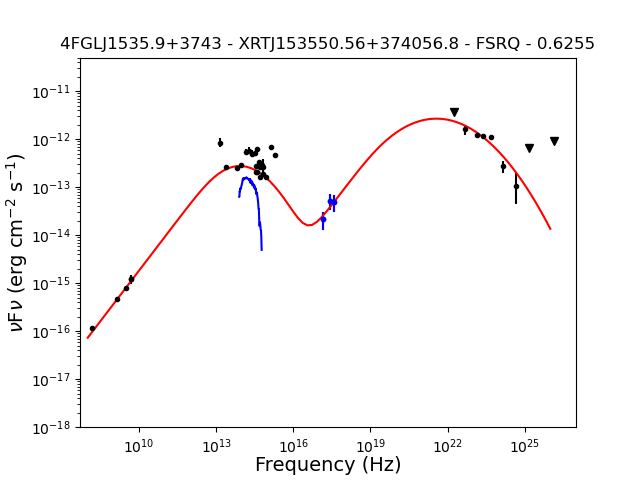}
\includegraphics[width=0.33\textwidth, angle=0]{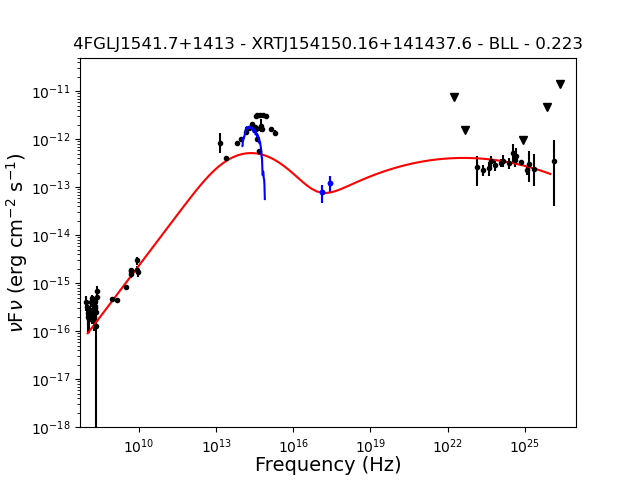}
\includegraphics[width=0.33\textwidth, angle=0]{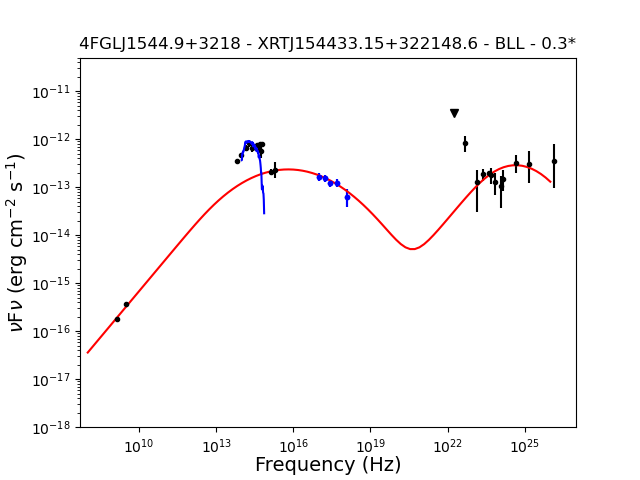}
\includegraphics[width=0.33\textwidth, angle=0]{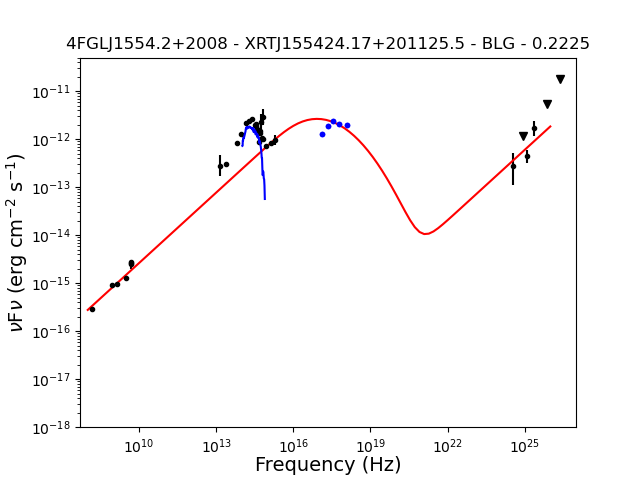}
\includegraphics[width=0.33\textwidth, angle=0]{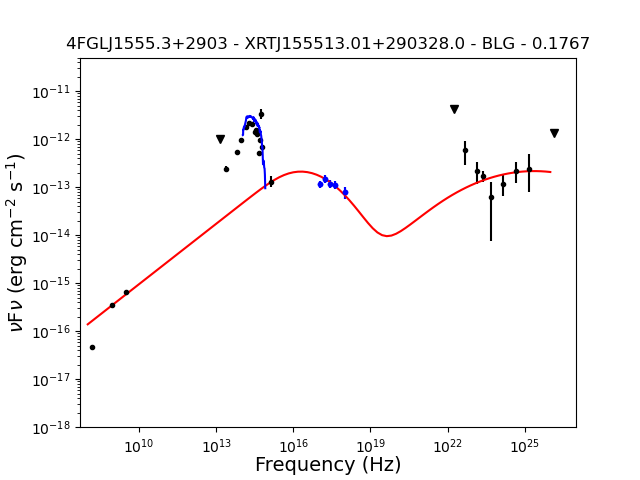}
\includegraphics[width=0.33\textwidth, angle=0]{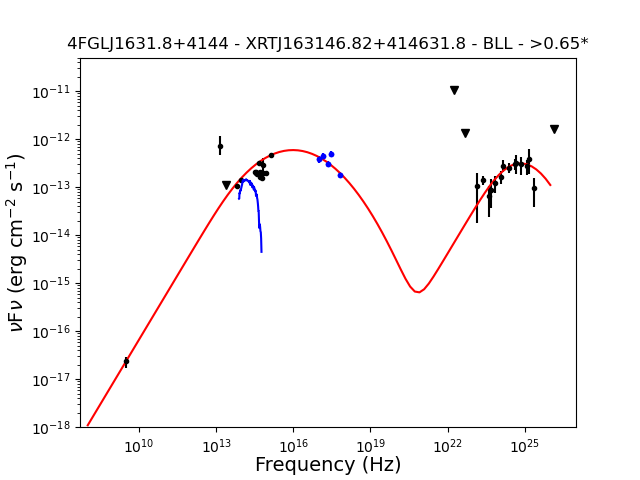}
\includegraphics[width=0.33\textwidth, angle=0]{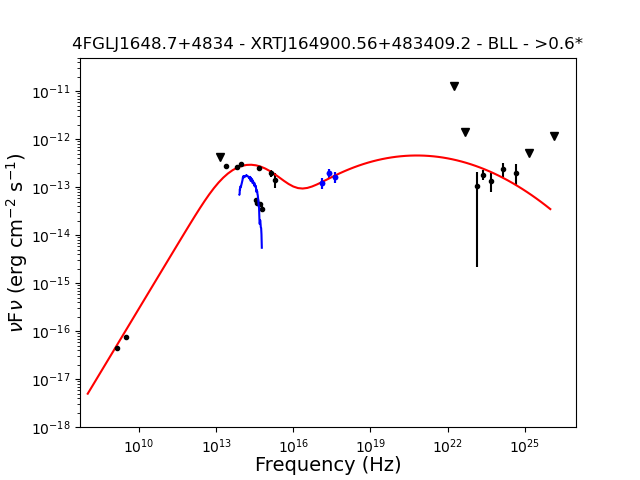}
\caption{Continued.}  
\label{fig:SED}
\end{figure*}

\setcounter{figure}{0}
\begin{figure*}
\center
\includegraphics[width=0.33\textwidth, angle=0]{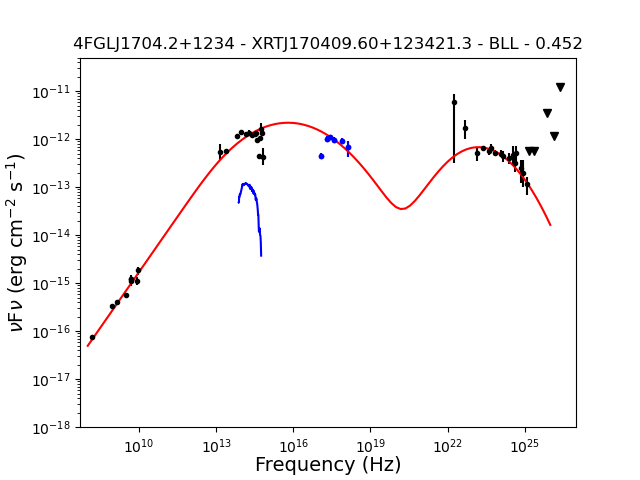}
\includegraphics[width=0.33\textwidth, angle=0]{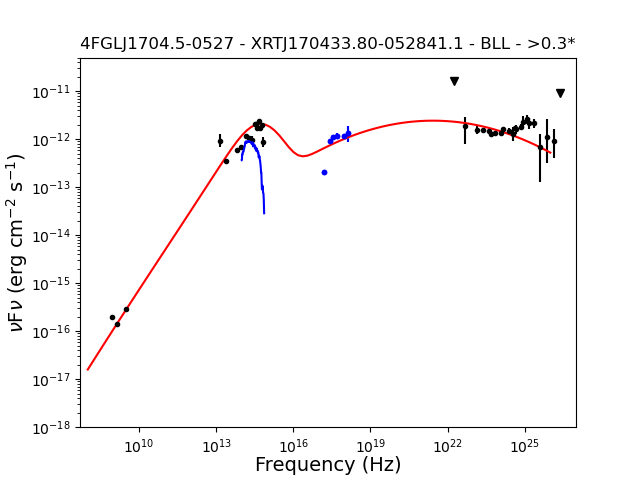}
\includegraphics[width=0.33\textwidth, angle=0]{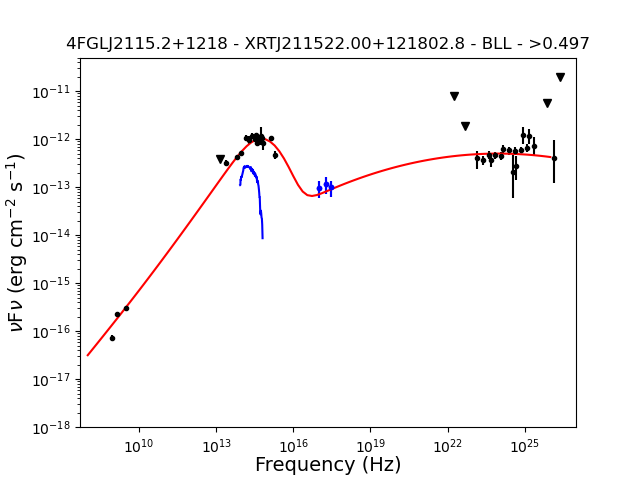}
\includegraphics[width=0.33\textwidth, angle=0]{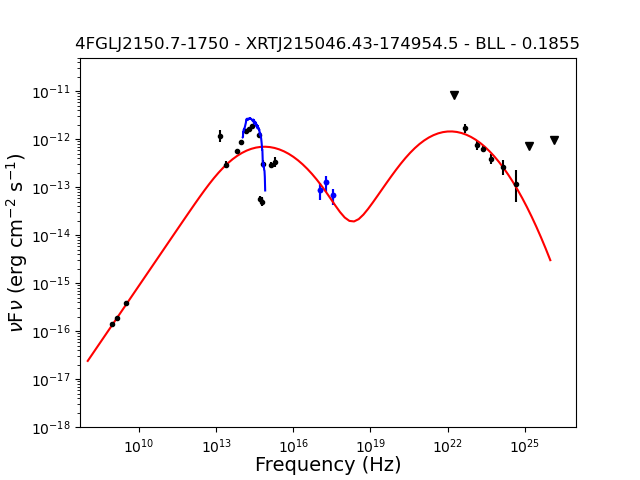}
\includegraphics[width=0.33\textwidth, angle=0]{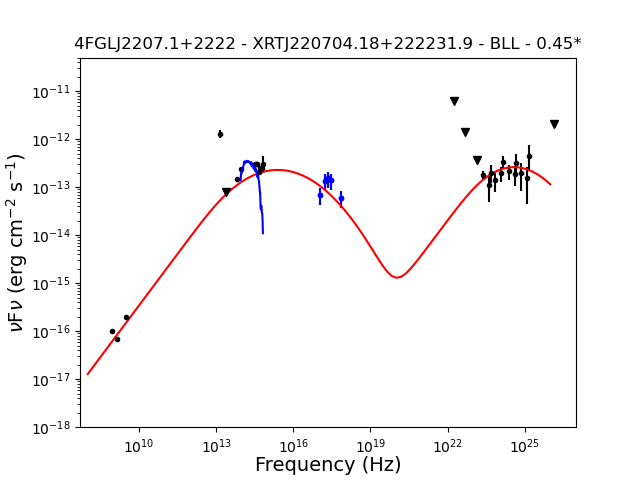}
\includegraphics[width=0.33\textwidth, angle=0]{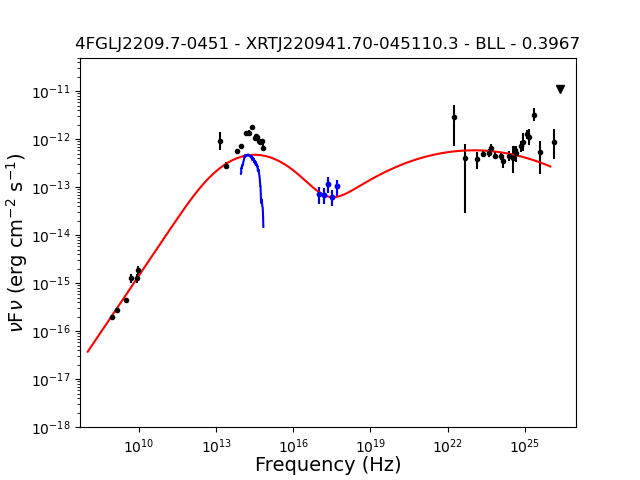}
\includegraphics[width=0.33\textwidth, angle=0]{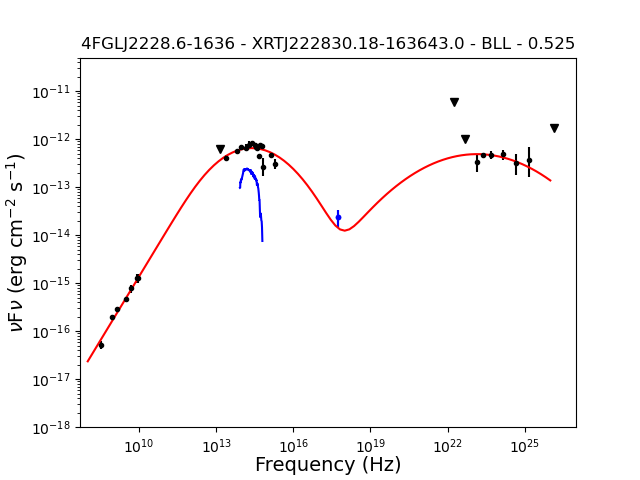}
\includegraphics[width=0.33\textwidth, angle=0]{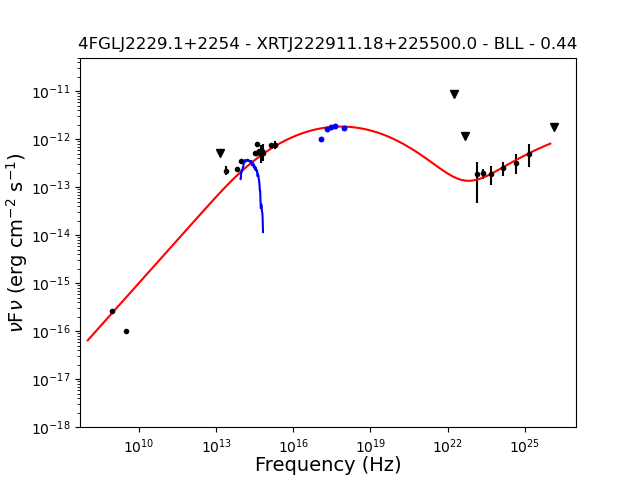}
\includegraphics[width=0.33\textwidth, angle=0]{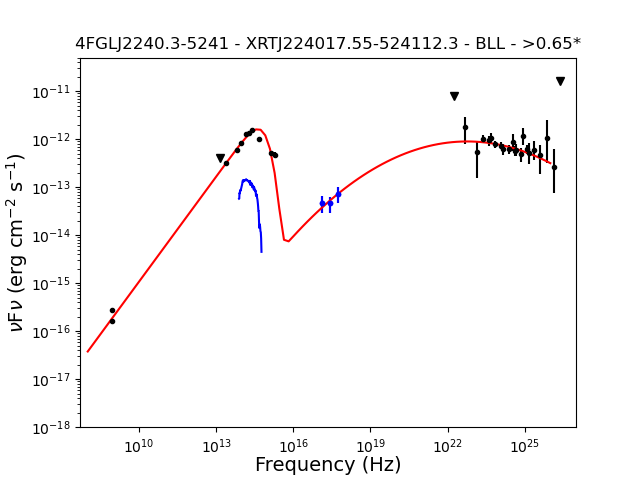}
\includegraphics[width=0.33\textwidth, angle=0]{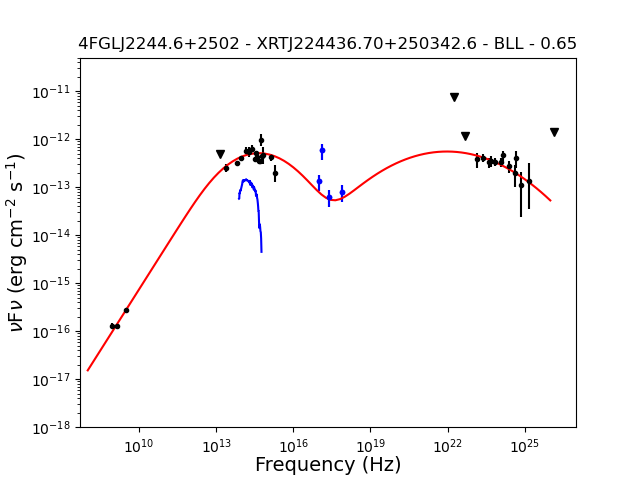}
\includegraphics[width=0.33\textwidth, angle=0]{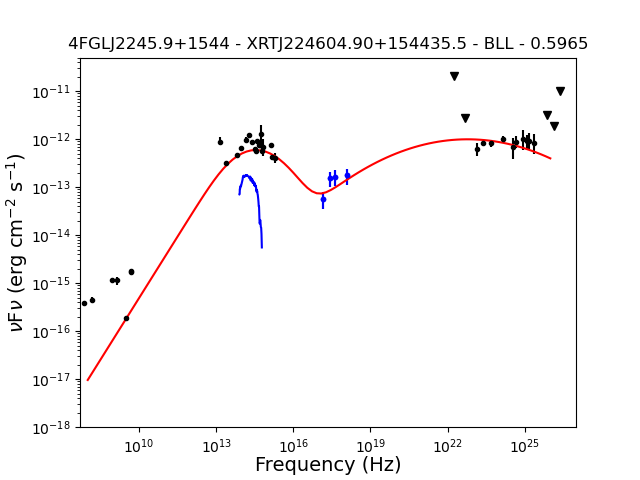}
\includegraphics[width=0.33\textwidth, angle=0]{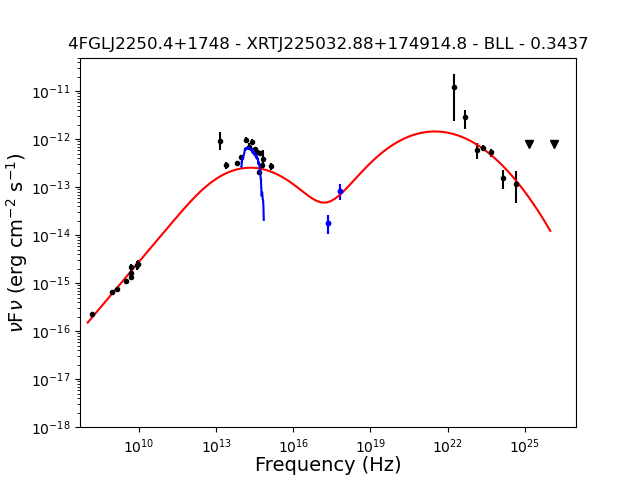}
\includegraphics[width=0.33\textwidth, angle=0]{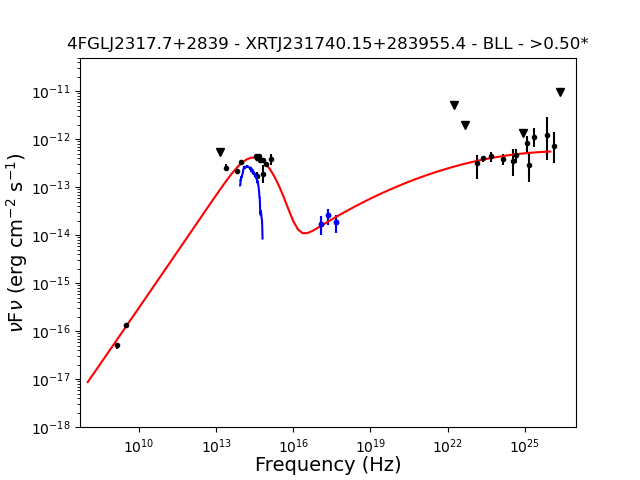}
\includegraphics[width=0.33\textwidth, angle=0]{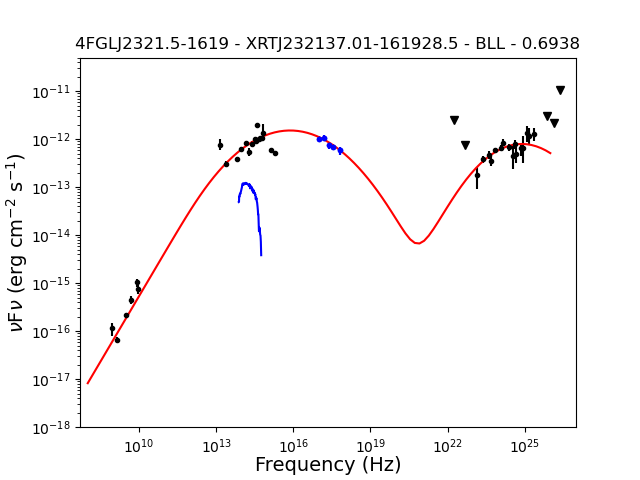}
\includegraphics[width=0.33\textwidth, angle=0]{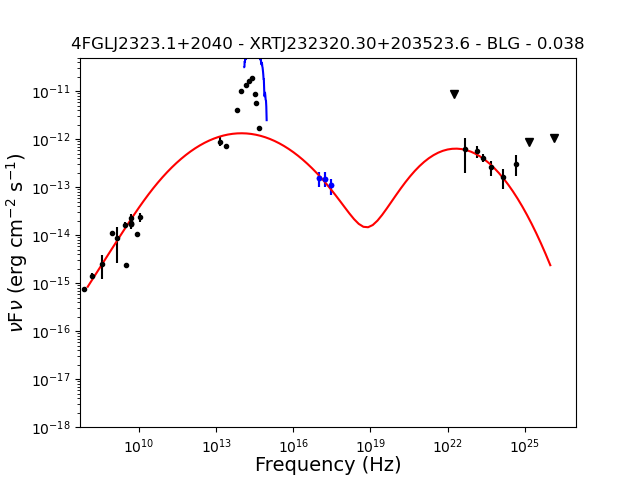}
\caption{Continued.}  
\label{fig:SED}
\end{figure*}

\setcounter{figure}{0}
\begin{figure*}
\center

\includegraphics[width=0.33\textwidth, angle=0]{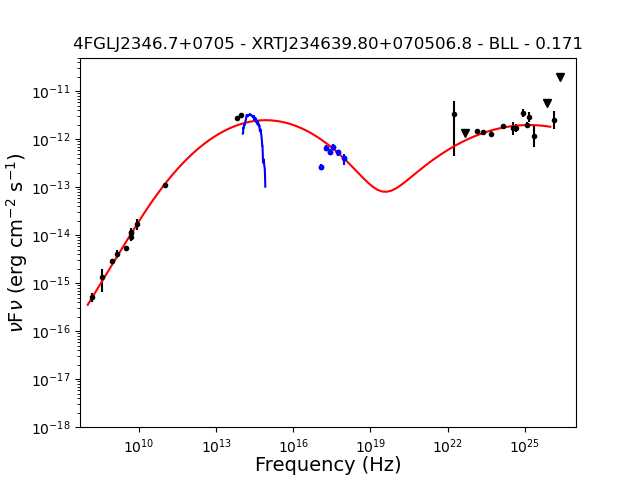}
\includegraphics[width=0.33\textwidth, angle=0]{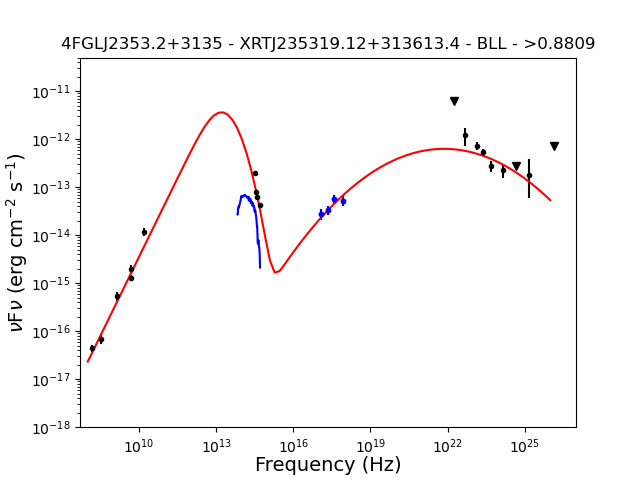}
\includegraphics[width=0.33\textwidth, angle=0]{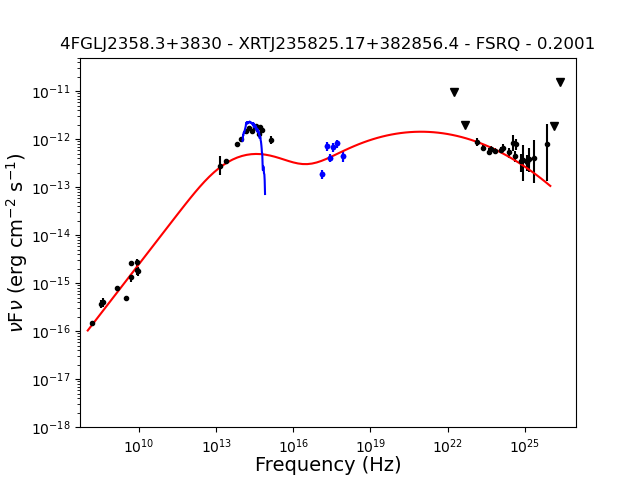}
\includegraphics[width=0.33\textwidth, angle=0]{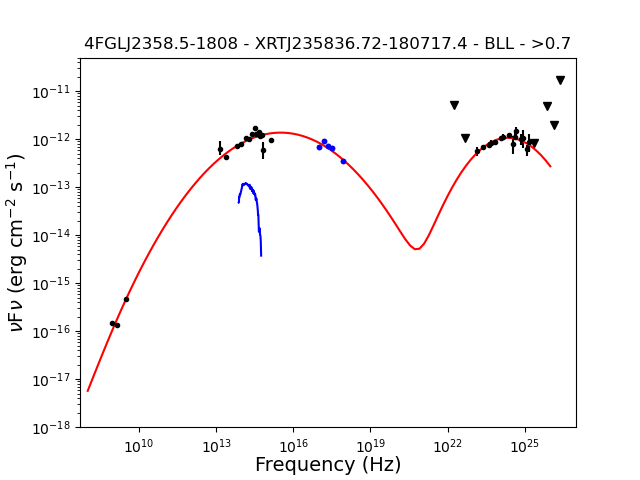}
\caption{Continued.}  
\label{fig:SED}
\end{figure*}

\setcounter{figure}{1}
\begin{figure*}
\center
\includegraphics[width=0.33\textwidth, angle=0]{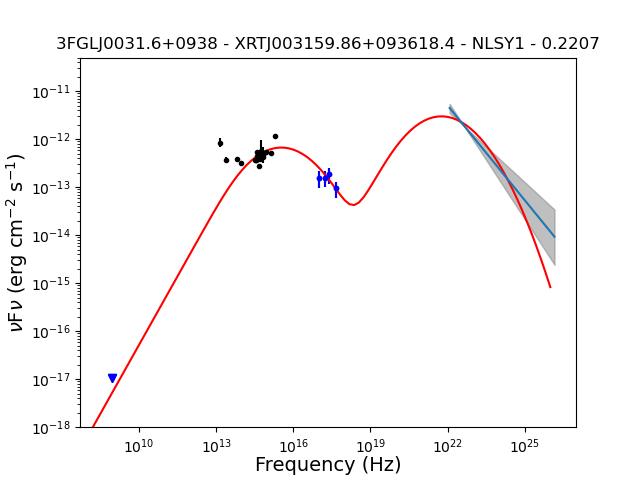}
\includegraphics[width=0.33\textwidth, angle=0]{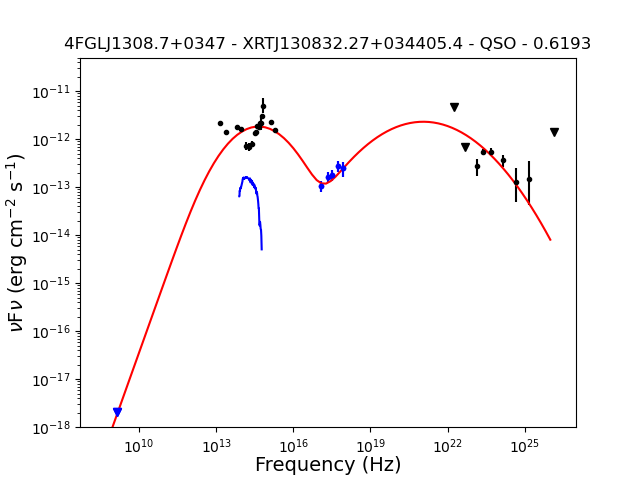}
\includegraphics[width=0.33\textwidth, angle=0]{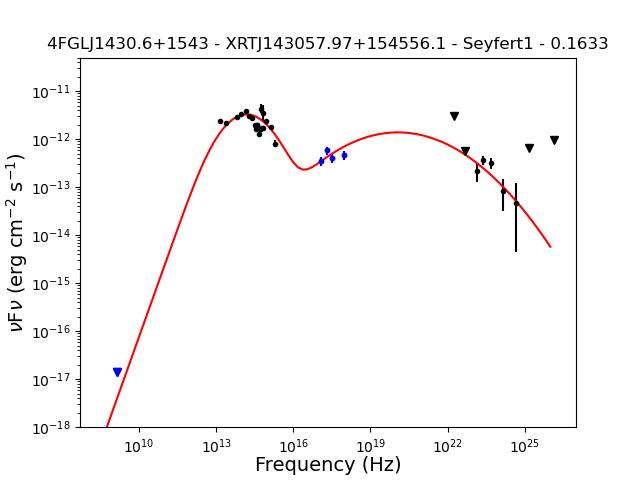}
\includegraphics[width=0.33\textwidth, angle=0]{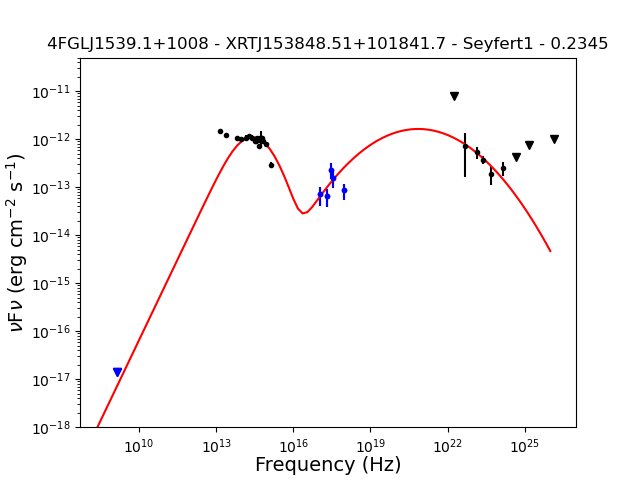}
\includegraphics[width=0.33\textwidth, angle=0]{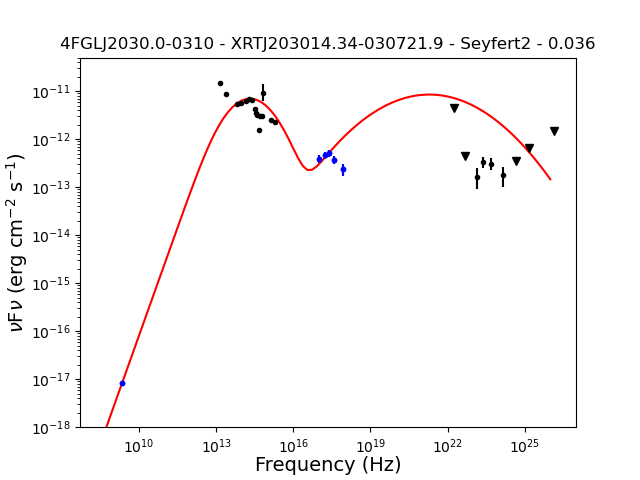}
\includegraphics[width=0.33\textwidth, angle=0]{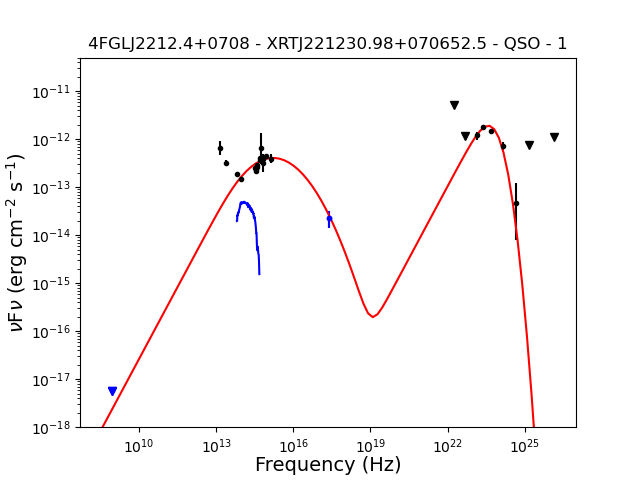}
\caption{Multi-wavelength SED of the UGS radio-quiet counterpart proposed in \citet[][]{Ulgiati_2024, paiano2017_ufo1, paiano2019_ufo2}. At the top of each figure, there is the UGS and counterparts name, the classification, and the redshift (if marked with a '*', the redshift is photometric, otherwise it is spectroscopic). Black points are data from VOU-Blazar, the blue points are X-ray data from our analysis \citep[][]{Ulgiati_2024b}. The triangle points indicates upper limits. The red curve emulates the typical double-peaked shape of blazars. The blue one is the template of a giant elliptical galaxy at the object redshift.}  
\label{fig:SED_RQ_noalt}
\end{figure*}

\setcounter{figure}{2}
\begin{figure*}
\center
\includegraphics[width=7.5truecm]{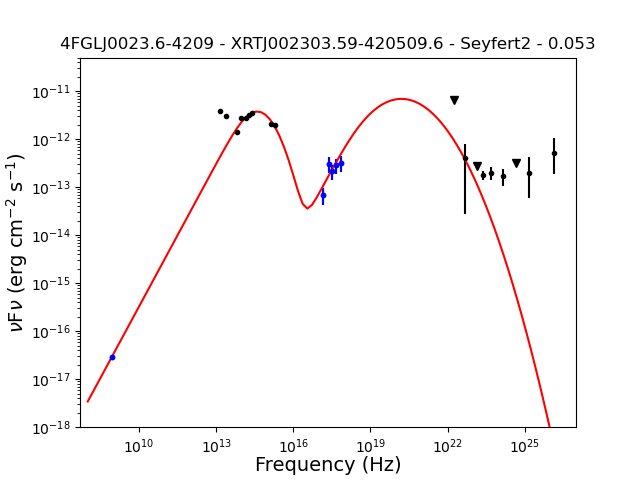}
\includegraphics[width=7.5truecm]{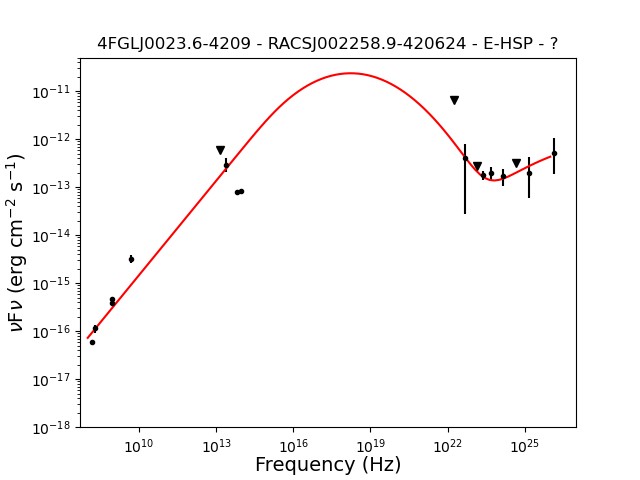}

\includegraphics[width=7.5truecm]{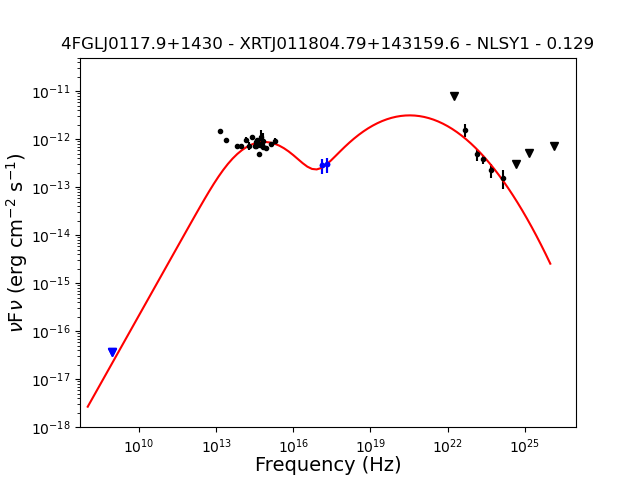}
\includegraphics[width=7.5truecm]{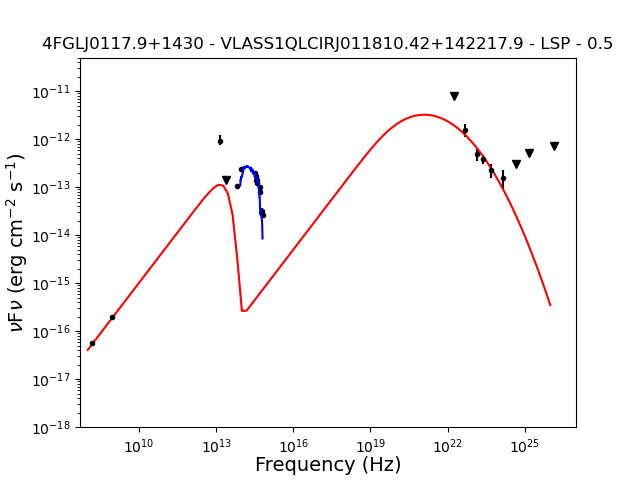}
\includegraphics[width=7.5truecm]{SED_3/SED_of_spettro_di_XRTJ064111p24+334502p0.png}
\includegraphics[width=7.5truecm]{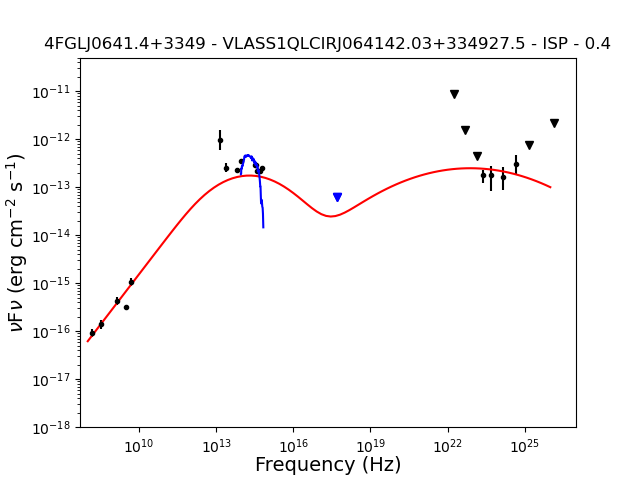}
\includegraphics[width=7.5truecm]{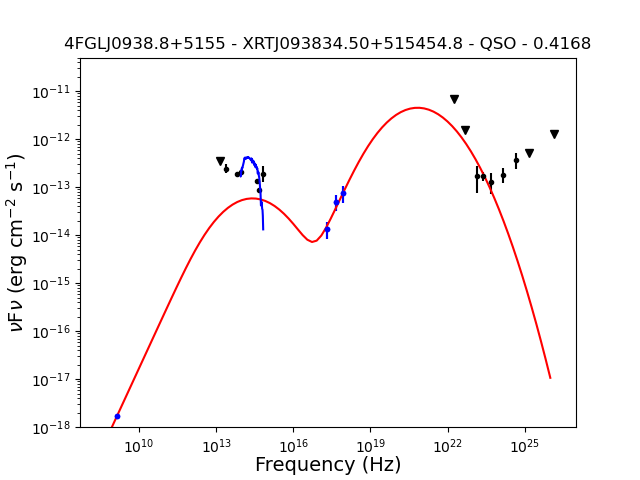}
\includegraphics[width=7.5truecm]{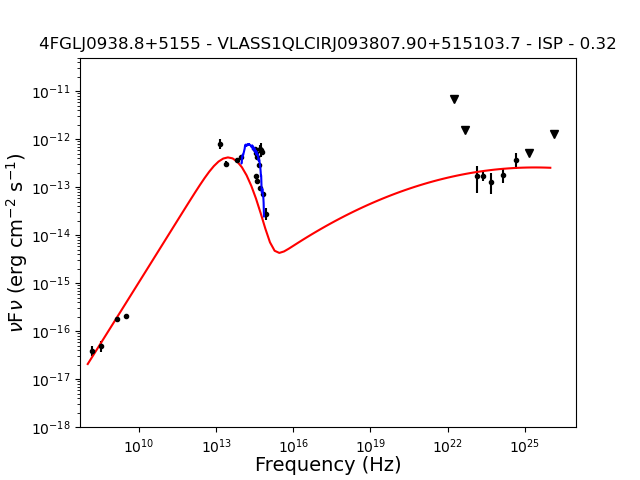}
\caption{Multi-wavelength SED of the UGS radio-quiet counterpart proposed in \citet[][]{Ulgiati_2024, paiano2017_ufo1, paiano2019_ufo2} (\textit{left panels}) and of their "alternative" counterparts found in this work (\textit{right panels}). At the top of each figure, there is the UGS and counterparts name, the classification, and the redshift (if marked with a '*', the redshift is photometric, otherwise it is spectroscopic). Black points are data from VOU-Blazar, the blue points are X-ray data from our analysis \citep[][]{Ulgiati_2024b}. The triangle points indicates upper limits. The red curve emulates the typical double-peaked shape of blazars. The blue one is the template of a giant elliptical galaxy at the object redshift.}  
\label{fig:SED_alt_all}
\end{figure*}

\setcounter{figure}{3}
\begin{figure*}
\center
\includegraphics[width=7.5truecm]{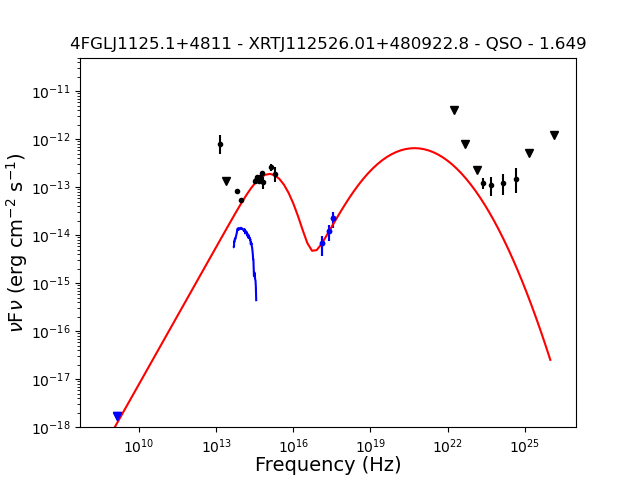}
\includegraphics[width=7.5truecm]{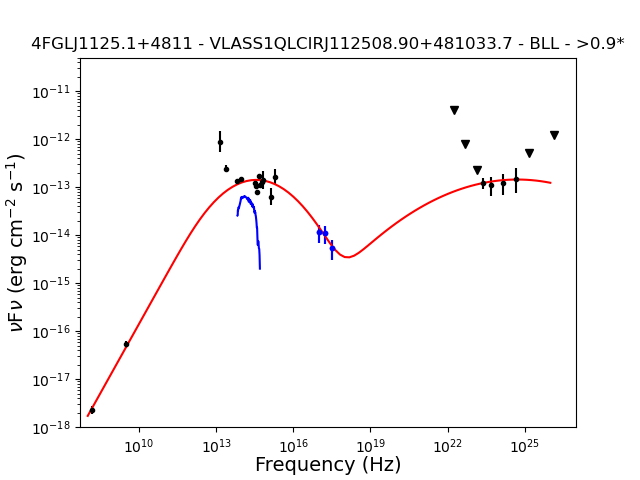}
\includegraphics[width=7.5truecm]{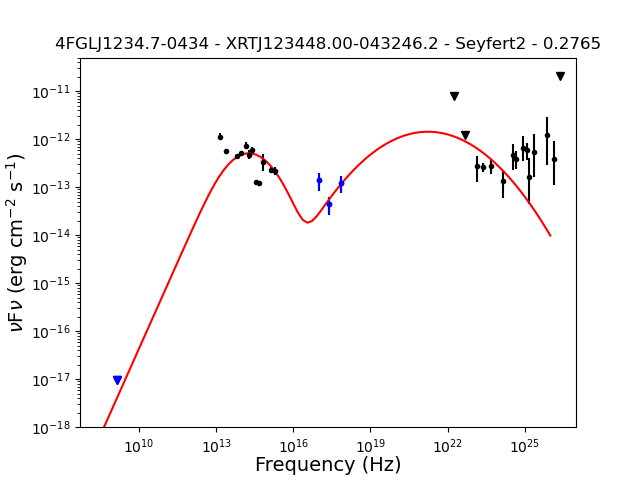}
\includegraphics[width=7.5truecm]{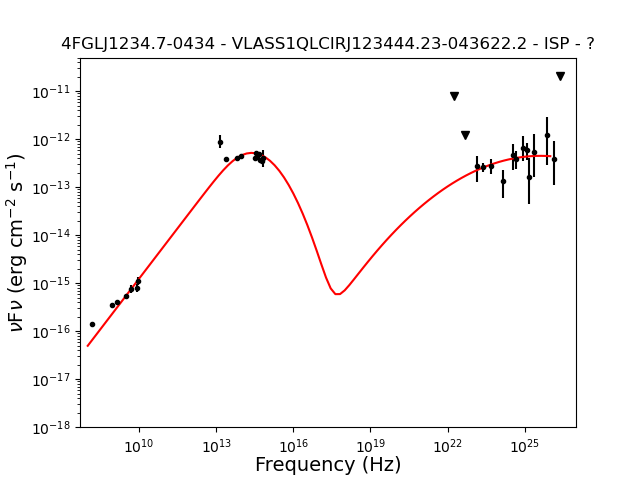}
\includegraphics[width=7.5truecm]{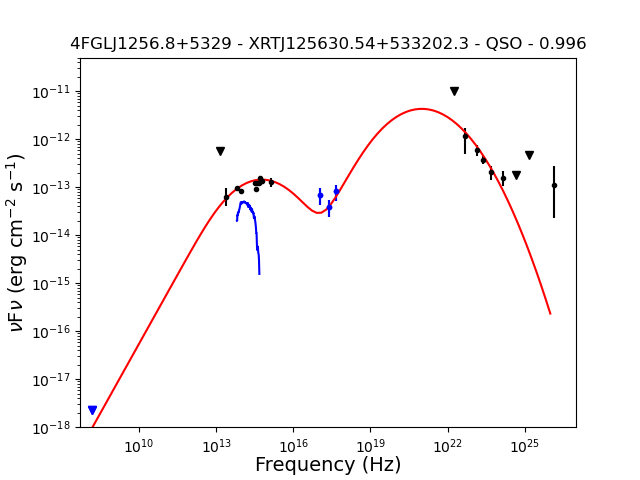}
\includegraphics[width=7.5truecm]{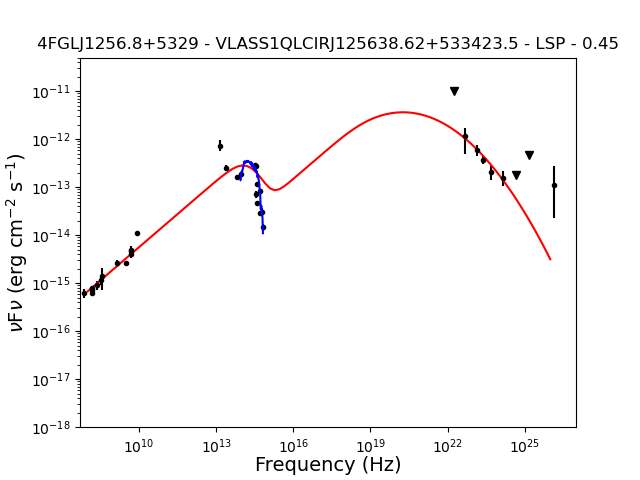}
\caption*{Fig: \ref{fig:SED_alt_all}: Continued.}  
\end{figure*}

%
%



\bsp	
\label{lastpage}
\end{document}